\documentclass{aa}  

\usepackage{graphicx, xcolor}
\usepackage[utf8]{inputenc}
\usepackage{newunicodechar}
\newunicodechar{⁻}{\textsuperscript{-}}
\usepackage{txfonts}
\usepackage{caption}
\usepackage{textcomp}
\usepackage{orcidlink}
\usepackage{natbib}
\usepackage{booktabs}
\usepackage{amsmath}
\usepackage{setspace}
\usepackage{hyperref}
\usepackage{subcaption}
\hypersetup{
    colorlinks=true,
    citecolor=black,
    linkcolor=blue,
    filecolor=blue,      
    urlcolor=blue
    }
\begin{document} 

   \title{A homogeneous three-dimensional view of molecular cloud kinematics out to 2.5 kpc}
   \subtitle{Using young stellar objects and open clusters as complementary tracers}

    \author{X. P\'erez-Couto\inst{1,2}\,\orcidlink{0000-0001-5797-252X}
    \and
          S. Torres\inst{3,4,5}\,\orcidlink{0000-0002-3150-8988}
          \and
          N. Miret-Roig\inst{6, 7}\,\orcidlink{0000-0001-5292-0421}
          \and
          F. Anders\inst{6, 7, 8}\,\orcidlink{0000-0003-4524-9363}
          \and
          E. Villaver\inst{4, 5}\,\orcidlink{0000-0003-4936-9418}
          \and
          A. J. Mustill\inst{9}\,\orcidlink{0000-0002-2086-3642}
          \and
          M. Manteiga\inst{2, 10}\,\orcidlink{0000-0002-7711-5581}
                    }

    \institute{Universidade da Coruña (UDC), Department of Computer Science and Information Technologies, Campus de Elviña s/n, 15071, A Coruña, Galiza, Spain
         \and
        CIGUS CITIC, Centre for Information and Communications Technologies Research, Universidade da Coruña, Campus de Elviña s/n, 15071 A Coruña, Galiza, Spain
        \and
        Institute of Science and Technology Austria (ISTA), Am Campus 1, 3400 Klosterneuburg, Austria
        \and
        Instituto de Astrofísica de Canarias, 38200 La Laguna, Tenerife, Spain
        \and 
        Universidad de La Laguna (ULL), Astrophysics Department, 38206 La Laguna, Tenerife, Spain
        \and
        Departament de Física Quàntica i Astrofísica (FQA), Universitat de Barcelona (UB), Martí i Franquès 1, 08028 Barcelona, Spain
        \and
        Institut de Ciències del Cosmos (ICCUB), Universitat de Barcelona (UB), Martí i Franquès 1, 08028 Barcelona, Spain
        \and
        Institut d’Estudis Espacials de Catalunya (IEEC), Edifici RDIT, Campus UPC, 08860 Castelldefels (Barcelona), Spain
        \and
        Lund Observatory, Division of Astrophysics, Department of Physics, Lund University, Box 118, 22100, Lund, Sweden
        \and 
        Universidade da Coruña (UDC), Department of Nautical Sciences and Marine Engineering, Paseo de Ronda 51, 15011, A Coruña, Galiza, Spain
        }

   \date{Received date / Accepted date }

 
  \abstract
   {Understanding the large-scale dynamics of molecular clouds (MCs) is crucial to constraining the processes that govern star formation and the structure and evolution of the Galaxy. While gas tracers have traditionally been used to map MC kinematics, stellar tracers such as young stellar objects (YSOs) and open clusters (OCs) provide a complementary approach that enables direct comparisons between the stellar and gaseous components.}
   {{We aim to {compare the kinematics of young ($\leq 30$~Myr) OCs with YSOs in the same star-forming region, the latter being assumed to trace the motions of the parent cloud gas}. Subsequently, we intend to reconstruct the three-dimensional (3D) motions of the main MC complexes within 2.5 kpc of the Sun using both YSOs and young OCs as tracers.}}
   {Using \textit{Gaia} Data Release 3 (DR3) astrometry and complementary spectroscopic surveys providing radial velocities (RVs), we compiled a unified sample of $24,146$ {young stars}, comprising YSOs and OC members. We applied robust clustering in proper-motion space to identify co-moving YSOs and derived cloud-averaged motions via Monte Carlo sampling. These motions were compared with the kinematics inferred from OCs younger than 30 Myr. Finally, we performed orbital integrations in a realistic Galactic potential to trace the past evolution of the clouds and to quantify their expansion and rotation.}
   {We derive homogeneous 3D kinematics for 15 star-forming regions within 2.5 kpc. YSOs and OCs exhibit strongly consistent kinematics {within each cloud}, {with a median spatial velocity offset of $\lesssim  2$ km s$^{-1}$, allowing us to combine them into a unified tracer sample.} {The resulting bulk cloud motions show a median peculiar velocity of {$\simeq 9$ km s$^{-1}$} with respect to the Galactic rotation.} We used our catalogue to trace back the Solar System's voyage through the Orion cloud and the common origin of Lupus, Ophiuchus, and Corona Australis in the Sco-Cen region. {Internally, we detected significant expansion in Orion and Ophiuchus (at a $5\sigma$ level) and coherent rotation in at least six complexes, likely reflecting local processes such as stellar feedback and angular-momentum redistribution}.}
   {{Young OCs and YSOs exhibit broadly consistent kinematics, and therefore OCs can also be used to trace the bulk motion of the star-forming regions}}
 


   \maketitle
%

\section{Introduction}

Molecular clouds (MCs) are cold, dense reservoirs of interstellar gas and dust where stars and planets are born. Understanding their three-dimensional (3D) structure and dynamics, which are influenced by processes such as supersonic turbulence, gravitational instability, magnetic fields and stellar feedback, is crucial for elucidating the mechanisms underlying star formation and galactic evolution \citep{Larson1981, Krumholz2011, Andre2014}.

Until recently, our understanding of the 3D structure of MCs was limited due to the difficulty of measuring reliable stellar distances, which are necessary to determine the extinction profile along a sight line (e.g. \citealt{Wolf1923, Wolf1924, Uranova1962, Bok1977}). Therefore, only a 2D (position--position) or a pseudo-3D (position--position--velocity) map of each MC was available \citep[e.g.][]{GregorioHetem1988, Lada2009, Rice2016}. Before the advent of high-precision astrometry, the radio emission of the gas was used to determine the radial velocity (RV) of several clouds in the literature \citep{Dame2001, Kohno2018, Soler2023}, but recovering the tangential components of the motion required point-like sources that preserved the proper motion of their birth cloud.

Indeed, it is generally accepted that Class II YSOs or earlier preserve the positions and overall motions of the gas in their parental clouds \citep[see e.g.][]{Heiderman2010, Hacar2016, Grossschedl2019, Grosschedl2021}, even under the influence of outflows, nearby massive stars, and supernova feedback \citep{Yang2025}.

With the launch of the \textit{Gaia} space mission \citep{Gaia_Collaboration2016, Gaia_Collaboration2023}, parallaxes for nearly 2 billion stars have been made available to the community, enabling the determination of precise distances \citep{Yan2019, Zucker2019, Zucker2020, Zhang2023} and true 3D shapes for most of the MCs in the solar neighbourhood \citep[$\lesssim 440$~pc,][]{Cahlon2024} as well as for a limited number at larger distances \citep{Chen2020, Dharmawardena2023}.

Building on this idea, several works have leveraged the high-quality \textit{Gaia} parallaxes, proper motions, and RVs to infer the kinematics of a few MCs from their young inhabitant stellar populations. \citet{Galli2019}, \citet{Grosschedl2021}, and \citet{Dong2024} used young stellar objects (YSOs) as tracers to reconstruct the internal kinematics of Taurus, Orion, and Canis Major complexes, respectively; while \citet{Lim2021} and \citet{Kounkel2022} combined YSOs and OB stars to study the motions of the Rosette Nebula and the Perseus--California clouds. Most recently, \citet{Zhou2025} used 102 YSO co-moving groups with 2D kinematics that are fully or partially associated with $^{12} \rm{CO}$ gas to provide 3D motions for the well-known Taurus, Perseus, Ophiuchus, Chamaeleon, Lupus, and Orion MCs.

Other authors have used young ($\leq 30$ Myr) open cluster (OC) catalogues and their dynamics to extrapolate the coherent motions of their parent clouds for larger scale kinematic studies. For instance, \citet{2024Natur.628...62K} used young OCs to model an oscillatory vertical motion in the Radcliffe Wave structure \citep{Alves2020}, while \citet{Maconi2025} performed orbital traceback analysis to show that the Solar System passed through the structure between 18.2 and 11.5 Myr ago. However, since OCs usually clean their natal cloud gas within a few megayears \citep[see e.g.][]{Lada2003}, the correlation between their kinematics and the current MCs is not straightforward and requires a dedicated study. Furthermore, despite significant progress in mapping the structure and properties of MCs, a unified catalogue that provides their full 3D kinematics is still lacking. To fill this gap, in this work we used Class I/II YSOs as primary kinematic tracers of the major molecular clouds out to 2.5 kpc, building on the well-established assumption that they preserve the bulk motions of their parent cloud gas. Independently, we derived cloud motions from young OCs ($\leq 30$~Myr) located in the same regions and compared them with the YSO kinematics. The aim is therefore to establish whether young OCs exhibit bulk kinematics consistent with the YSOs identified in the same region. If such a consistency was found, we combined both populations into a single, larger tracer sample for orbital reconstruction and internal kinematic analysis.

The paper is structured as follows. Section \ref{sec:data} describes the data used for both MCs and for their tracers and the methodology used to cluster and obtain mean motions for each cloud. Section \ref{sec:results} presents and discusses the results of this study. Finally, in Sect. \ref{sec:conclusions} we present a summary of conclusions of the paper.

\section{Data and methodology}\label{sec:data}

\subsection{Cloud boundaries}\label{molecularclouds_sample}

To obtain the physical three-dimensional (3D) extent of MCs, we used the molecular cloud catalogue of \citet{Dharmawardena2023}, {which provides spatial boundaries in Galactic coordinates and distance for 16 MCs within 2.5~kpc}, with a spatial resolution of 1 pc. To derive those 3D contours, the authors used the \texttt{Dustribution} algorithm \citep{Dharmawardena2022} and the {\tt astrodendro} package\footnote{\url{https://dendrograms.readthedocs.io/en/stable/}}, in combination with {\it Gaia} DR3 distances and interstellar extinctions \citep{Fouesneau2023}.  For our purposes, the extended environment of each cloud is defined as a trapezoidal region bounded by lower and upper limits in each dimension: galactic longitude ($l$), galactic latitude ($b$), and distance to the Sun ($d$). 

Because the well-known cloud complexes of Ophiuchus and Corona Australis are not included in this catalogue, we supplemented their cloud contours using the data from \citet{Zucker2020}. In that work, the authors provided extinction sight lines and derived posterior distance estimates for the clouds listed in the Star Formation Handbook \citep{Reipurth2008a, Reipurth2008b}, following the methodology described in \citet{Zucker2019}. 
{We therefore set the angular limits for Ophiuchus and Corona Australis from their extreme sight-line coordinates and extended them by twice the mean sight-line spacing in each coordinate separately to account for the extended environment of the complexes.} Similarly, the limits in $d$ were taken as the $2 \sigma$ confidence interval of the posterior distributions provided in that work. {As a result, Ophiuchus spans $\left[96,       177      \right]$~pc and Corona Australis $\left[137, 172 \right]$~pc. By dividing the depth ($d_{\rm max} - d_{\rm min}$) obtained with this procedure by that provided in \citet{Dharmawardena2023} for the remaining 13 clouds, we find a median ratio of $1.1_{-0.6}^{+0.1}$ (16--84$^{\rm th}$ percentiles). Therefore, we consider our approximation to be reasonable, and it will be used for Ophiuchus and Corona Australis throughout the rest of this work.}

To determine whether a source lies within a MC, we used the coordinates $l$ and $b$ and distances $d$ \citep[computed with the estimator of][]{Weiler2025} for each YSO and OC member. A source was considered to belong to a given cloud if its coordinates satisfied the condition $l_{\rm min} < l < l_{\rm max}$, where $l_{\rm min}$ and $l_{\rm max}$ denote the lower and upper galactic longitude limits of that cloud; analogous criteria were applied for $b$ and $d$).

\begin{figure*}
    \centering
    \includegraphics[width=0.95\textwidth]{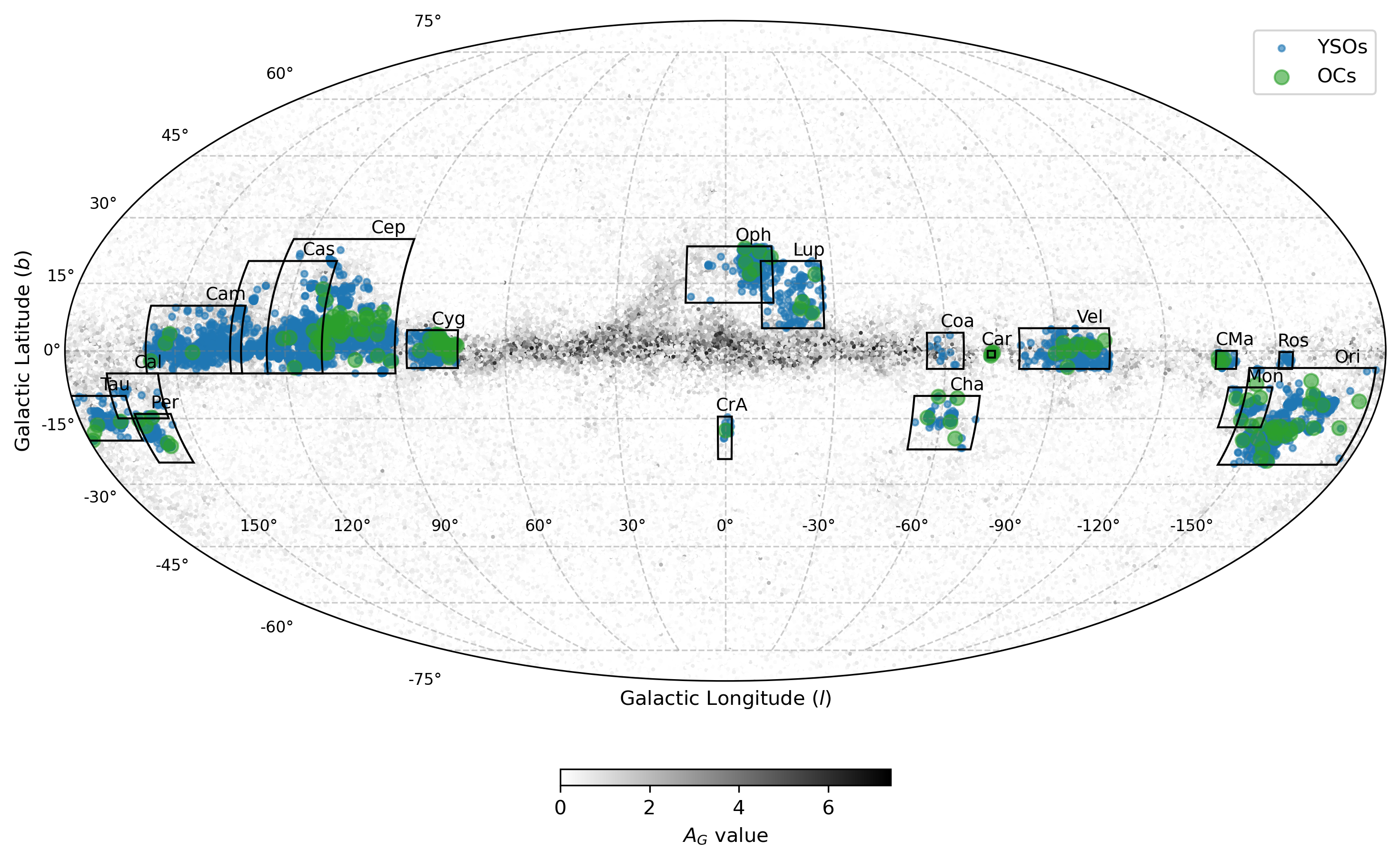}
    \caption{\label{fig:dharma_map} Galactic map of the molecular clouds from \citet{Dharmawardena2023} and \citet{Zucker2020} used in this study and the YSOs (blue) and OCs (green) identified within each cloud; these are overlaid on a reference $A_G$ extinction map derived with GSP-Phot \citep{Andrae2023} from $10^7$ randomly selected \textit{Gaia} DR3 sources.}
\end{figure*}

Figure \ref{fig:dharma_map} presents the outcome of the membership selection showing YSOs (blue) and open cluster (green) members associated with each MC. These sources are overplotted on a reference extinction map $A_G$ computed via the Generalized Stellar Parametrizer from Photometry (GSP-Phot; \citealt{Andrae2023}) for a random selection of $10^7$ {\it Gaia} DR3 sources.

\subsection{The YSO sample}\label{subsec:yso_sample}

For the purposes of this study, we used several catalogues of YSO and YSO candidates; hereafter, the term `YSO' refers to both. First, the catalogue of \citet{Zhang2023} contains nearly 25,000 Class I/II YSO candidates identified using the classification scheme of \citet{2014ApJ...791..131K} combined with the infrared excesses in the AllWISE catalogue, and further incorporating the previous YSO catalogues of \citet{Marton2016, Marton2019}. Second, the Konkoly Optical YSO (KYSO) catalogue presented by \citet{Marton2023} comprises about 12,000 optically detected and spectroscopically confirmed YSOs and was compiled through an extensive literature search. {We discarded YSOs classified as weak-line T Tauri stars and post-T Tauri stars (WTT and PTT), and thus Class III YSOs, retaining 7384 Class I/II objects.} Third, we included 660 Class I/II YSOs out of the approximately 1,000 YSOs compiled in \citet{Yang2025}. These YSOs were selected for a kinematic study of the three major molecular cloud complexes: Orion, Perseus, and Taurus. Finally, whenever possible, we supplemented the list of {Class I/II YSOs (hereafter, `YSOs' for simplicity)} in clouds with limited numbers of YSOs using three additional catalogues: \citet{Dong2024} for Canis Major, \citet{Zeidler2016} for the Carina Nebula, and \citet{Cambresy2013} for the Rosette Nebula.


We used the limits $l$, $b$, and $d$ of each cloud as derived in Section \ref{molecularclouds_sample} to determine YSO membership. To ensure the use of YSOs with optimal astrometry, we corrected the proper motions following \citet{Cantat-Gaudin2021} and the parallax using the zero point of \citet{Lindegren2021} (through the \texttt{get\_zpt()} function of its corresponding library\footnote{\url{https://pypi.org/project/gaiadr3-zeropoint/}}). Subsequently, we inflated the parallax errors following \citet{MaizApellaniz2022} and exclusively preserved YSOs with a positive parallax and a S/N>3. While this relative error in the parallax can be considered too high to obtain accurate distances, we are using the formalism of \citet{Weiler2025}, implemented through the \texttt{weiler2025} library\footnote{\url{https://github.com/fjaellet/weiler2025/}}, which provides a very good estimator for the true distances even at  { parallax signal-to-noise levels of S/N$ \simeq 2$}. Our kinematic analysis is based on a final sample of {9722 YSOs.} 

We supplemented this sample with RVs from the {\it Gaia} DR3 Radial Velocity Spectrometer \citep[\textit{Gaia RVS,}][]{Katz2023}, with APOGEE-2 DR17 \citep{Abdurrouf2022}, LAMOST DR11, \citep{Luo2026}, GALAH DR4 \citep{Buder2025}, RAVE DR6 \citep{Steinmetz2020}, and the \textit{Gaia}-ESO DR5 surveys \citep{Hourihane2023}. We note that it is important to consider the systematics between these surveys before combining them. The median offset between the different RV surveys used in this work ranges from 0.04 to 4.97~km~s$^{-1}$ \citep[see][for a further treatment of these systematics]{Tsantaki2022}. We excluded RV values with errors above $5$ km s$^{-1}$, and when a source showed multiple RV measurements we retained the one with the smallest uncertainty only if its value is within $5$ km s$^{-1}$ of the median RV for that source. Otherwise, we preserved the {\it Gaia} RV. These RV cuts keep the offsets between surveys below the maximum allowed RV uncertainty {and below the inter-survey threshold stated above.}

In the end, {we preserved 2326 RVs}. The final number of YSOs ($N_\textup{YSO}$) and YSOs with available RV ($N_\textup{YSO, RV}$) per cloud is summarised in Table \ref{ref:table_N}.

\begin{table}
\setlength{\tabcolsep}{5pt}
\caption{\label{ref:table_N} Number of YSOs ($N_{\textup{YSO}}$), YSO with available RVs ($N_{\textup{YSO,RV}}$), OCs ($N_{\textup{OC}}$), OC members ($N_{\rm OC_{members}}$), and OC members with available RV ($N_{\rm OC_{members}, RV}$) per each cloud.}
{\renewcommand{\arraystretch}{1.15}
{\fontsize{7}{9}\selectfont
    \centering
    \begin{tabular}{lccccccc}
    \toprule
\textbf{Complex} & $N_{\rm  YSO}$ & $N_{\rm  YSO, RV}$ & $N_{\rm  OC}$ & $N_{\rm  OC_{members}}$ & $N_{\rm  OC_{members}, RV}$ \\ 
\midrule
\textbf{California} [Cal]      &    32  &          10   &        -        &            -    &       -  \\
\textbf{Camelopardalis} [Cam]  &   765  &           4   &        4        &          253    &      14  \\
\textbf{Canis Major} [CMa]     &   123  &           -   &        4        &          600    &       7  \\
\textbf{Carina} [Car]          &    75  &          21   &        4        &          435    &      21  \\
\textbf{Cassiopeia} [Cas]      &   364  &           2   &        -        &            -    &       -  \\
\textbf{Cepheus} [Cep]         &  2565  &          24   &       45        &         4875    &     293  \\
\textbf{Chamaeleon} [Cha]       &   153 &         54    &        5        &          557    &     137  \\
\textbf{Coalsack} [Coa]        &    11  &           1   &        -        &            -    &       -  \\
\textbf{Corona Australis} [CrA] &    37 &          21   &        1        &          222    &      56  \\
\textbf{Cygnus} [Cyg]          &   476  &          63   &       21        &         1426    &      48  \\
\textbf{Lupus} [Lup]           &   179  &          24   &        4        &          262    &      75  \\
\textbf{MonR2} [Mon]           &   107  &           5   &        2        &           72    &       -  \\
\textbf{Ophiuchus} [Oph]        &   478 &         262   &        6        &          947    &     498  \\
\textbf{Orion} [Ori]           &  3070  &        1387   &       27        &         4219    &    1250  \\
\textbf{Perseus} [Per]         &   387  &         234   &        5        &          432    &     98  \\
\textbf{Rosette} [Ros]         &    40  &          21   &        -        &            -    &       -  \\
\textbf{Taurus} [Tau]          &   171  &         108   &        4        &          138    &     103  \\
\textbf{Vela} [Vel]            &   689  &          85   &       12        &         2192    &     334  \\ \midrule
\textbf{Total}   &   9722   &  2326     &    144    &   16630     &   2934
\\  

\bottomrule
\end{tabular}
}}
\end{table}

\subsection{The open-cluster sample}\label{oc_sample}

Our working sample of open clusters (OCs) is drawn from the catalogue of \citet[hereafter HR24]{Hunt2024}. The HR24 catalogue builds upon and extends the catalogue presented by \citet{Hunt2023}, where the authors used Hierarchical Density-Based Spatial Clustering of Applications with Noise (HDBSCAN; \citealt{McInnes2017}) to identify $7\,167$ OC candidates over an all-sky uniform sample of 729 million {\it Gaia} DR3 sources with exquisite astrometry. Furthermore, in \citet{Hunt2023} the authors evaluated the reliability of each of their cluster candidates using a density-based statistical test on the astrometric data and a Bayesian convolutional neural network to evaluate the morphology of the colour--magnitude diagrams.

In the HR24 catalogue, the authors refined their previous work by calculating the cluster masses and Jacobi radii for each of the cluster candidates. They then divided them into bound clusters (OCs), unbound moving groups (MGs), and globular clusters, and they excluded some dwarf galaxies or clustering errors that appeared in \citet{Hunt2023} as OC impostors. 

The HR24 catalogue provides key parameters for our research, such as Gaia astrometry and RVs, and estimated cluster distances and ages. {For our analysis, we selected members of both OCs and MGs, since even unbound groups young enough are expected to preserve their birth kinematics. Hereafter, we refer to both collectively as `OC members' for brevity. Additionally, we required positive parallaxes and parallax} S/N>3 and imposed some quality cuts for the OCs to which they belonged, namely an astrometric S/N ({\tt CST}) greater than 5, and a {\tt CMDCl50} (probability of the OC members sharing the same isochrone) greater than 0.5. Furthermore, we supplemented the OC members' RVs with the same criteria as done for the YSOs in Sect. \ref{subsec:yso_sample}. 

Finally, we imposed an age cut ($\leq 30$ Myr) to maintain only young OCs, as older ones have been observed to drift away from their birthplaces \citep{Dias2005, Piecka2021, Castro-Ginard2021}. We ended up with {144 OCs}, with a total of {16 630 members}. The number of OCs ($N_{\rm OC}$), OC members ($N_{\rm OC_{members}}$), and OC members with available RVs ($N_{\rm OC_{members}, RV}$) is given in Table \ref{ref:table_N}. We found OCs in 14 of the 18 clouds, with California, Cassiopeia, Coalsack, and Rosette being the only MCs without any OC (see Sect. \ref{subsec:missing_ocs} for a discussion).

{We note that $1417$ sources are common to both data sets, corresponding to $14.6\%$ of the YSO sample and $8.5\%$ of the OC members. We kept these sources in both samples rather than removing them from either one. Discarding them would amount to excluding the youngest members of the OCs (or the youngest OCs altogether) from the OC sample, which would bias the very comparison we set out to make.}

\subsection{Clustering}\label{sec:method}

Having assembled homogeneous samples of YSOs and OC members associated with each MC, we then extracted their collective kinematic signatures. To this end, we isolated the dominant co-moving group within each complex and derived robust mean space motions representative of the underlying dynamics.

{To remove kinematic outliers (i.e. field contaminants or unrelated sources) from the YSO sample associated with each cloud, we employed the Density-Based Spatial Clustering of Applications with Noise (DBSCAN) algorithm \citep{1996kddm.conf..226E} as implemented in the \texttt{scikit-learn} library \citep{Pedregosa2012} in the 2D ($\mu_{\alpha}^\ast = \mu_{\alpha}\cos{\delta}$, $\mu_{\delta}$) proper-motion space. The goal is not to discover a new substructure, but to isolate the dominant co-moving population in each cloud and to reject sources whose proper motions are inconsistent with the bulk cloud motion.}

DBSCAN is well known for efficiently identifying clusters\footnote{Here, we use the term `cluster' more generically to refer to sets of elements that are close together according to a certain metric (e.g. the Euclidean metric) in the parameter space.} of arbitrary shapes \citep[e.g.][]{Castro-Ginard2018}. The algorithm works by searching for core samples that are points in the parameter space that have a minimum of points ($\texttt{min\_samples}$) within a pre-defined radius ($\epsilon$), such that a cluster is finally defined as a set of core samples. Moreover, DBSCAN does not require the number of clusters to be specified in advance.

A modified version of DBSCAN is HDBSCAN \citep{McInnes2017}, the preferred cluster search method of \citet{Hunt2021}, which eliminates the need to tune the $\epsilon$ hyperparameter by using a wide range of $\epsilon$ values, building a hierarchical cluster tree, and subsequently extracting the most stable clusters from this tree using a stability measure. {However, it has been noticed that HDBSCAN exhibits a higher false-positive rate (i.e. the membership probabilities tend to be overestimated) than the standard DBSCAN \citep[e.g.,][]{Hunt2021}.}

We therefore adopted DBSCAN to identify the co-moving YSOs in each cloud. To this end, we first scaled each ($\mu_{\alpha}^\ast$, $\mu_{\delta}$) array using the \texttt{RobustScaler} function from \texttt{scikit-learn}. Subsequently, we applied the DBSCAN algorithm to our rescaled dataset. We chose $\texttt{min\_samples} = 6$ after visual inspection of the resulting clusters in order to maximise the number of YSOs considered as inliers and thus remove only those sources classified as noise. To find the optimal $\epsilon$, we used the $k-$distance method \citep{Rahmah2016} by plotting the $k$th nearest-neighbour distance for each point in ascending order and determining $\epsilon$ as the knee point (i.e. the point where the slope shows a sharp change; \citealt{Zhang2023}). To this end, we used the \texttt{NearestNeighbors} object from $\texttt{scikit-learn}$ to compute the $k$th distance (where $k=\texttt{min\_samples}=6$) and the $\texttt{KneeLocator}$ function of the $\texttt{kneed}$ Python library to detect knee points after specifying a sensitivity value of $S=10$, which acts as a measure of how many flat points the algorithm detects before declaring a knee \citep{Satopaa2011}.

Finally, we used $10^5$ Monte Carlo trials to sample the proper motions of the resulting co-moving YSOs, assuming a multivariate Gaussian distribution from the 2x2 \textit{Gaia} covariance matrix to account for their uncertainties. To sample their RVs, we assumed an independent Gaussian distribution centred in the RV value, and we adopted the survey-provided uncertainties as the standard deviations. Therefore, we obtained an estimate of the proper motion and RV of each cloud and their uncertainty by taking the mean and standard error ($e_x = \sigma_x/\sqrt{N-1}$) of the resulting distributions. {To avoid distortions caused by sources with very discrepant velocities, we first applied $2\sigma$ clipping around the median RV of each complex.}

At the same time, we used the mean proper motions and RV of the OC members selected in Sect. \ref{oc_sample} from the \citet{Hunt2024} catalogue to obtain the mean and the standard error of the motion of each cloud by conducting $10^5$ trials of the corresponding parameters. Finally, we applied the same quality filters for the RVs as described above for the YSOs.

\section{Results and discussion}\label{sec:results}

\subsection{Comparing YSOs and OCs to trace 3D motions}\label{subsec:ysos_vs_ocs}

{As stated in Sect. 1, our working assumption is that YSOs trace the bulk gas motion of their parent clouds. The purpose of this comparison is not to establish the YSO–gas link, which rests on the body of evidence cited in the introduction, but rather to test whether young OCs provide consistent kinematics and can therefore be combined with the YSOs in a single, larger tracer sample.}

After applying the DBSCAN algorithm to the YSOs in each cloud, we successfully retrieved refined samples of comoving YSOs for 17 MCs. Only the Coalsack cloud was excluded from the following analysis due to insufficient data.

\begin{figure*}
\centering

  \begin{minipage}[b]{0.33\linewidth}
    \includegraphics[width=\linewidth]{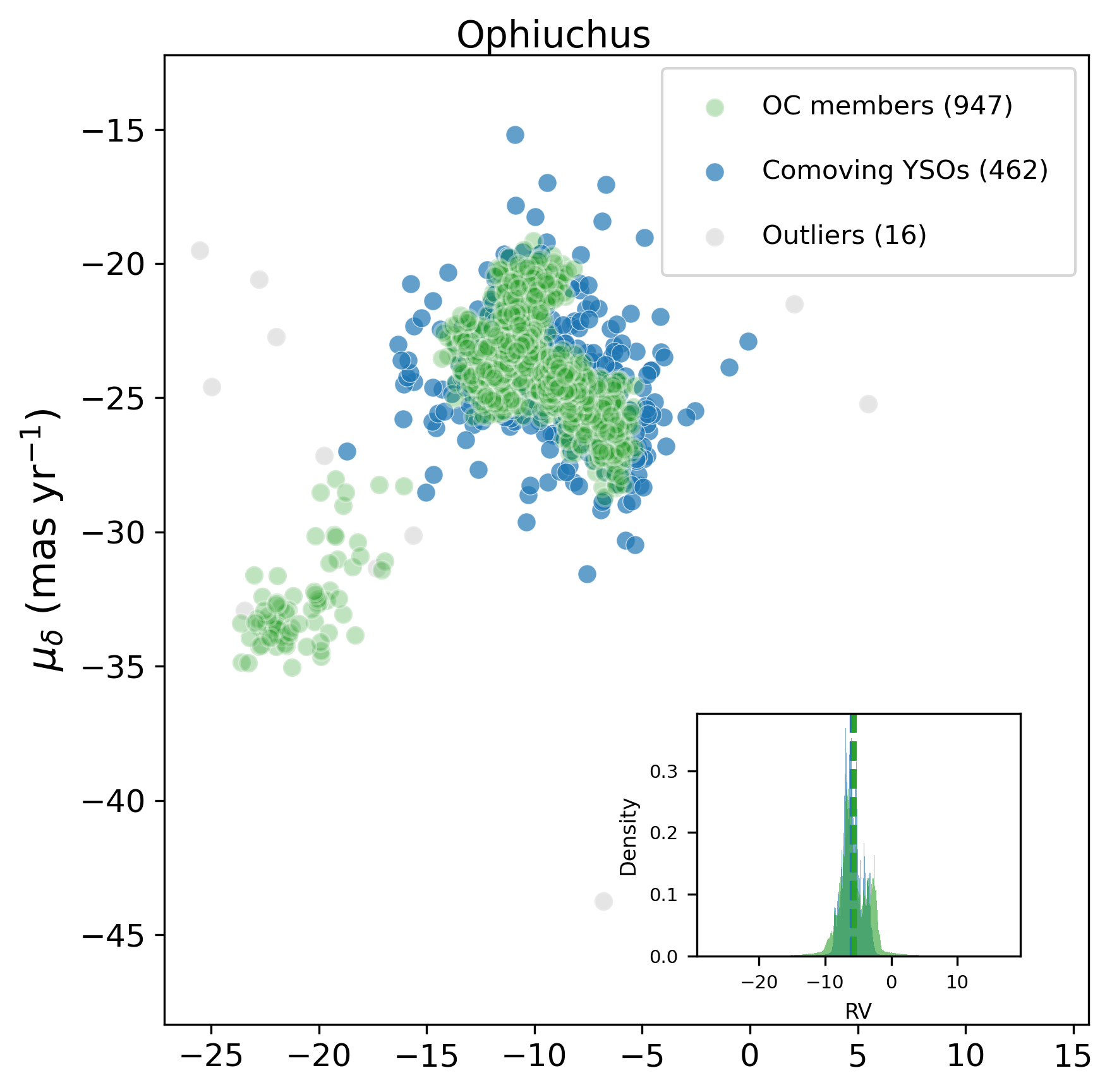}
  \end{minipage}\hfill
  \begin{minipage}[b]{0.33\linewidth}
    \includegraphics[width=\linewidth]{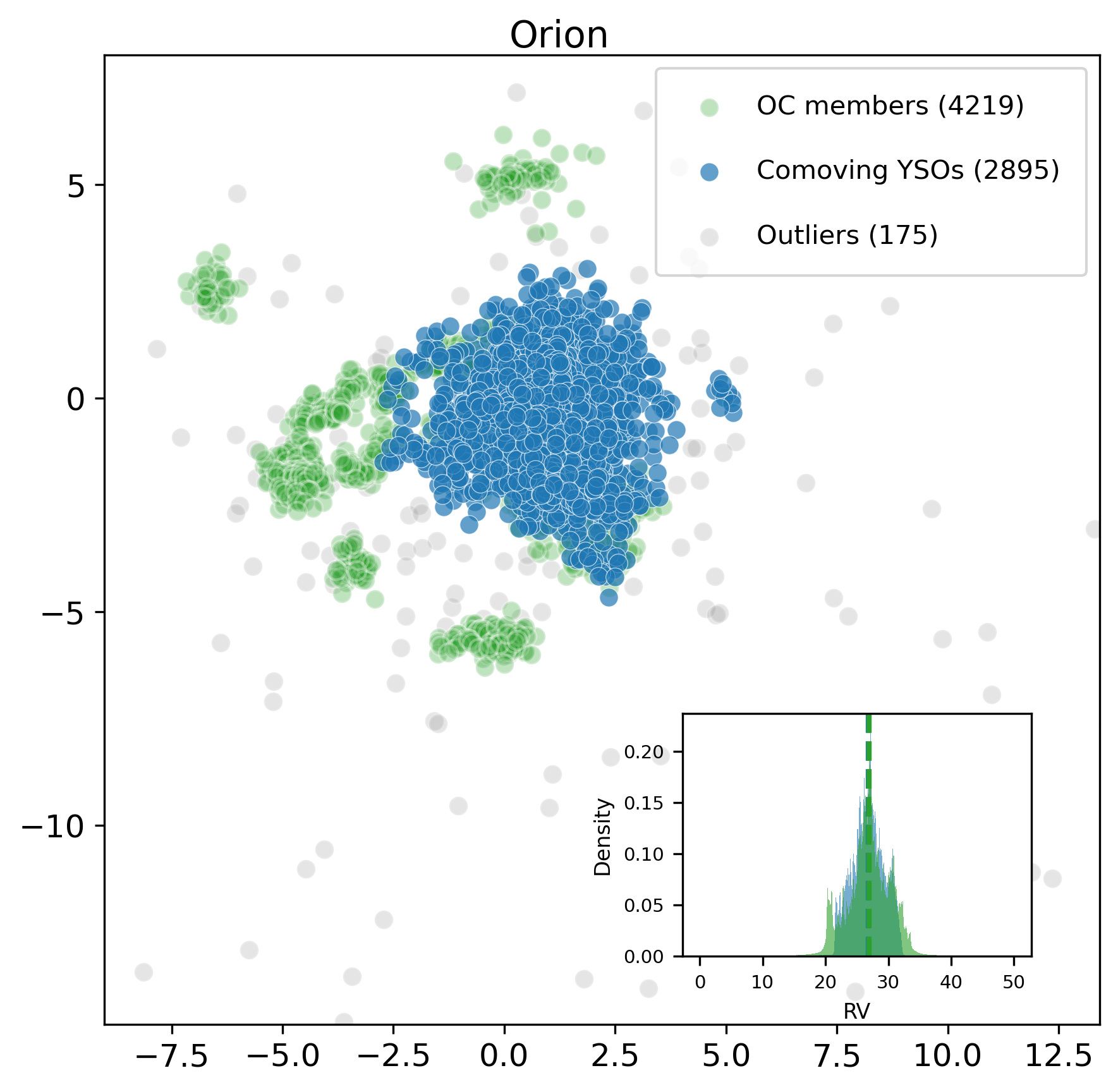}
  \end{minipage}\hfill
  \begin{minipage}[b]{0.33\linewidth}
    \includegraphics[width=\linewidth]{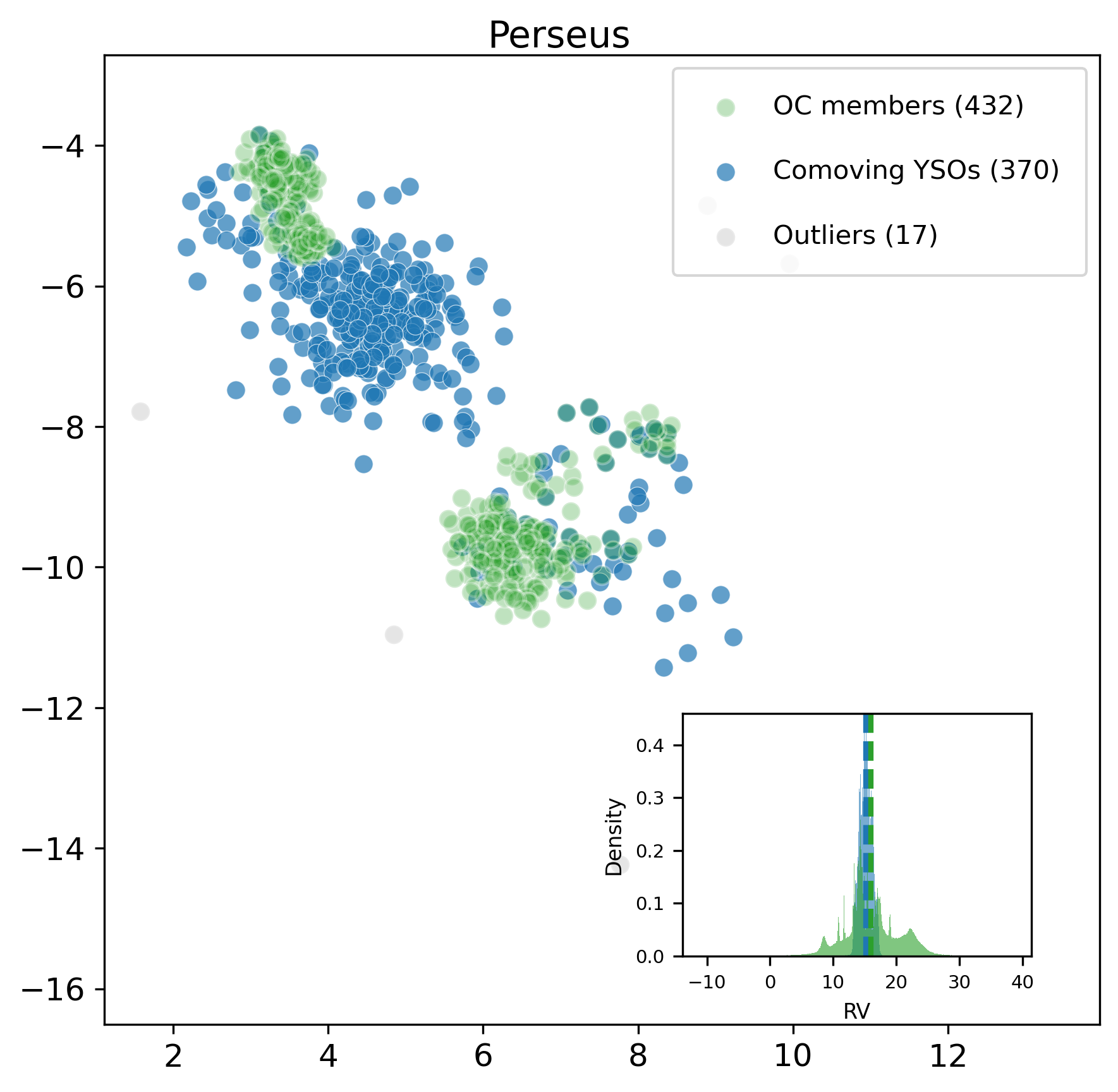}
  \end{minipage}

  \caption{\label{fig:DBSCAN_clusters}Resulting clusters of co-moving YSOs (blue) in the 2D ($\mu_{\alpha}^\ast$, $\mu_{\delta}$) parameter space for Ophiuchus, Orion, and Perseus, with the OC members from the HR24 catalogue shown in green for comparison. When available, an inset in the lower right corner of each plot shows the RV distributions of both populations based on $10^5$ Monte Carlo samplings; the dotted lines indicate the mean RV of the YSO and OC members. Similar figures for the remaining clouds can be found in Fig. \ref{fig:DBSCAN_clusters_complete}}
\end{figure*}

The resulting co-moving YSOs for Ophiuchus, Orion, and Perseus are shown in Fig. \ref{fig:DBSCAN_clusters}, with outliers in grey and those in some DBSCAN cluster in blue. We additionally plotted the OC members according to their proper motions in green. Furthermore, for YSOs and OC members with available RVs, we show their normalised density histograms as insets in each panel. Similar diagrams for the remaining clouds can be found in Appendix \ref{sec:appendix_clusters}. The proper motions and RVs of the OCs and YSOs are consistent in the 14 molecular clouds where both populations are present.

To gain more insight into the offset between YSOs and OCs, we transformed the proper motions into physical units. {More specifically, we calculated the tangential velocity by means of $v_{\alpha*} = 4.74 \mu_{\alpha}^{\ast} d$, with $d$ being the median distance of the cloud. Analogously, $v_{\delta} = 4.74 \mu_{\delta} d$.}

{Subsequently, by computing each of the aforementioned quantities with both YSOs and OC members, we were able to compute the signed offset for each component of the motion: $\Delta v_{\alpha*} = v_{\alpha*,\rm{YSO}} - v_{\alpha*,\rm{OC}}$, $\Delta v_{\delta} = v_{\delta,\rm{YSO}} - v_{\delta,\rm{OC}}$, and $\Delta \rm{RV} = RV_{YSO} - RV_{OC}$. {Finally}, we were able to obtain the magnitude of the total velocity-vector difference using $\Delta V_{\rm T} = \sqrt{{\Delta v_{\alpha*}}^2 + {\Delta v_{\delta}}^2 + {\Delta \rm{RV}}^2}$.}

{We found a median total velocity offset across the clouds of $\Delta V_{\rm T} =2_{-0.7}^{+3.6}$~km s$^{-1}$ (16th and 84th percentiles). This $\Delta V_{\rm T}$ implies a median deviation in orbital integrations of around 2 pc per megayear, which is much smaller than the typical size of a Galactic MC for orbits shorter than 10 Myr \citep{Dharmawardena2023}.}

{We interpret this as evidence that the YSO and OC populations share consistent bulk kinematics on the scale of each star-forming region and therefore can be treated together to assess the bulk motion of each MC complex. We stress that this neither asserts nor requires the present-day gas to be the natal cloud of the OCs; in fact, young OCs typically disperse the gas in which they formed within a few megayears \citep{Lada2003}. Rather, it is precisely because the OC motions match those of the YSOs (which do preserve the kinematics of the current gas) that both can be combined into a single tracer sample without assuming any direct OC--gas link.}

{Interestingly, the largest inconsistencies are found in the farthest clouds. In Cygnus, a $13.3 \pm 0.9$~km~s$^{-1}$ offset is coherent across the OC population, with 19 out of 21 OCs in that region sharing a common bulk motion systematically offset from the YSOs, including well-known OCs such as Berkeley 86/87 and NGC 6913 \citep{Boeche2004, Costado2017, Raudeli2026}. Since most of the discrepancy comes from the tangential component, we restricted both samples to a matched distance to discard a line-of-sight artefact, but that reduced the offset by only $\sim 1$~km~s$^{-1}$. Furthermore, we find that the offset is independent of the age of the cluster. Therefore, we attribute this large offset to the well-documented kinematic substructure of Cygnus OB associations driven by strong stellar feedback \citep{Berlanas2019, Quintana2022} or to an interloping population.}

{Analogously, we ascribed a genuine kinematic {offset of $8 \pm 2$~km~s$^{-1}$ to the OCs in Carina} instead of a distance artefact due to YSOs and OCs sharing a similar median distance and the strong feedback of $\eta$ Carinae and the Trumpler 14/16 OB population \citep{Smith2006, Damiani2017}. This complexity is also reflected in the multimodality shown by Carina and Cygnus RV histograms, as can be seen in Fig. \ref{fig:DBSCAN_clusters_complete}.} {Furthermore, in this case, we note an age difference among the three clusters concentrated around ($\mu_{\alpha^\ast}, \mu_\delta) \simeq (-6.5, 2.0)$~mas~yr$^{-1}$, with ages $\leq 8$~Myr \citep{Hunt2023}; and the cluster with ($\mu_{\alpha^\ast}, \mu_\delta) \simeq (-7.5, 3)$~mas~yr$^{-1}$, with an age of $18.1_{-6.9}^{+7.7}$~Myr \citep{Hunt2023}. As a consequence, this offset can also be explained by age difference.}

{While there is not a large offset {($\Delta V_\mathrm{T} = 1.1 \pm 0.3$~km~s$^{-1}$)}, it is worth mentioning} that the proper motion diagram of Chamaeleon in Fig. \ref{fig:DBSCAN_clusters_complete} shows two moving groups, HSC 2453 and HSC 2523, with $\mu_\alpha^\ast$ of  $-13$ and $-40$ mas yr$^{-1}$, respectively. As can be seen, they do not match any YSO clustered by the algorithm. We identified HSC 2453 as an outlier since the median age computed in \citet{Hunt2023} for that OC is 23 Myr. Combined with the dispersion of its members in the proper-motion space, this suggests that the OC is {either too old to preserve the motion of its parent cloud or that it originated from clouds that were dispersed from and are now unrelated to the present-day gas structure}. {In the case of HSC 2523, however, the median age is $\simeq 7$~Myr, and it matches with some YSOs classified as outliers by DBSCAN. Therefore, we considered these outliers as false negatives, and the group to be physical.}

The members of the only OC candidate found in Corona Australis, HSC 2986 \citep[which is actually a moving group according to][]{Hunt2024} shows a larger dispersion in the $\mu_{\alpha}^\ast$ axis than the YSOs. However, since this group is very young ($\simeq 7 \pm 3$~Myr), we consider it a likely intrinsic property of the complex.

{Another tentative outlier is the moving group CWNU 1143 in Ophiuchus, with $\mu_{\alpha}^\ast \simeq -20$~mas~yr$^{-1}$ and $\mu_\delta \simeq -33$~mas~yr$^{-1}$. In addition to the large separation in the proper-motion space with respect to the centre of the distribution, its age ($18.5^{+15}_{-8.3}$~Myr) is an order of magnitude older than the other two OCs found in that region ($\simeq 4 \pm 2$~Myr), indicating dynamical decoupling. The specific total velocity offset for Ophiuchus, however, is $\Delta V_{\rm T,\,  \rm Oph} = 1.5 \pm 0.2$~km~s$^{-1}$.}

{Similarly, the seven OCs and moving groups surrounding the centre of the distribution of Orion in the proper-motion space without YSO matches also show a median age ($\simeq 18$~Myr) systematically older than those that are in close agreement with the YSOs ($\simeq 7$~Myr). Even so, $\Delta V_{\rm T,\, \rm Ori} = 1.88 \pm 0.06$~km~s$^{-1}$.}

{With the aforementioned OC outliers in mind, we conclude that despite applying a $30$~Myr age filter for the OCs, some of the oldest ones still exhibit systematic offsets with respect to the youngest OCs and YSOs. Nonetheless, we remark that even when including these candidate outliers in the comparison we found kinematic offsets ($\simeq 2$~km~s$^{-1}$) at the level of the typical internal dispersion in these associations \citep{Kuhn2019}. Consequently, we decided to maintain those OCs in our final tracer sample in order to avoid losing any potential substructure and because the added dispersion in velocity is not meaningfully large.}

{We note that the regions defined here are not homogeneous in  terms of physical scale, ranging from sub-components of Sco-Cen such as Ophiuchus and Lupus \citep{Miret-Roig2022, Grossschedl2025} to entire giant molecular cloud (GMC) complexes such as Orion and Cygnus. The homogeneity claimed in this work is therefore methodological; i.e. the same tracer selection, clustering, and uncertainty treatment are applied identically to every region.} 

{To obtain some insights into how the offsets behave at different scales, we also used contours from the literature to constrain the mean motion of their different substructures wherever possible. Further information on these sub-divisions, along with a detailed comparison with the literature, can be found in Table \ref{tab:YSOvsOCmotions}. Overall, we find that our results regarding the general YSO-OC agreement are scale-insensitive.}

{Finally, we note that Cepheus is an interesting case in a different manner. The complex contains two substructures, Cepheus OB2 and the Cepheus Flare, with clearly distinct kinematics (Table \ref{tab:YSOvsOCmotions}), {yet the global offset ($\Delta V_\mathrm{T} = 2.9 \pm 0.8$~km~s$^{-1}$)} is moderate. This may look contradictory at first, but it is not; as long as both substructures are well sampled in the YSO and OC populations, each contributes to the mean in proportion to its size, and the pooled value still reflects the true YSO–OC kinematic relation in the complex as a whole.}

\subsection{Catalogue of 3D molecular cloud kinematics}\label{sec:catalogue}

{Given the small kinematic offsets reported in Sect. \ref{subsec:ysos_vs_ocs}, we proceeded to combine YSO and OCs in a single, unified dataset for the bulk dynamical analysis that follows.} After merging both datasets and dropping duplicate sources, we used a total of {$24,146$} tracers ($8923$ YSOs and $16,630$ OC members, with $1407$ sources common to both datasets), $3618$ of which have high-quality RVs. We note that the census of known YSOs continues to expand \citep[see e.g. ][]{Roquette2025}, and therefore our catalogue of tracers is expected to receive updates in the future. Applying the methodology described in Sect. \ref{sec:method}, we retrieved the mean 3D motions of 15 molecular cloud complexes, as shown in Table \ref{tab:final_motions}. 

\begin{table*}[h!]
\caption{\label{tab:final_motions} 3D positions and motions in the Cartesian heliocentric reference frame of the 15 molecular cloud complexes.}
{\renewcommand{\arraystretch}{1.5}
{\fontsize{7}{9}\selectfont
    \centering
    \begin{tabular}{lccccccc}
    \toprule
        Complex & $X\pm e_{X}\:(\sigma_X)$ pc & $Y\pm e_{Y}\:(\sigma_Y)$ pc & $Z\pm e_{Z}\:(\sigma_Z)$ pc & $U\pm e_{U}\:(\sigma_U)$ km s$^{-1}$ & $V\pm e_{V}\:(\sigma_V)$ km s$^{-1}$ & $W\pm e_{W}\:(\sigma_W)$ km s$^{-1}$ & $N_\mu | N_{RV}$ \\ \midrule
        \textbf{California} & -476.4 ± 19.0 (91.4) & 124.0 ± 5.1 (24.6) & -78.2 ± 3.2 (15.3) & 0.4 ± 0.8 (2.0) & 1.9 ± 0.7 (1.7) & 1.9 ± 0.5 (1.1) & 24 | 7 \\ 
        \textbf{Camelopardalis} & -531.0 ± 7.9 (239.9) & 352.7 ± 5.2 (159.1) & 6.0 ± 0.9 (28.8) & 22.5 ± 3.7 (14.1) & 1.0 ± 2.8 (11.0) & -2.1 ± 0.7 (2.9) & 925 | 16 \\ 
        \textbf{Canis Major} & -802.3 ± 5.6 (143.0) & -803.0 ± 5.6 (143.2) & -43.1 ± 0.5 (12.1) & -17.6 ± 2.2 (5.5) & 5.4 ± 1.8 (4.5) & -8.3 ± 0.7 (1.8) & 661 | 7 \\ 
        \textbf{Carina} & 707.3 ± 5.0 (109.1) & -2291.4 ± 15.9 (350.4) & -19.4 ± 0.5 (10.8) & -65.7 ± 2.8 (13.2) & -2.1 ± 1.3 (6.2) & -5.3 ± 0.7 (3.3) & 485 | 24 \\ 
        \textbf{Cepheus} & -188.5 ± 1.2 (102.4) & 686.2 ± 3.1 (261.7) & 53.5 ± 0.5 (45.2) & 9.5 ± 0.7 (10.0) & -0.3 ± 0.4 (5.7) & 0.1 ± 0.2 (3.2) & 6974 | 236 \\ 
        \textbf{Chamaeleon} & 88.6 ± 1.0 (23.4) & -178.5 ± 1.4 (33.6) & -51.7 ± 0.5 (12.2) & 0.1 ± 0.3 (2.7) & -7.6 ± 0.3 (2.4) & -4.2 ± 0.2 (1.4) & 570 | 81 \\ 
        \textbf{Corona Australis} & 143.3 ± 0.3 (5.1) & -1.2 ± 0.1 (2.1) & -38.8 ± 0.4 (6.0) & 7.2 ± 0.5 (3.1) & -5.1 ± 0.1 (0.7) & -1.5 ± 0.2 (1.1) & 229 | 44 \\ 
        \textbf{Cygnus} & 389.4 ± 2.6 (110.8) & 1720.7 ± 7.6 (326.0) & 36.7 ± 1.0 (41.4) & 52.1 ± 1.2 (10.7) & -9.3 ± 1.2 (10.4) & 3.1 ± 0.4 (3.6) & 1852 | 76 \\ 
        \textbf{Lupus} & 134.8 ± 1.4 (28.0) & -54.0 ± 0.5 (10.2) & 31.1 ± 0.5 (9.1) & 6.0 ± 0.5 (4.2) & -6.9 ± 0.3 (2.5) & 1.7 ± 0.2 (1.9) & 382 | 73 \\ 
        \textbf{Ophiuchus} & 133.2 ± 0.4 (14.9) & -18.8 ± 0.2 (5.5) & 46.3 ± 0.2 (7.3) & 5.6 ± 0.1 (1.9) & -3.8 ± 0.1 (1.4) & -0.4 ± 0.1 (1.7) & 1266 | 506 \\ 
        \textbf{Orion} & -328.6 ± 0.6 (50.9) & -149.8 ± 0.5 (42.4) & -107.4 ± 0.4 (29.0) & -11.7 ± 0.1 (3.1) & -0.2 ± 0.1 (2.5) & 0.1 ± 0.0 (2.2) & 6633 | 1878 \\ 
        \textbf{Perseus} & -294.1 ± 1.9 (51.1) & 103.1 ± 0.7 (19.0) & -94.2 ± 0.5 (14.9) & -6.0 ± 0.2 (2.4) & 5.1 ± 0.2 (2.9) & -0.6 ± 0.1 (1.4) & 757 | 242 \\ 
        \textbf{Rosette} & -1162.4 ± 47.6 (256.5) & -583.7 ± 23.9 (128.6) & -43.6 ± 2.6 (13.9) & -17.5 ± 0.9 (3.4) & 5.1 ± 0.9 (3.4) & -3.2 ± 0.8 (3.1) & 30 | 15 \\ 
        \textbf{Taurus} & -137.0 ± 1.3 (20.8) & 14.9 ± 0.6 (9.4) & -40.7 ± 0.5 (8.2) & -5.5 ± 0.1 (1.4) & -0.4 ± 0.2 (2.3) & -1.7 ± 0.2 (1.9) & 239 | 124 \\ 
        \textbf{Vela} & -98.0 ± 1.1 (56.0) & -848.6 ± 3.7 (191.0) & -0.9 ± 0.6 (30.5) & -20.7 ± 0.3 (4.4) & -5.0 ± 0.1 (2.1) & 0.5 ± 0.2 (2.6) & 2607 | 284 \\ 
       \bottomrule
    \end{tabular}
    \tablefoot{Velocities were calculated with respect to the LSR. A machine-readable version of this table and the tracers used to build it are available as part of the online supplementary material.}
}}
\end{table*}

We used the \texttt{astropy} package \citep{The_Astropy_Collaboration2022} to provide Cartesian heliocentric positions and velocities, with the latter expressed with respect to the local standard of rest (LSR). We adopted a solar galactocentric distance of $R_0 = 8.122$ kpc \citep{GRAVITYCollaboration2019}, an LSR velocity of $V_C(R_0) = 229$ km s$^{-1}$ \citep{Eilers2019}, the solar velocity relative to the LSR as $(U_\odot,V_\odot,W_\odot) = (11.1, 12.24, 7.25)$ km s$^{-1}$ \citep{Schonrich2010}, and a solar height above the Galactic plane of $Z_\odot = 20.8$ pc \citep{Bennett2019}. 

To compute uncertainties, we used the {\it Gaia} covariance matrix of $\alpha$ and $\delta$, as well as the posterior distances of \citet{Weiler2025}, to calculate the positions. For velocities, we used the objects with complete 3D kinematic information and the 4x4 covariance matrix for positions and proper motions, assuming the RVs and distances as independent distributions. We show \href{https://lia2.es/astro/molecular_cloud_kinematics.html}{here} an interactive visualisation of our catalogue, with the XYZ coordinates, spherical approximation for each cloud (left), and a zoomed-in view of the Orion cloud (right), with the UVW velocities of each tracer overlaid as red arrows.

We compared the motions of the clouds in common with \citet{Zhou2025} by subtracting the solar motion from their values. We found a median offset of $\sim 3 $~km~s$^{-1}$, which is slightly larger than the uncertainties provided by \citet{Zhou2025} ($\sim 1.3$~km~s$^{-1}$). We note, however, that for the same clouds our uncertainties are approximately an order of magnitude smaller ($\sim 0.3$~km~s$^{-1}$), which we attribute to our larger number of tracers ($24,146$) compared to those of \citep[]{Zhou2024, Zhou2025} ($6,271$). 
Although the authors did not provide the IDs of the YSOs belonging to their 102 associations, we replicated the initial sample they obtained from Gaia DR2 and \citet{Marton2016} for their complete set of 150 YSO groups, as described in \citet{Zhou2022}. We obtained $18,246$ sources, only $2,125$ (11 \%) of which are included in our sample. We therefore consider this value as the upper limit of the overlap between the two samples, and consequently our works can be regarded as complementary.

\subsection{The Solar System's past}\label{sec:ism_enrichment}

Our catalogue opens up many opportunities to explore how the local interstellar medium (ISM) has been influenced by its surroundings, i.e. the recent history of our Solar 
System \citep[see e.g.][]{Maconi2025}. To this end, we computed the orbit of each MC 20~Myr back in time using the 3D positions and velocities provided in Table \ref{tab:final_motions} as initial conditions and the \texttt{Gala} package \citep{gala,adrian_price_whelan_2020_4159870} with the \texttt{MilkyWayPotential2022} model. This model combines a spherical nucleus and bulge, a sum of Miyamoto-Nagai disks, and a spherical NFW halo, fitted to the \cite{Eilers2019} rotation curve. Each MC was integrated in steps of $0.01$ Myr up to $-20$~Myr.

\begin{figure}[h]
\centering

\begin{subfigure}{0.71\columnwidth}
    \centering
    \includegraphics[width=\linewidth]{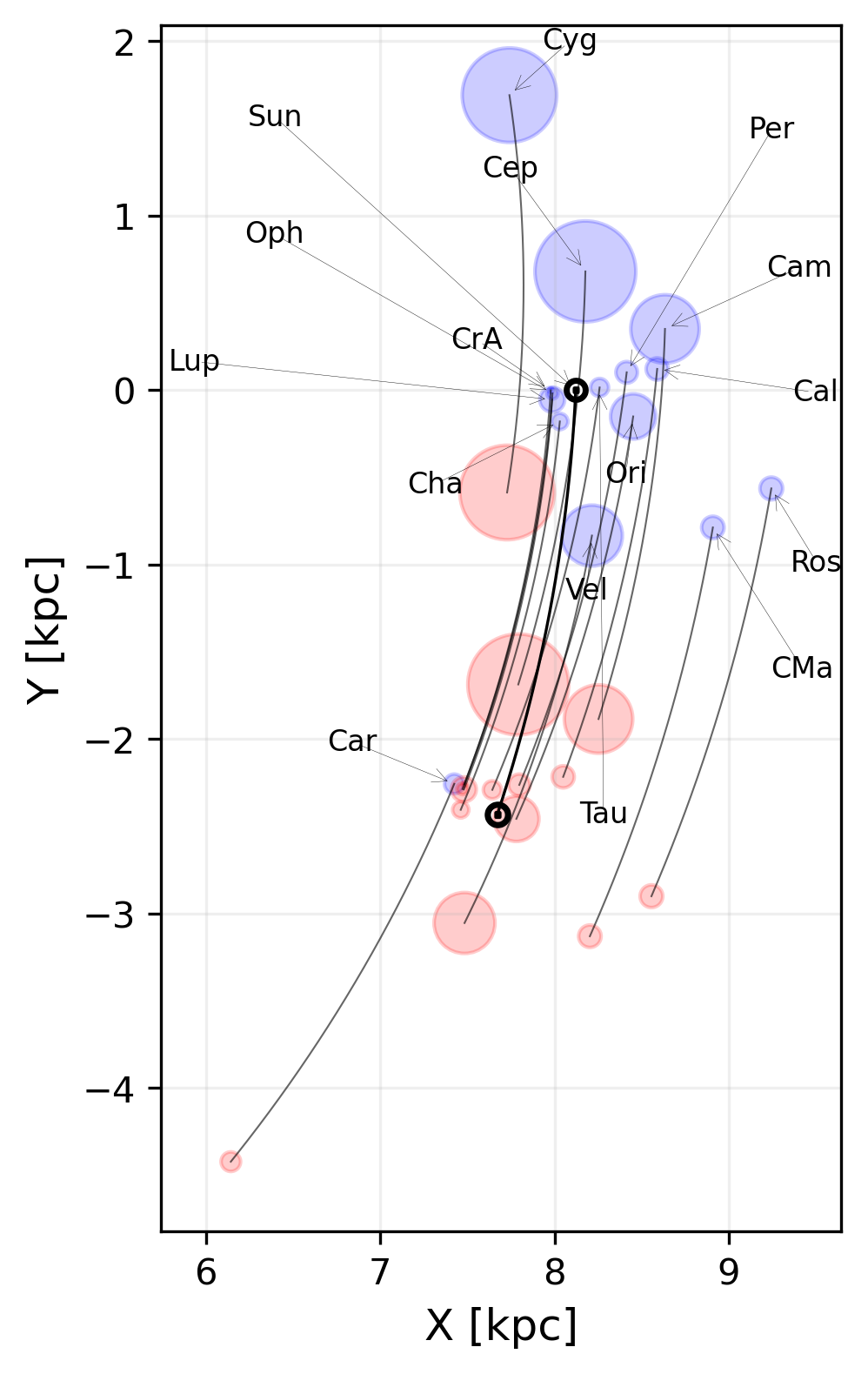}
    \label{fig:xy}
\end{subfigure}
\hfill
\begin{subfigure}{0.271\columnwidth}
    \centering
    \includegraphics[width=\linewidth]{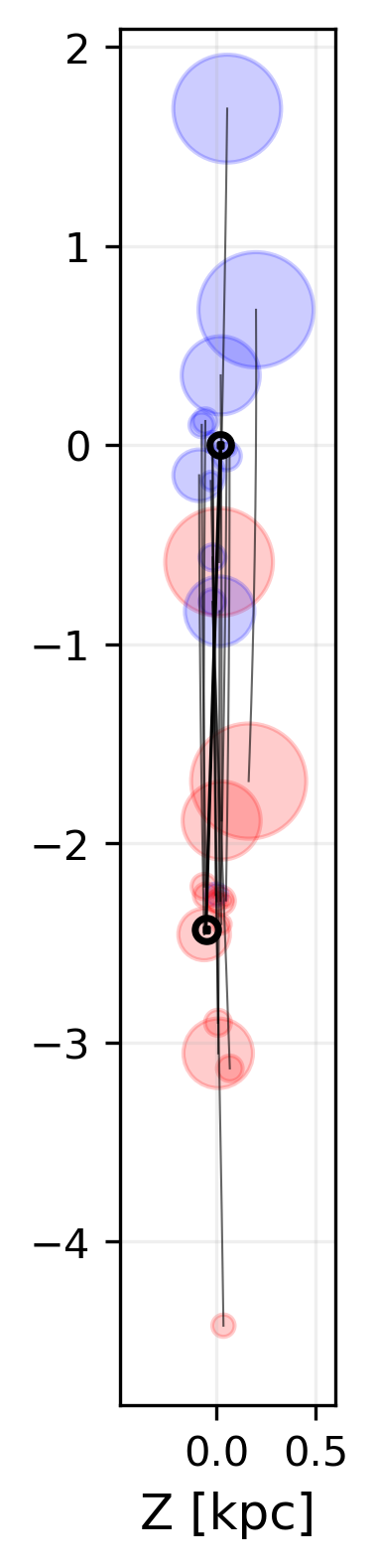}
    \label{fig:zy}
\end{subfigure}

\begin{subfigure}{0.98\columnwidth}
    \centering
    \includegraphics[width=\linewidth]{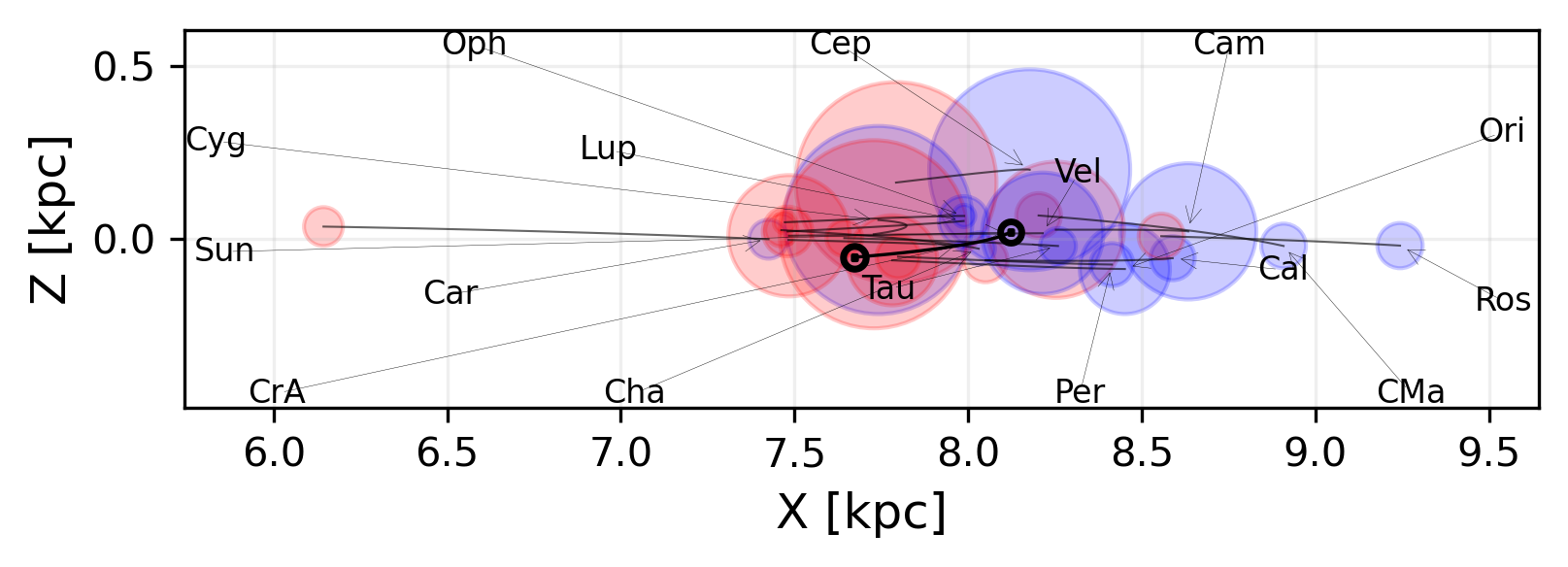}
    \label{fig:xz}
\end{subfigure}

\caption{Galactic projections of the molecular cloud complexes
with their present positions (purple circles) and projections 10 Myr ago (in pink).}
\label{fig:galactic_panels}
\end{figure}

{In Fig. \ref{fig:galactic_panels} we show the positions of MCs in the present (purple circles) and 10 Myr ago (in pink). To account for encounters between the clouds and Solar System in the 3D space, we approximated the shapes of MCs as spheres, since the true morphology of the clouds in the past cannot be reconstructed. Therefore, the radius, $R$, of each cloud is computed so that the resulting sphere contains the same volume as the 3D region enclosed in the $l$, $b$, and $d$ limits given by \citet{Dharmawardena2023}.}

By using those radii and $10^4$ Monte Carlo orbits by sampling the 6D initial phase space of the MCs from our catalogue, we checked the relative separation between the Sun and the centroid of the clouds at each integration step and each realisation. Subsequently, we define the encounter probability, $P_{\rm encounter}$, as the fraction of realisations that produce at least one hit within the integration window. By imposing a minimum $P_{\rm encounter}>0.5$ our orbital traceback indicates that the Solar System traversed the Orion cloud complex between approximately 18.6 and 9.4 Myr ago, assuming $R\simeq 130$~pc, and thus agreeing with the results recently reported by \citet{Maconi2025}. {Furthermore, using the $R\simeq 50$~pc value assumed in their work, the encounter occurred with $P_{\rm encounter} = 98.5 \%$, and between $14.6$ and $13.4$~Myr ago, in excellent agreement with the $14.8-12.4$~Myr window reported by them.}

{\subsection{Cloud--cloud encounters}}

We can also use the orbits of the MCs and their estimated radii to search for cloud--cloud encounters in the past. Such events are of particular interest because they may point to a shared dynamical history for clouds that are now widely separated. Furthermore, encounters may trigger star formation via shocks that externally increase the gas density locally \citep{Fukui2021}.

{For each cloud, we drew $10^{4}$ Monte Carlo realisations of its 6D phase-space coordinates from our catalogue uncertainties and integrated the resulting orbits from $t=0$ to $t=-30$~Myr with a time step of $0.01$~Myr. For each pair of clouds, we registered a {close encounter whenever their relative separation ($d_{\rm rel}$) fell within the mean of the estimated containment radii of the clouds (i.e. when $d_{\rm rel} < (R_i + R_j)/2$). }

\begin{table}[!ht]
    \centering
    \caption{\label{tab:encounters} Significant cloud--cloud close encounters.}
    {\renewcommand{\arraystretch}{1.5}
    \setlength{\tabcolsep}{4pt}
    {\fontsize{8}{9}\selectfont
    \begin{tabular}{lccccc}
            \toprule
            Encounter & $P_{\rm encounter}$& $t_{\rm closest}$~[Myr] & $d_{\rm rel}$~[pc] & $v_{\rm rel}$~[km s$^{-1}$] & $R_i + R_j$~[pc] \\  \midrule

            Lup -- Oph & 1.000 & $7^{+1}_{-1}$ & $30^{+2}_{-2}$  & $3.5^{+0.3}_{-0.3}$  & $90$  \\
            Lup -- CrA & 0.989 & $14.3^{+0.6}_{-0.6}$ & $26^{+5}_{-5}$  & $6.6^{+0.2}_{-0.2}$  & $83$  \\

 \bottomrule
    \end{tabular}
    \tablefoot{For each encounter, we list the MCs involved; encounter probability; time of closest approach, $t_{\rm closest}$, in past time; median relative distance, $d_{\rm rel}$; and velocity, $v_{\rm rel}$ with the $16th and 84th$ percentiles. The last column provides the sum of the radius of both clouds.}
    }}
\end{table}

{In Table \ref{tab:encounters} we list the two cloud--cloud encounters with $P_{\rm encounter} > 0.50$, Lupus--Ophiuchus and Lupus--Corona Australis, around 7 and 14 Myr ago, respectively. In fact, the wide percentile range in $t_{\rm closest}$ for the Lupus--Ophiuchus encounter suggests co-movement.}

{Indeed, it has been previously argued that these clouds were formed within the feedback environment of the Sco-Cen star-forming region \citep{Miret-Roig2022, Posch2023}, potentially shaped by an expanding H~{\sc i} shell driven by supernova explosions in the Local Bubble $\sim 14$~Myr ago \citep{Zucker2022}. Hence, our results corroborate this scenario.}

{Furthermore, the inferred relative velocities for these encounters ($v_{\rm rel}\simeq 3$--$6$~km~s$^{-1}$) {lie below the canonical cloud–cloud collision velocity of $\sim 10$~km~s$^{-1}$ \citep{Fukui2021}}, which is compatible with gas coupling and a common origin. Moreover, these $v_{\rm rel}$ are fully consistent with the internal kinematics reported for Sco--Cen by \citet{Grossschedl2025}, where the authors found coherent velocities of $5-6$~km s$^{-1}$ and a total velocity dispersion of $4-5$~km~s$^{-1}$.}

Although most of the OCs found in these clouds are younger than the Lup-Oph-CrA encounter time, CWNU 1143{, a moving group identified as an OC in \citet{He2022}, and belonging to the Ophiuchus cloud \citep{Zucker2020}, has a median age of $18^{+15}_{-8}$~Myr \citep{Hunt2023},} which is consistent with the interval during which those clouds shared the same 3D phase space. However, the uncertainties are too large to establish a direct relation, and deriving more accurate ages is beyond the scope of this study.

\subsection{Internal kinematics: Expansion and rotation}
Our tracer sample can also be used to investigate the internal kinematics of each molecular cloud complex \citep[see e.g.][]{Rivera2015}. The radial component of the motion of each tracer relative to the cloud centre can be obtained as, 
$v_{r,\,i} = \left(\boldsymbol{r_i} \cdot \boldsymbol{v_i}\right)/r_i$, where $\boldsymbol{r_i}$ and $\boldsymbol{v_i}$ are the position and velocity vectors of the tracer $i$ with respect to the centre of the cloud, and $r_i$ is the absolute value of $\boldsymbol{r_i}$. 

Similarly, the tangential velocity of tracer $i$ with respect to the cloud centre is $\boldsymbol{v_{rot, \, i}} =  \left(\boldsymbol{r_i} \times \boldsymbol{v_i}\right)
/r_i$. We therefore estimated the mean expansion velocity of each cloud, $v_{\rm exp}$, as the average of the dot product for all of its members and the mean tangential velocity of the cloud (hereafter, the rotational velocity, or $\boldsymbol{v_{\rm rot}}$) from the cross product. By definition, the modulus ($v_{\rm rot} = |\boldsymbol{v_{\rm rot}}|$) indicates the magnitude of the mean tangential motion of the cloud, while the direction of the unit vector defines the rotation axis, $\hat{\boldsymbol{v_{\rm rot}}}$, and therefore it is a highly informative quantity. 

It is important to note that clouds located further out are subject to the so-called fingers-of-God (FoG) effect, observed when the distance errors are comparable to the cloud size measured through the line of sight (LOS). Thus, we only studied the internal kinematics of the clouds where at least 75\% of the tracers with available RVs have a distance error (computed as the half-width of the 68\% interpercentile range of the distance posterior distribution) that is less than $10\%$ of the MC size measured along the LOS. This excludes California, Canis Major, Carina, Cygnus, Rosette, and Vela.

The results for the nine remaining cloud complexes with 
$N_{RV}\geq 10$ are summarised in Table \ref{tab:v_rot}, including the angle, $\theta$, between the rotation axis and the Galactic rotation axis, $\boldsymbol{Z}$.

\begin{table}[!ht]
\caption{\label{tab:v_rot} Mean expansion and rotational velocities for the nine clouds with $N_{RV}\geq 10$.}
    {\renewcommand{\arraystretch}{1.25}
    {\fontsize{7}{9}\selectfont
    \centering
    \begin{tabular}{lccccc}
    \toprule
        Complex & $v_{\rm exp}$ [km s$^{-1}$] & $v_{\rm rot}$ [km s$^{-1}$] & $\theta$ [$^\circ$] &  $v_{\rm rot, corr}$ [km s$^{-1}$]\\ \midrule
        \textbf{Camelopardalis} & 1.26 ± 3.85 & 3.54 ± 2.21 & 175.3 & 0 ± 2\\ 
        \textbf{Cepheus} & 0.08 ± 0.49 & 3.30 ± 0.54$^{5\sigma}$ & 170.7 & 2.1 ± 0.5$^{4\sigma}$\\ 
        \textbf{Chamaeleon} & 0.00 ± 0.29 & 0.94 ± 0.23$^{4\sigma}$ & 146.6 & 0.8 ± 0.2$^{4\sigma}$\\ 
        \textbf{Corona Australis} & 0.47 ± 0.27 & 0.63 ± 0.33 & 109.1 & -\\ 
        \textbf{Lupus} & 0.02 ± 0.35 & 1.23 ± 0.32$^{3\sigma}$ & 61.7 & -\\ 
        \textbf{Ophiuchus} & 0.56 ± 0.08$^{5\sigma}$ & 0.30 ± 0.06$^{5\sigma}$ & 143.3 & 0.17 ± 0.06$^{2\sigma}$\\ 
        \textbf{Orion} & 1.52 ± 0.06$^{5\sigma}$ & 0.32 ± 0.05$^{5\sigma}$ & 20.7 & 0.77 ± 0.06$^{5\sigma}$\\ 
        \textbf{Perseus} & 0.46 ± 0.16$^{2\sigma}$ & 0.55 ± 0.14$^{3\sigma}$ & 147.2 & 0.12 ± 0.14\\ 
        \textbf{Taurus} & -0.03 ± 0.13 & 1.38 ± 0.14$^{5\sigma}$ & 59.1 & -\\ 
        \bottomrule
    \end{tabular}
    \tablefoot{Superscripts indicate the significance of each calculation, from $2\sigma$ to $5\sigma$.}
    }}
\end{table}
Only three clouds show expansion velocities significantly different from zero. Orion and Ophiuchus exhibit the most significant expansion (above $5\sigma$), followed by Perseus. For these clouds, the values reported in Table \ref{tab:v_rot} indicate the presence of an organised large-scale expansion. In Orion the values are consistent with the Orion Big Blast scenario suggested by \citet{Grosschedl2021} and with studies that identified expanding shells from the gas \citep{Feddersen2018, Pabst2020}. 

Moreover, most of the clouds, with the only exception of Camelopardalis and Corona Australis, show a significant non-zero $v_{\rm rot}$. Such significant, large-scale rotations have, to our knowledge, only been previously reported for Taurus \citep{Rivera2015}, with the most recent mean value ($1.5 \pm 0.1$~km~s$^{-1}$, \citealt{Galli2019}) being in excellent agreement with our result.

{{Regarding the rotation-axis orientation, Orion is the only cloud whose axis is approximately aligned with the Galactic North pole, with a tilt of $\theta \simeq 21^{\circ}$}. Conversely, four clouds (Cepheus, Chamaeleon, Ophiuchus, and Perseus) exhibit significant rotation with their axes oriented towards the South Pole. These non-isotropic orientations suggest that Galactic shear may play some role in the inferred rotation. Using the Oort constant in the Solar neighbourhood, $A = 14.79 \pm 0.11$ km s$^{-1}$ kpc$^{-1}$ \citep{Akhmetov2024}, and the radial extent of each cloud in the direction of the Galactic centre from its tracers ($R_{X} = \sqrt{\langle X^2 \rangle }$), we estimated a rough first approximation of the Galactic shear-induced differential rotation as $v_{\rm rot, shear} \simeq A \, R_X$ km s$^{-1}$. Subsequently, we corrected $v_{\rm rot}$ by subtracting or adding this magnitude depending on the value of $\theta$ {(Galactic-shear-induced rotation should have $\theta \simeq 180^{\circ}$)}. We provide the result as $v_{\rm rot,corr}$ in Table \ref{tab:v_rot}.}

{As a result, we found that Orion rotates faster than observed, while a significant part of the rotation noticed in Cepheus, Chamaeleon, Ophiuchus, and Perseus comes from the Galactic shear. However, even after correcting from the shear, the coherent rotation remains statistically significant for all of the clouds except Perseus, which does not show a significant intrinsic pattern.}

{The coherent rotation we detect is dynamically negligible as a support mechanism; in all cases, $v_{\rm rot}$ lies well below the internal velocity dispersion of the tracers, so rotation cannot counteract collapse. Its interest lies instead in the information it records in terms of cloud assembly. The orientation of the rotation axis is informative in this respect, with the retrograde sense measured for Orion requiring an additional source, such as turbulence acquired at assembly or subsequent feedback. Plausibly, the same event is responsible for the $5\sigma$ expansion detected in that complex as shown in Table \ref{tab:v_rot}.}

{We consider it important to note that Galactic shear introduces a velocity gradient across a cloud rather than a single bulk rotational velocity. Therefore, we only used $v_{\rm rot, shear}$ as a rough order-of-magnitude reference for the expected shear-induced amplitude.}

\subsection{Kinematic ages}\label{subsec:orbit_evol}

The kinematic or dynamical traceback age, often defined as the time of highest stellar concentration, is useful as an age constraint independent of evolutionary models and as a tool for studying the dynamics of young star clusters and star-forming regions (e.g. \citealt{Brown1997, Asiain1999, Miret-Roig2020}). {Here, we study the kinematic ages of the clouds with the most significant expansion motion: Orion and Ophiuchus.}

To measure how concentrated each system is at each epoch, we first used the full 6D phase-space information of each source as the initial condition to integrate its orbit backwards and forward in time, as explained in Sect. \ref{sec:ism_enrichment}. Each orbit was integrated in steps of $0.01$ Myr, spanning $20$ Myr backwards and $10$ Myr forwards in time. The associated movies are available online.

Subsequently, we computed the median pairwise separation between tracers (YSOs and OC members) and used it as a proxy for the characteristic size of the population. This provides a simple, model-independent indicator of when the group was most compact. At each time step, we extracted the 3D positions of all tracers and computed the median pairwise separation.  For each population, we quantified the spatial compactness by computing the full set of Euclidean pairwise separations among its members. A `pairwise' separation refers to the Euclidean distance between every possible pair of stars in the group. For a population of $N$ tracers, this results in $N(N-1)/2$ unique separations. 

To propagate the measurement errors, we performed a Monte Carlo analysis with $1000$ realisations per cloud. The ensemble of realisations yields, at every epoch, a posterior distribution of compactness values. From these, we report the median and the associated $1\sigma$ (16$^{
\rm th}$--84$^{
\rm th}$ percentile).

{We examined, for each cloud, the time at which the global stellar separation reaches a minimum. The morphology of the compactness curves varies from cloud to cloud (see Fig. \ref{Orion_comp}). As expected for the expansion motion found in the previous section, the time of maximum concentration for Orion and Ophiuchus is in the past.}

For the Orion complex, we find that the minimum global stellar separation occurs at $t \simeq -6.6^{+0.3}_{-0.3}~\mathrm{Myr}$ (Fig. \ref{Orion_comp}), with values that depend on the tracer population: $-1.1^{+0.2}_{-0.2}$ Myr
when using only YSOs and $-6.2^{+0.3}_{-0.3}$ Myr when using only OC members. This difference is naturally explained by the age offset between the two populations. Class II YSOs typically have ages in the 1--3 Myr range \citep{Kucuk2010}, whereas the Orion OC members have a weighted mean age of $\simeq 6.9$ Myr based on the isochrone ages reported by \citet{Hunt2023}. This result is consistent with \citet{Grosschedl2021}, who inferred a minimum OC member age of $\sim 6$--$7~\mathrm{Myr}$ based on the summed separations among the 14 sub-regions they analysed (see their Table 4). This illustrates that the inferred kinematic age of a large complex depends strongly on the age distribution of the adopted tracers. 

{This implies that a single kinematic age is not well defined for a complex hosting populations of different ages and that traceback ages quoted in the literature should be interpreted with reference to the tracer used.}
Together, these results pinpoint the convergence epoch to within $\sim 1$--$7~\mathrm{Myr}$, supporting the view that the present-day Orion complex originates from a common dynamical configuration several megayears ago.

{In the case of Ophiuchus, the median kinematic age obtained in this work {($\simeq 1.6 \pm 0.1$~Myr; see Fig. \ref{Orion_comp})} is in excellent agreement with the 1-3 Myr age obtained through evolutionary models of the pre-main sequence in \citet{Greene1995}. To our knowledge, this represents the first kinematic age determination for the Ophiuchus cloud.} However, we note that the accuracy of the compactness of a population as a function of time is strongly dominated by the precision in the present velocities (see e.g. \citealt{Miret-Roig2018}). Therefore, an exhaustive analysis of the error propagation in the orbital analysis is necessary before interpreting the results in these regions. Additionally, the parent molecular cloud might have an impact on the early dynamical evolution of the system (see e.g. \citealt{Miret-Roig+2024}), which is not accounted for by our orbital integration based on the Galactic potential alone.

\begin{figure*}[!h]
    \centering
    \includegraphics[width=0.49\textwidth]{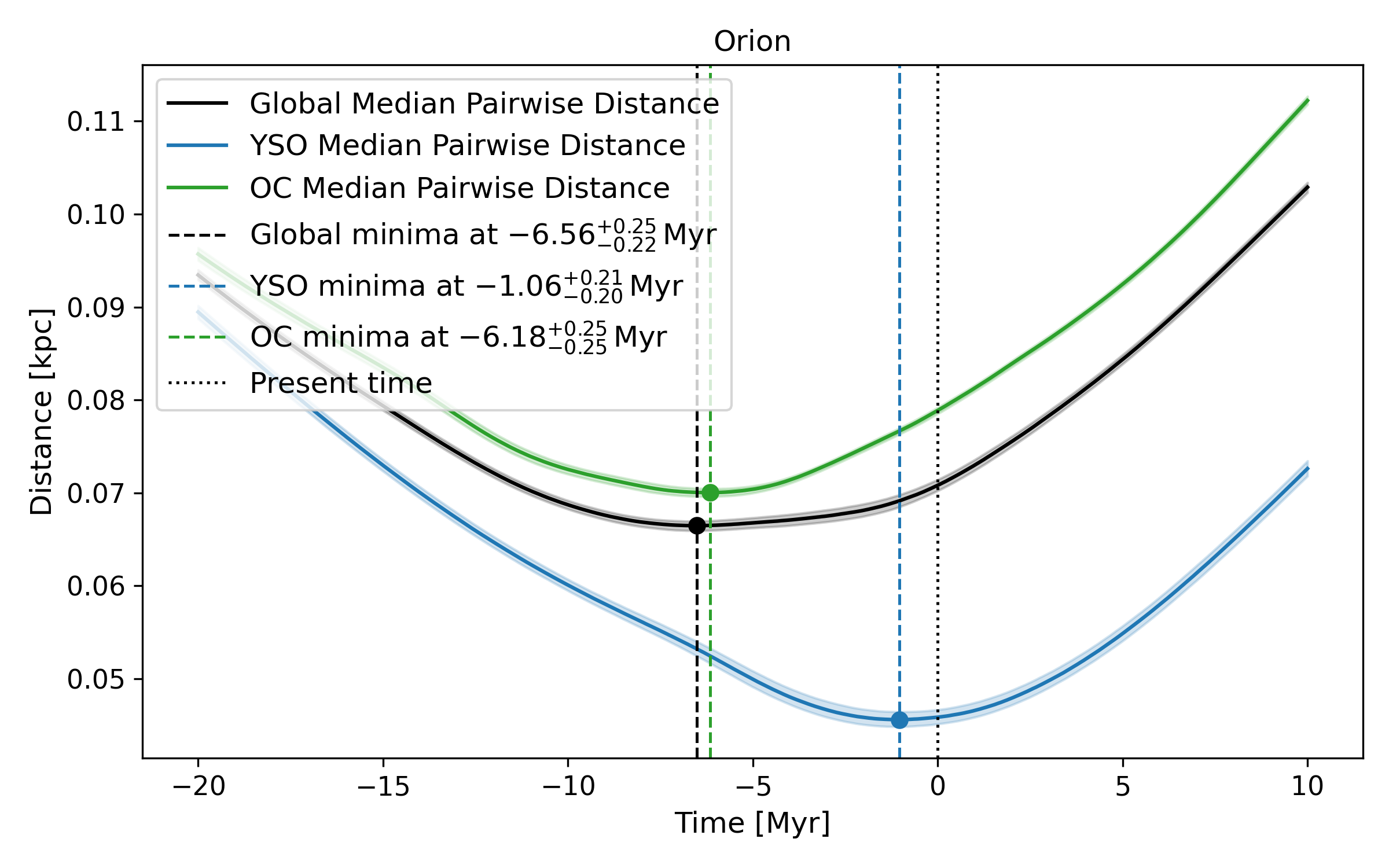}
    \includegraphics[width=0.49\textwidth]{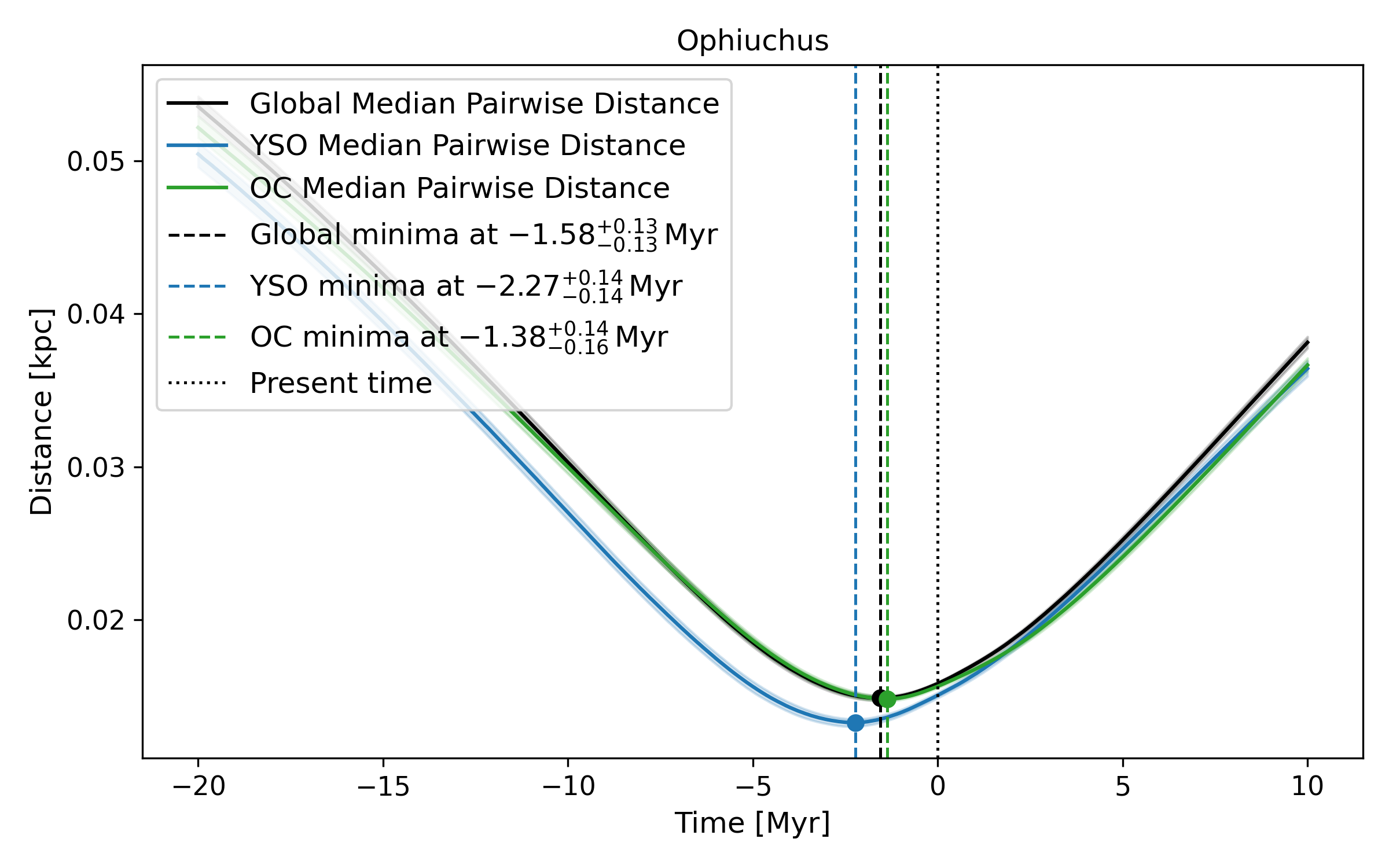}
    \caption{
    Evolution of the spatial compactness of the Orion (\textit{left}) and Ophiuchus (\textit{right}) complexes, showing the median pairwise separations for all tracers (black), YSOs (blue), and OC members (green). Vertical dashed lines mark the epochs of minimum separation for each population, indicating the times of maximum compactness.}
    \label{Orion_comp}
\end{figure*}

\subsection{Peculiar velocities relative to Galactic rotation}\label{subsec:internal_kinematics}

{To quantify the peculiarity of the clouds' motions with respect to the Galactic rotation, we transformed the 3D LSR velocity of each cloud to galactocentric cylindrical velocities ($V_{R}$, $V_{\phi}$, $V_z$). We then subtracted the Galactic circular velocity, $V_C$, at the position of each cloud given by the Milky Way model adopted so far. We obtain the peculiar velocity relative to the local circular flow of each cloud as $V_{\rm pec} = \sqrt{{V_R}^2 + {V_Z}^2 + \left({V_\phi - {V_C}}\right)^2}$. } 

{We find that all the clouds show a peculiar motion different from zero at the $5\sigma$ level, ranging from $V_{\rm pec,\, Cal} = 4.0 \pm 0.6$~km s$^{-1}$ in California up to $V_{\rm pec, \, CMa} = 11^{+2}_{-1}$~km s$^{-1}$ in Canis Major. The median $V_{\rm pec}$ across all clouds is $9^{+1}_{-3}$~km s$^{-1}$ ($16^{\rm th}-84^{\rm th}$ percentile), in agreement with \citet{Zhou2025}.}

{Such a peculiar velocity displaces a cloud by $\simeq 90$~pc in 10~Myr, which is comparable to the size of the complexes themselves.
Present-day positions are therefore poor indicators of birth sites over the lifetime of a GMC, which is precisely why the orbital reconstruction of §3.3–3.4 is required to connect clouds to their origins.}

\subsection{Open cluster detection limits}\label{subsec:missing_ocs}

\begin{figure}
    \centering
    \includegraphics[width=0.49\textwidth]{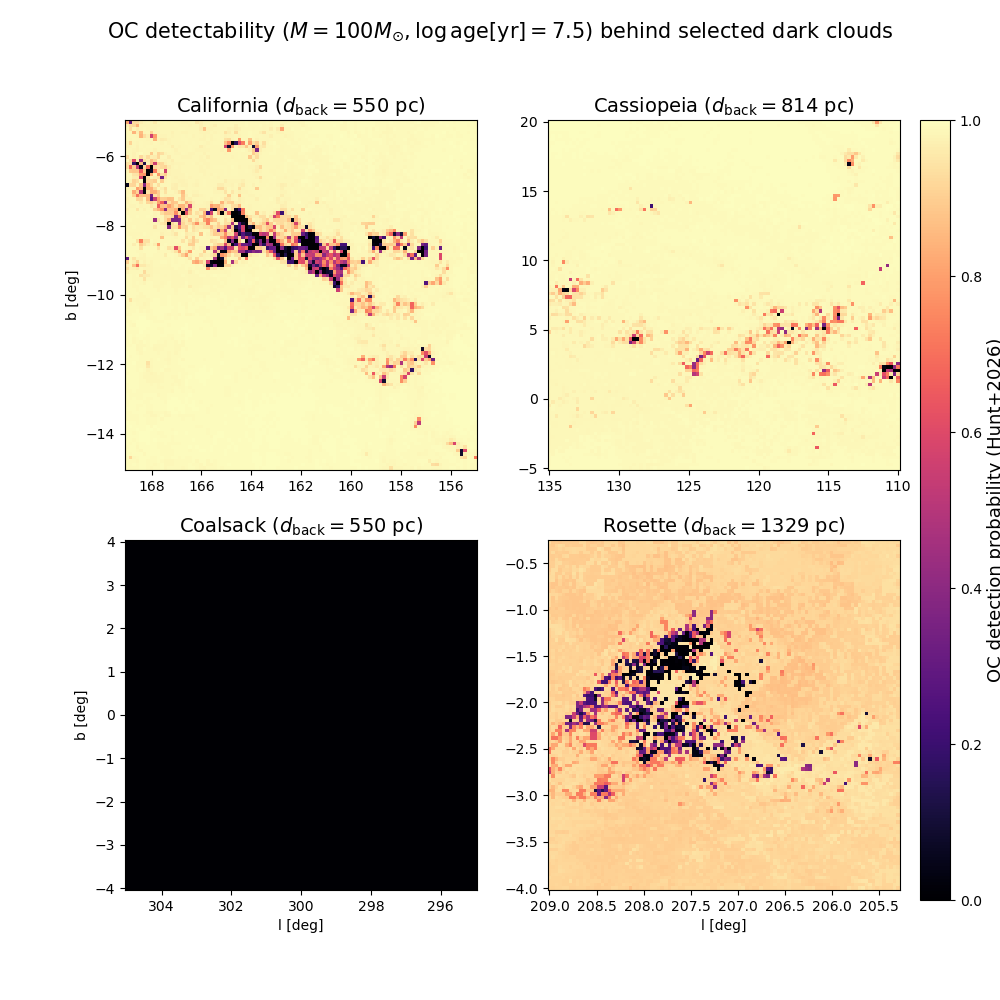}
    \caption{Detection probability of a 30 Myr, 100 $M_{\odot}$ OC located behind the four molecular clouds without any associated OC (see Table \ref{ref:table_N}), as a function of Galactic coordinates. The colour-scale shows the detection probability computed with HDBSCAN using {\tt hr\_selection\_function} \citep{Hunt2026}, adopting the CST threshold of $\tau=5$ used in this work.}
    \label{fig:missing_ocs}
\end{figure}

We further explored the fact that in some MCs no associated OCs were identified in Sect. \ref{sec:data}, i.e. California, Cassiopeia, Coalsack, and Rosette. To this end, we used the {\tt hr\_selection\_function} package\footnote{\url{https://github.com/emilyhunt/hr_selection_function}} \citep{Hunt2026}, which provides a complete implementation of the selection function of the \citet{Hunt2024} catalogue, which is valid for distances of $>500$ pc. This allowed us to estimate the probability that an OC with a given mass, age, position in the Galaxy, and so on would be recovered from the \citet{Hunt2024} catalogue. 

Figure \ref{fig:missing_ocs} shows an example of {\tt hr\_selection\_function} being used for the four MC regions for which no OC members were listed in Table \ref{ref:table_N}. The age was fixed to 30 Myr (an upper limit for the ages of the studied regions) and the mass of the injected OCs to 100 $M_{\odot}$ (a rather low value for a newborn OC). Because the detectability of an OC increases with mass and decreases with age \citep{Hunt2025}, the inferred detection probabilities correspond to a pessimistic case. Even so, we find that the probability of detecting low-mass OCs behind the California, Cassiopeia, and Rosette MCs is generally high; this drops mainly towards the highest extinction (densest) parts of the filaments. When we repeated the calculation for a cluster of mass 500 $M_{\odot}$, the detection probability was very close to 100\% across the full extent of the three clouds. 

Our test indicates that the lack of {\it Gaia}-detected OCs in California, Cassiopeia, and Rosette most likely reflects a genuine absence of clustered star formation above a total mass of $500 M_{\odot}$. Lower mass embedded clusters may still be present in the centre of these clouds, as indicated by the presence of the loosely clustered YSOs. In contrast, for the Coalsack region {\tt hr\_selection\_function} shows that {\it Gaia} is essentially insensitive to the detection of OCs along this LOS. The non-detection places an upper limit on the mass of potentially hidden young clusters of $(5000\pm2000)~M_{\odot}$.

\section{Conclusions}\label{sec:conclusions}

We present the first homogeneous catalogue of 3D kinematics for 15 major molecular cloud complexes within 2.5~kpc of the Sun, based on $\approx$ 24,000 stellar tracers. Our sample combines YSOs and OC members {located in the same star-forming region.}

{We confirm that YSOs and young OCs ($\leq 30$ Myr) exhibit highly consistent kinematics (median $\Delta V_T \simeq 2$~km s$^{-1}$). This agreement, together with the well-established link between YSOs and their parent gas, can therefore be used to trace the galactic trajectories of present-day MCs using a combined sample of YSOs and young OC members.} Indeed, we show two examples of applications of our catalogue by effectively reconstructing the Solar System's past voyage through the Orion cloud as well as the common origin of Lupus, Ophiuchus, and Corona Australis in Sco-Cen.

Beyond their bulk motions, we also characterised internal dynamical trends and detected statistically significant signatures of rotation and expansion in several regions. In particular, Orion and Ophiuchus show  $5\sigma$ expansion signals consistent with cloud dispersal, while coherent rotation is detected in at least six of the MCs, even after correcting from the Galactic shear. These non-zero rotation signatures likely reflect the imprint of stellar feedback and local redistribution of angular momentum.

This work and the accompanying catalogue, based on stellar rather than gaseous tracers, provide a new and complementary view of the large-scale dynamics and feedback history of the interstellar medium. They enable direct comparisons between stellar and gaseous components of molecular clouds, facilitate orbital reconstructions to trace cloud trajectories and possible origins, and provide a benchmark for testing numerical simulations of star formation and Galactic structure. {Overall, our results establish YSOs and young OC members as a powerful, homogeneous sample to characterise the large-scale kinematics of molecular cloud complexes in the Milky Way.}
\begin{acknowledgements}

Scientific progress thrives on discussion and collaboration, and this paper is no exception. We warmly thank the anonymous referee whose insightful comments have greatly improved the overall quality of this paper. We thank Efrem Maconi and Emily Hunt for detailed discussions and suggestions regarding their methodologies. XPC acknowledges financial support from the Spanish National Programme for the Promotion of Talent and its Employability grant PRE2022-104959 cofunded by the European Social Fund. ST acknowledges the funding from the European Union’s Horizon 2020 research and innovation program under the Marie Sk\l{}odowska-Curie grant agreement No 101034413.

XPC, ST, EV and MM acknowledge support from the Spanish Ministerio de Ciencia, Innovación y Universidades under grants PID2021122842OB-C22, PID2021-127289NB-I00  and PID2024157964OB-C22; XPC and MM also from the Xunta de Galicia and the European Union (FEDER Galicia 2021-2027 Program) Ref. ED431B 2024/21, ED431B 2024/02, and CITIC ED431G 2023/01. XPC and MM also acknowledge the support from EU HORIZON-CL4-2023-SPACE-01-71 SPACIOUS project, Ref. 101135205. NMR and FA acknowledge support from the Spanish MICIN/AEI/10.13039/501100011033 and "ERDF, A way of making Europe" by the European Union through grant PID2021-122842OB-C21 and PID2024-157964OB-C21, and the Institute of Cosmos Sciences University of Barcelona (ICCUB, Unidad de Excelencia Mar\'{\i}a de Maeztu) through grant CEX2024-001451-M and the project 2021-SGR-00679 GRC de l'Agència de Gestió d'Ajuts Universitaris i de Recerca (Generalitat de Catalunya). FA acknowledges the grant RYC2021-031683-I funded by MCIN/AEI/10.13039/501100011033 and by the European Union's NextGenerationEU/PRTR. AJM acknowledges support from the Swedish National Space Agency (career grant 2023-00146).
This work has made use of data from the European Space Agency (ESA) Gaia mission and processed by the Gaia Data Processing and Analysis Consortium (DPAC). Funding for the DPAC has been provided by national institutions, in particular, the institutions participating in the {\it Gaia} Multilateral Agreement.

\end{acknowledgements}

\bibliography{aa}

@ARTICLE{Castro-Ginard2018,
       author = {{Castro-Ginard}, A. and {Jordi}, C. and {Luri}, X. and {Julbe}, F. and {Morvan}, M. and {Balaguer-N{\'u}{\~n}ez}, L. and {Cantat-Gaudin}, T.},
        title = "{A new method for unveiling open clusters in Gaia. New nearby open clusters confirmed by DR2}",
      journal = {\aap},
         year = 2018,
        month = oct,
       volume = {618},
          eid = {A59},
        pages = {A59},
          doi = {10.1051/0004-6361/201833390},
archivePrefix = {arXiv},
       eprint = {1805.03045},
 primaryClass = {astro-ph.GA},
       adsurl = {https://ui.adsabs.harvard.edu/abs/2018A&A...618A..59C}
}

@ARTICLE{Hunt2021,
       author = {{Hunt}, Emily L. and {Reffert}, Sabine},
        title = "{Improving the open cluster census. I. Comparison of clustering algorithms applied to Gaia DR2 data}",
      journal = {\aap},
         year = 2021,
        month = feb,
       volume = {646},
          eid = {A104},
        pages = {A104},
          doi = {10.1051/0004-6361/202039341},
archivePrefix = {arXiv},
       eprint = {2012.04267},
 primaryClass = {astro-ph.GA},
       adsurl = {https://ui.adsabs.harvard.edu/abs/2021A&A...646A.104H}
}

@ARTICLE{Miret-Roig2018,
       author = {{Miret-Roig}, N. and {Antoja}, T. and {Romero-G{\'o}mez}, M. and {Figueras}, F.},
        title = "{Dynamical ages of the young local associations with Gaia}",
      journal = {\aap},
         year = 2018,
        month = jul,
       volume = {615},
          eid = {A51},
        pages = {A51},
          doi = {10.1051/0004-6361/201731976},
archivePrefix = {arXiv},
       eprint = {1803.00573},
 primaryClass = {astro-ph.GA},
       adsurl = {https://ui.adsabs.harvard.edu/abs/2018A&A...615A..51M}
}

@ARTICLE{Miret-Roig2020,
       author = {{Miret-Roig}, N. and {Galli}, P.~A.~B. and {Brandner}, W. and {Bouy}, H. and {Barrado}, D. and {Olivares}, J. and {Antoja}, T. and {Romero-G{\'o}mez}, M. and {Figueras}, F. and {Lillo-Box}, J.},
        title = "{Dynamical traceback age of the {\ensuremath{\beta}} Pictoris moving group}",
      journal = {\aap},
         year = 2020,
        month = oct,
       volume = {642},
          eid = {A179},
        pages = {A179},
          doi = {10.1051/0004-6361/202038765},
archivePrefix = {arXiv},
       eprint = {2007.10997},
 primaryClass = {astro-ph.GA},
       adsurl = {https://ui.adsabs.harvard.edu/abs/2020A&A...642A.179M}
}

@ARTICLE{Miret-Roig+2024,
       author = {{Miret-Roig}, N{\'u}ria and {Alves}, Jo{\~a}o and {Barrado}, David and {Burkert}, Andreas and {Ratzenb{\"o}ck}, Sebastian and {Konietzka}, Ralf},
        title = "{Insights into star formation and dispersal from the synchronization of stellar clocks}",
      journal = {Nat. Astron.},
         year = 2024,
        month = feb,
       volume = {8},
        pages = {216-222},
          doi = {10.1038/s41550-023-02132-4},
archivePrefix = {arXiv},
       eprint = {2311.13042},
 primaryClass = {astro-ph.SR},
       adsurl = {https://ui.adsabs.harvard.edu/abs/2024NatAs...8..216M}
}

@ARTICLE{Hunt2025,
       author = {{Hunt}, Emily L. and {Cantat-Gaudin}, Tristan and {Anders}, Friedrich and {Spina}, Lorenzo and {Cavallo}, Lorenzo and {Castro-Ginard}, Alfred and {Belokurov}, Vasily and {Brown}, Anthony G.~A. and {Casey}, Andrew R. and {Drimmel}, Ronald and {Fouesneau}, Morgan and {Reffert}, Sabine},
        title = "{The completeness of the open cluster census towards the Galactic anticentre}",
      journal = {\aap},
         year = 2025,
        month = jul,
       volume = {699},
          eid = {A273},
        pages = {A273},
          doi = {10.1051/0004-6361/202452614},
archivePrefix = {arXiv},
       eprint = {2506.18708},
 primaryClass = {astro-ph.GA},
       adsurl = {https://ui.adsabs.harvard.edu/abs/2025A&A...699A.273H}
}

@ARTICLE{Hunt2026,
       author = {{Hunt}, Emily L. and {Cantat-Gaudin}, Tristan and {Anders}, Friedrich and {Malhotra}, Sagar and {Spina}, Lorenzo and {Castro-Ginard}, Alfred and {Cavallo}, Lorenzo},
        title = "{The selection function of the Gaia DR3 open cluster census}",
      journal = {\aap},
         year = 2026,
        month = feb,
       volume = {706},
          eid = {A341},
        pages = {A341},
          doi = {10.1051/0004-6361/202557781},
archivePrefix = {arXiv},
       eprint = {2510.18343},
 primaryClass = {astro-ph.GA},
       adsurl = {https://ui.adsabs.harvard.edu/abs/2026A&A...706A.341H}
}

@ARTICLE{Asiain1999,
       author = {{Asiain}, R. and {Figueras}, F. and {Torra}, J.},
        title = "{On the evolution of moving groups: an application to the Pleiades moving group}",
      journal = {\aap},
         year = 1999,
        month = oct,
       volume = {350},
        pages = {434-446},
          doi = {10.48550/arXiv.astro-ph/9909050},
archivePrefix = {arXiv},
       eprint = {astro-ph/9909050},
 primaryClass = {astro-ph},
       adsurl = {https://ui.adsabs.harvard.edu/abs/1999A&A...350..434A}
}

@ARTICLE{Brown1997,
       author = {{Brown}, A.~G.~A. and {Dekker}, G. and {de Zeeuw}, P.~T.},
        title = "{Kinematic ages of OB associations}",
      journal = {\mnras},
         year = 1997,
        month = mar,
       volume = {285},
       number = {3},
        pages = {479-492},
          doi = {10.1093/mnras/285.3.479},
       adsurl = {https://ui.adsabs.harvard.edu/abs/1997MNRAS.285..479B}
}

@ARTICLE{GRAVITYCollaboration2019,
       author = {{GRAVITY Collaboration} and {Abuter}, R. and {Amorim}, A. and {Baub{\"o}ck}, M. and {Berger}, J.~P. and {Bonnet}, H. and {Brandner}, W. and {Cl{\'e}net}, Y. and {Coud{\'e} Du Foresto}, V. and {de Zeeuw}, P.~T. and {Dexter}, J. and {Duvert}, G. and {Eckart}, A. and {Eisenhauer}, F. and {F{\"o}rster Schreiber}, N.~M. and {Garcia}, P. and {Gao}, F. and {Gendron}, E. and {Genzel}, R. and {Gerhard}, O. and {Gillessen}, S. and {Habibi}, M. and {Haubois}, X. and {Henning}, T. and {Hippler}, S. and {Horrobin}, M. and {Jim{\'e}nez-Rosales}, A. and {Jocou}, L. and {Kervella}, P. and {Lacour}, S. and {Lapeyr{\`e}re}, V. and {Le Bouquin}, J.-B. and {L{\'e}na}, P. and {Ott}, T. and {Paumard}, T. and {Perraut}, K. and {Perrin}, G. and {Pfuhl}, O. and {Rabien}, S. and {Rodriguez Coira}, G. and {Rousset}, G. and {Scheithauer}, S. and {Sternberg}, A. and {Straub}, O. and {Straubmeier}, C. and {Sturm}, E. and {Tacconi}, L.~J. and {Vincent}, F. and {von Fellenberg}, S. and {Waisberg}, I. and {Widmann}, F. and {Wieprecht}, E. and {Wiezorrek}, E. and {Woillez}, J. and {Yazici}, S.},
        title = "{A geometric distance measurement to the Galactic center black hole with 0.3\% uncertainty}",
      journal = {\aap},
         year = 2019,
        month = may,
       volume = {625},
          eid = {L10},
        pages = {L10},
          doi = {10.1051/0004-6361/201935656},
archivePrefix = {arXiv},
       eprint = {1904.05721},
 primaryClass = {astro-ph.GA},
       adsurl = {https://ui.adsabs.harvard.edu/abs/2019A&A...625L..10G}
}

@ARTICLE{Abdurrouf2022,
  title     = "The seventeenth data release of the Sloan Digital Sky Surveys:
               Complete release of {MaNGA}, {MaStar}, and {APOGEE-2} data",
  author    = "{Abdurro'uf} and Accetta, Katherine and Aerts, Conny and Silva
               Aguirre, V{\'\i}ctor and Ahumada, Romina and Ajgaonkar, Nikhil
               and Filiz Ak, N and Alam, Shadab and Allende Prieto, Carlos and
               Almeida, Andr{\'e}s and Anders, Friedrich and Anderson, Scott F
               and Andrews, Brett H and Anguiano, Borja and Aquino-Ort{\'\i}z,
               Erik and Arag{\'o}n-Salamanca, Alfonso and Argudo-Fern{\'a}ndez,
               Maria and Ata, Metin and Aubert, Marie and Avila-Reese, Vladimir
               and Badenes, Carles and Barb{\'a}, Rodolfo H and Barger, Kat and
               Barrera-Ballesteros, Jorge K and Beaton, Rachael L and Beers,
               Timothy C and Belfiore, Francesco and Bender, Chad F and
               Bernardi, Mariangela and Bershady, Matthew A and Beutler,
               Florian and Bidin, Christian Moni and Bird, Jonathan C and
               Bizyaev, Dmitry and Blanc, Guillermo A and Blanton, Michael R
               and Boardman, Nicholas Fraser and Bolton, Adam S and Boquien,
               M{\'e}d{\'e}ric and Borissova, Jura and Bovy, Jo and Brandt, W N
               and Brown, Jordan and Brownstein, Joel R and Brusa, Marcella and
               Buchner, Johannes and Bundy, Kevin and Burchett, Joseph N and
               {Martin Bureau} and Burgasser, Adam and Cabang, Tuesday K and
               Campbell, Stephanie and Cappellari, Michele and Carlberg, Joleen
               K and Wanderley, F{\'a}bio Carneiro and Carrera, Ricardo and
               Cash, Jennifer and Chen, Yan-Ping and Chen, Wei-Huai and
               Cherinka, Brian and Chiappini, Cristina and Choi, Peter Doohyun
               and Chojnowski, S Drew and Chung, Haeun and Clerc, Nicolas and
               Cohen, Roger E and Comerford, Julia M and Comparat, Johan and da
               Costa, Luiz and Covey, Kevin and Crane, Jeffrey D and
               Cruz-Gonzalez, Irene and Culhane, Connor and Cunha, Katia and
               Dai, Y Sophia and Damke, Guillermo and Darling, Jeremy and
               Davidson, Jr, James W and Davies, Roger and Dawson, Kyle and De
               Lee, Nathan and Diamond-Stanic, Aleksandar M and Cano-D{\'\i}az,
               Mariana and S{\'a}nchez, Helena Dom{\'\i}nguez and Donor, John
               and Duckworth, Chris and Dwelly, Tom and Eisenstein, Daniel J
               and Elsworth, Yvonne P and Emsellem, Eric and Eracleous, Mike
               and Escoffier, Stephanie and Fan, Xiaohui and Farr, Emily and
               Feng, Shuai and Fern{\'a}ndez-Trincado, Jos{\'e} G and Feuillet,
               Diane and Filipp, Andreas and Fillingham, Sean P and Frinchaboy,
               Peter M and Fromenteau, Sebastien and Galbany, Llu{\'\i}s and
               Garc{\'\i}a, Rafael A and Garc{\'\i}a-Hern{\'a}ndez, D A and Ge,
               Junqiang and Geisler, Doug and Gelfand, Joseph and G{\'e}ron,
               Tobias and Gibson, Benjamin J and Goddy, Julian and
               Godoy-Rivera, Diego and Grabowski, Kathleen and Green, Paul J
               and Greener, Michael and Grier, Catherine J and Griffith, Emily
               and Guo, Hong and Guy, Julien and Hadjara, Massinissa and
               Harding, Paul and Hasselquist, Sten and Hayes, Christian R and
               Hearty, Fred and Hern{\'a}ndez, Jes{\'u}s and Hill, Lewis and
               Hogg, David W and Holtzman, Jon A and Horta, Danny and Hsieh,
               Bau-Ching and Hsu, Chin-Hao and Hsu, Yun-Hsin and Huber, Daniel
               and {Marc Huertas-Company} and Hutchinson, Brian and Hwang, Ho
               Seong and Ibarra-Medel, H{\'e}ctor J and Chitham, Jacob Ider and
               Ilha, Gabriele S and Imig, Julie and Jaekle, Will and
               Jayasinghe, Tharindu and Ji, Xihan and Johnson, Jennifer A and
               Jones, Amy and J{\"o}nsson, Henrik and Katkov, Ivan and
               Khalatyan, Dr Arman and Kinemuchi, Karen and Kisku, Shobhit and
               Knapen, Johan H and Kneib, Jean-Paul and Kollmeier, Juna A and
               Kong, Miranda and Kounkel, Marina and Kreckel, Kathryn and
               Krishnarao, Dhanesh and Lacerna, Ivan and Lane, Richard R and
               Langgin, Rachel and Lavender, Ramon and Law, David R and Lazarz,
               Daniel and Leung, Henry W and Leung, Ho-Hin and Lewis, Hannah M
               and Li, Cheng and Li, Ran and Lian, Jianhui and Liang, Fu-Heng
               and Lin, Lihwai and Lin, Yen-Ting and Lin, Sicheng and Lintott,
               Chris and Long, Dan and Longa-Pe{\~n}a, Pen{\'e}lope and
               L{\'o}pez-Cob{\'a}, Carlos and Lu, Shengdong and Lundgren, Britt
               F and Luo, Yuanze and Mackereth, J Ted and de la Macorra, Axel
               and Mahadevan, Suvrath and Majewski, Steven R and Manchado,
               Arturo and Mandeville, Travis and Maraston, Claudia and
               Margalef-Bentabol, Berta and Masseron, Thomas and Masters, Karen
               L and Mathur, Savita and McDermid, Richard M and Mckay, Myles
               and Merloni, Andrea and Merrifield, Michael and Meszaros,
               Szabolcs and Miglio, Andrea and Di Mille, Francesco and Minniti,
               Dante and Minsley, Rebecca and Monachesi, Antonela and Moon,
               Jeongin and Mosser, Benoit and Mulchaey, John and Muna, Demitri
               and Mu{\~n}oz, Ricardo R and Myers, Adam D and Myers, Natalie
               and Nadathur, Seshadri and Nair, Preethi and Nandra, Kirpal and
               Neumann, Justus and Newman, Jeffrey A and Nidever, David L and
               Nikakhtar, Farnik and Nitschelm, Christian and O'Connell, Julia
               E and Garma-Oehmichen, Luis and Luan Souza de Oliveira, Gabriel
               and Olney, Richard and Oravetz, Daniel and Ortigoza-Urdaneta,
               Mario and Osorio, Yeisson and Otter, Justin and Pace, Zachary J
               and Padilla, Nelson and Pan, Kaike and Pan, Hsi-An and Parikh,
               Taniya and Parker, James and Peirani, Sebastien and Pe{\~n}a
               Ram{\'\i}rez, Karla and Penny, Samantha and Percival, Will J and
               Perez-Fournon, Ismael and Pinsonneault, Marc and Poidevin,
               Fr{\'e}d{\'e}rick and Poovelil, Vijith Jacob and Price-Whelan,
               Adrian M and B{\'a}rbara de Andrade Queiroz, Anna and Raddick, M
               Jordan and Ray, Amy and Rembold, Sandro Barboza and Riddle,
               Nicole and Riffel, Rogemar A and Riffel, Rog{\'e}rio and Rix,
               Hans-Walter and Robin, Annie C and Rodr{\'\i}guez-Puebla, Aldo
               and Roman-Lopes, Alexandre and Rom{\'a}n-Z{\'u}{\~n}iga, Carlos
               and Rose, Benjamin and Ross, Ashley J and Rossi, Graziano and
               Rubin, Kate H R and Salvato, Mara and S{\'a}nchez, Seb{\'a}stian
               F and S{\'a}nchez-Gallego, Jos{\'e} R and Sanderson, Robyn and
               Santana Rojas, Felipe Antonio and Sarceno, Edgar and Sarmiento,
               Regina and Sayres, Conor and Sazonova, Elizaveta and Schaefer,
               Adam L and Schiavon, Ricardo and Schlegel, David J and
               Schneider, Donald P and Schultheis, Mathias and Schwope, Axel
               and Serenelli, Aldo and Serna, Javier and Shao, Zhengyi and
               Shapiro, Griffin and Sharma, Anubhav and Shen, Yue and Shetrone,
               Matthew and Shu, Yiping and Simon, Joshua D and Skrutskie, M F
               and Smethurst, Rebecca and Smith, Verne and Sobeck, Jennifer and
               Spoo, Taylor and Sprague, Dani and Stark, David V and Stassun,
               Keivan G and Steinmetz, Matthias and Stello, Dennis and
               Stone-Martinez, Alexander and Storchi-Bergmann, Thaisa and
               Stringfellow, Guy S and Stutz, Amelia and Su, Yung-Chau and
               Taghizadeh-Popp, Manuchehr and Talbot, Michael S and Tayar,
               Jamie and Telles, Eduardo and Teske, Johanna and Thakar, Ani and
               Theissen, Christopher and Tkachenko, Andrew and Thomas, Daniel
               and Tojeiro, Rita and Hernandez Toledo, Hector and Troup,
               Nicholas W and Trump, Jonathan R and Trussler, James and Turner,
               Jacqueline and Tuttle, Sarah and Unda-Sanzana, Eduardo and
               V{\'a}zquez-Mata, Jos{\'e} Antonio and Valentini, Marica and
               Valenzuela, Octavio and Vargas-Gonz{\'a}lez, Jaime and
               Vargas-Maga{\~n}a, Mariana and Alfaro, Pablo Vera and Villanova,
               Sandro and Vincenzo, Fiorenzo and Wake, David and Warfield, Jack
               T and Washington, Jessica Diane and Weaver, Benjamin Alan and
               Weijmans, Anne-Marie and Weinberg, David H and Weiss, Achim and
               Westfall, Kyle B and Wild, Vivienne and Wilde, Matthew C and
               Wilson, John C and Wilson, Robert F and Wilson, Mikayla and
               Wolf, Julien and Wood-Vasey, W M and Yan, Renbin and Zamora,
               Olga and Zasowski, Gail and Zhang, Kai and Zhao, Cheng and
               Zheng, Zheng and Zheng, Zheng and Zhu, Kai",
  journal   = {\apjs},
  publisher = "American Astronomical Society",
  volume    =  259,
  number    =  2,
  pages     = "35",
  month     =  apr,
  year      =  2022,
  copyright = "http://creativecommons.org/licenses/by/4.0/"
}

@ARTICLE{Alves2020,
  title     = "A Galactic-scale gas wave in the solar neighbourhood",
  author    = "Alves, Jo{\~a}o and Zucker, Catherine and Goodman, Alyssa A and
               Speagle, Joshua S and Meingast, Stefan and Robitaille, Thomas
               and Finkbeiner, Douglas P and Schlafly, Edward F and Green,
               Gregory M",
  journal   = "Nature",
  publisher = "Springer Science and Business Media LLC",
  volume    =  578,
  number    =  7794,
  pages     = "237--239",
  month     =  feb,
  year      =  2020,
  language  = "en"
}

@INCOLLECTION{Andre2014,
       author = {{Andr{\'e}}, P. and {Di Francesco}, J. and {Ward-Thompson}, D. and {Inutsuka}, S.-I. and {Pudritz}, R.~E. and {Pineda}, J.~E.},
        title = "{From Filamentary Networks to Dense Cores in Molecular Clouds: Toward a New Paradigm for Star Formation}",
    booktitle = {Protostars and Planets VI},
         year = 2014,
       editor = {{Beuther}, Henrik and {Klessen}, Ralf S. and {Dullemond}, Cornelis P. and {Henning}, Thomas},
        month = jan,
        pages = {27-51},
          doi = {10.2458/azu_uapress_9780816531240-ch002},
archivePrefix = {arXiv},
       eprint = {1312.6232},
 primaryClass = {astro-ph.GA},
       adsurl = {https://ui.adsabs.harvard.edu/abs/2014prpl.conf...27A}
}

@ARTICLE{Andrae2023,
  title     = "\textit{Gaia} Data Release 3",
  author    = "Andrae, R and Fouesneau, M and Sordo, R and Bailer-Jones, C A L
               and Dharmawardena, T E and Rybizki, J and De Angeli, F and
               Lindstr{\o}m, H E P and Marshall, D J and Drimmel, R and Korn, A
               J and Soubiran, C and Brouillet, N and Casamiquela, L and Rix,
               H-W and Abreu Aramburu, A and {\'A}lvarez, M A and Bakker, J and
               Bellas-Velidis, I and Bijaoui, A and Brugaletta, E and Burlacu,
               A and Carballo, R and Chaoul, L and Chiavassa, A and Contursi, G
               and Cooper, W J and Creevey, O L and Dafonte, C and Dapergolas,
               A and de Laverny, P and Delchambre, L and Demouchy, C and
               Edvardsson, B and Fr{\'e}mat, Y and Garabato, D and
               Garc{\'\i}a-Lario, P and Garc{\'\i}a-Torres, M and Gavel, A and
               Gomez, A and Gonz{\'a}lez-Santamar{\'\i}a, I and Hatzidimitriou,
               D and Heiter, U and Jean-Antoine Piccolo, A and Kontizas, M and
               Kordopatis, G and Lanzafame, A C and Lebreton, Y and Licata, E L
               and Livanou, E and Lobel, A and Lorca, A and Magdaleno Romeo, A
               and Manteiga, M and Marocco, F and Mary, N and Nicolas, C and
               Ordenovic, C and Pailler, F and Palicio, P A and
               Pallas-Quintela, L and Panem, C and Pichon, B and Poggio, E and
               Recio-Blanco, A and Riclet, F and Robin, C and Santove{\~n}a, R
               and Sarro, L M and Schultheis, M S and Segol, M and Silvelo, A
               and Slezak, I and Smart, R L and S{\"u}veges, M and
               Th{\'e}venin, F and Torralba Elipe, G and Ulla, A and Utrilla, E
               and Vallenari, A and van Dillen, E and Zhao, H and Zorec, J",
  journal   = {\aap},
  publisher = "EDP Sciences",
  volume    =  674,
  pages     = "A27",
  month     =  jun,
  year      =  2023,
  copyright = "https://creativecommons.org/licenses/by/4.0"
}

@ARTICLE{The_Astropy_Collaboration2022,
  title     = "The Astropy Project: Sustaining and growing a community-oriented
               open-source Project and the latest major release (v5.0) of the
               core package",
  author    = "{Astropy Collaboration} and Price-Whelan, Adrian M and Lim,
               Pey Lian and Earl, Nicholas and Starkman, Nathaniel and Bradley,
               Larry and Shupe, David L and Patil, Aarya A and Corrales, Lia
               and Brasseur, C E and N{\"o}the, Maximilian and Donath, Axel and
               Tollerud, Erik and Morris, Brett M and Ginsburg, Adam and Vaher,
               Eero and Weaver, Benjamin A and Tocknell, James and Jamieson,
               William and van Kerkwijk, Marten H and Robitaille, Thomas P and
               Merry, Bruce and Bachetti, Matteo and G{\"u}nther, H Moritz and
               Authors, Paper and Aldcroft, Thomas L and Alvarado-Montes, Jaime
               A and Archibald, Anne M and B{\'o}di, Attila and Bapat, Shreyas
               and Barentsen, Geert and Baz{\'a}n, Juanjo and Biswas, Manish
               and Boquien, M{\'e}d{\'e}ric and Burke, D J and Cara, Daria and
               Cara, Mihai and Conroy, Kyle E and Conseil, Simon and Craig,
               Matthew W and Cross, Robert M and Cruz, Kelle L and D'Eugenio,
               Francesco and Dencheva, Nadia and Devillepoix, Hadrien A R and
               Dietrich, J{\"o}rg P and Eigenbrot, Arthur Davis and Erben,
               Thomas and Ferreira, Leonardo and Foreman-Mackey, Daniel and
               Fox, Ryan and Freij, Nabil and Garg, Suyog and Geda, Robel and
               Glattly, Lauren and Gondhalekar, Yash and Gordon, Karl D and
               Grant, David and Greenfield, Perry and Groener, Austen M and
               Guest, Steve and Gurovich, Sebastian and Handberg, Rasmus and
               Hart, Akeem and Hatfield-Dodds, Zac and Homeier, Derek and
               Hosseinzadeh, Griffin and Jenness, Tim and Jones, Craig K and
               Joseph, Prajwel and Kalmbach, J Bryce and Karamehmetoglu, Emir
               and Ka{\l}uszy{\'n}ski, Miko{\l}aj and Kelley, Michael S P and
               Kern, Nicholas and Kerzendorf, Wolfgang E and Koch, Eric W and
               Kulumani, Shankar and Lee, Antony and Ly, Chun and Ma, Zhiyuan
               and MacBride, Conor and Maljaars, Jakob M and Muna, Demitri and
               Murphy, N A and Norman, Henrik and O'Steen, Richard and Oman,
               Kyle A and Pacifici, Camilla and Pascual, Sergio and
               Pascual-Granado, J and Patil, Rohit R and Perren, Gabriel I and
               Pickering, Timothy E and Rastogi, Tanuj and Roulston, Benjamin R
               and Ryan, Daniel F and Rykoff, Eli S and Sabater, Jose and
               Sakurikar, Parikshit and Salgado, Jes{\'u}s and Sanghi, Aniket
               and Saunders, Nicholas and Savchenko, Volodymyr and Schwardt,
               Ludwig and Seifert-Eckert, Michael and Shih, Albert Y and Jain,
               Anany Shrey and Shukla, Gyanendra and Sick, Jonathan and
               Simpson, Chris and Singanamalla, Sudheesh and Singer, Leo P and
               Singhal, Jaladh and Sinha, Manodeep and Sip{\H o}cz, Brigitta M
               and Spitler, Lee R and Stansby, David and Streicher, Ole and {\v
               S}umak, Jani and Swinbank, John D and Taranu, Dan S and Tewary,
               Nikita and Tremblay, Grant R and Val-Borro, Miguel de and Van
               Kooten, Samuel J and Vasovi{\'c}, Zlatan and Verma, Shresth and
               de Miranda Cardoso, Jos{\'e} Vin{\'\i}cius and Williams, Peter K
               G and Wilson, Tom J and Winkel, Benjamin and Wood-Vasey, W M and
               Xue, Rui and Yoachim, Peter and Zhang, Chen and Zonca, Andrea
               and {Astropy Project Contributors}",
  journal   = {\apj},
  publisher = "American Astronomical Society",
  volume    =  935,
  number    =  2,
  pages     = "167",
  month     =  aug,
  year      =  2022,
  copyright = "http://creativecommons.org/licenses/by/4.0/"
}

@ARTICLE{Berlanas2023,
  title     = "{\textit{Gaia}-ESO} survey: Massive stars in the Carina Nebula",
  author    = "Berlanas, S R and Ma{\'\i}z Apell{\'a}niz, J and Herrero, A and
               Mahy, L and Blomme, R and Negueruela, I and Dorda, R and
               Comer{\'o}n, F and Gosset, E and Pantaleoni Gonz{\'a}lez, M and
               Molina Lera, J A and Sota, A and Furst, T and Alfaro, E J and
               Bergemann, M and Carraro, G and Drew, J E and Morbidelli, L and
               Vink, J S",
  journal   = {\aap},
  publisher = "EDP Sciences",
  volume    =  671,
  pages     = "A20",
  month     =  mar,
  year      =  2023,
  copyright = "https://creativecommons.org/licenses/by/4.0"
}

@ARTICLE{Cahlon2024,
  title     = "A parsec-scale catalog of molecular clouds in the solar
               neighborhood based on {3D} dust mapping: Implications for the
               mass--size relation",
  author    = "Cahlon, Shlomo and Zucker, Catherine and Goodman, Alyssa and
               Lada, Charles and Alves, Jo{\~a}o",
  journal   = {\apj},
  publisher = "American Astronomical Society",
  volume    =  961,
  number    =  2,
  pages     = "153",
  month     =  feb,
  year      =  2024,
  copyright = "http://creativecommons.org/licenses/by/4.0/"
}

@ARTICLE{Cambresy2013,
  title     = "Young stellar clusters in the Rosette molecular cloud",
  author    = "Cambrésy, L and Marton, G and Feher, O and T{\'o}th, L V and
               Schneider, N",
  journal   = {\aap},
  publisher = "EDP Sciences",
  volume    =  557,
  pages     = "A29",
  month     =  sep,
  year      =  2013
}

@ARTICLE{Castro-Ginard2021,
  title     = "Milky Way spiral arms from open clusters in Gaia {EDR3}",
  author    = "Castro-Ginard, A and McMillan, P J and Luri, X and Jordi, C and
               Romero-G{\'o}mez, M and Cantat-Gaudin, T and Casamiquela, L and
               Tarricq, Y and Soubiran, C and Anders, F",
  journal   = {\aap},
  publisher = "EDP Sciences",
  volume    =  652,
  pages     = "A162",
  month     =  aug,
  year      =  2021,
  copyright = "https://www.edpsciences.org/en/authors/copyright-and-licensing"
}

@ARTICLE{Chen2020,
  title     = "A large catalogue of molecular clouds with accurate distances
               within 4 kpc of the Galactic disc",
  author    = "Chen, B-Q and Li, G-X and Yuan, H-B and Huang, Y and Tian, Z-J
               and Wang, H-F and Zhang, H-W and Wang, C and Liu, X-W",
  journal   = {\mnras},
  publisher = "Oxford University Press (OUP)",
  volume    =  493,
  number    =  1,
  pages     = "351--361",
  month     =  mar,
  year      =  2020,
  copyright = "https://academic.oup.com/journals/pages/open\_access/funder\_policies/chorus/standard\_publication\_model",
  language  = "en"
}

@ARTICLE{Dharmawardena2022,
  title     = "Three-dimensional dust density structure of the Orion, Cygnus X,
               Taurus, and Perseus star-forming regions",
  author    = "Dharmawardena, T E and Bailer-Jones, C A L and Fouesneau, M and
               Foreman-Mackey, D",
  journal   = {\aap},
  publisher = "EDP Sciences",
  volume    =  658,
  pages     = "A166",
  month     =  feb,
  year      =  {2022},
  copyright = "https://creativecommons.org/licenses/by/4.0"
}

@ARTICLE{Dharmawardena2023,
       author = {{Dharmawardena}, T.~E. and {Bailer-Jones}, C.~A.~L. and {Fouesneau}, M. and {Foreman-Mackey}, D. and {Coronica}, P. and {Colnaghi}, T. and {M{\"u}ller}, T. and {Henshaw}, J.},
        title = "{The three-dimensional structure of galactic molecular cloud complexes out to 2.5 kpc}",
      journal = {\mnras},
         year = 2023,
        month = feb,
       volume = {519},
       number = {1},
        pages = {228-247},
          doi = {10.1093/mnras/stac2790},
archivePrefix = {arXiv},
       eprint = {2210.03615},
 primaryClass = {astro-ph.GA},
       adsurl = {https://ui.adsabs.harvard.edu/abs/2023MNRAS.519..228D}
}

@ARTICLE{Dias2005,
  title     = "Direct determination of the spiral pattern rotation speed of the
               galaxy",
  author    = "Dias, Wilton S and Lepine, J R D",
  journal   = {\apj},
  publisher = "American Astronomical Society",
  volume    =  629,
  number    =  2,
  pages     = "825--831",
  month     =  aug,
  year      =  2005,
  language  = "en"
}

@ARTICLE{Digel1996,
  title     = "A large-scale {CO} survey toward W3, W4, and {W5}",
  author    = "Digel, Seth W and Lyder, David A and Philbrick, Amy J and Puche,
               Daniel and Thaddeus, Patrick",
  journal   = {\apj},
  publisher = "IOP Publishing",
  volume    =  458,
  pages     = "561",
  month     =  feb,
  year      =  1996
}

@ARTICLE{Dong2024,
  title     = "{3D} morphology and motions of the Canis Major region from Gaia
               {DR3}",
  author    = "Dong, Yiwei and Xu, Ye and Hao, Chaojie and Li, Yingjie and Liu,
               Dejian and Sun, Yan and Lin, Zehao",
  journal   = {\aj},
  publisher = "American Astronomical Society",
  volume    =  168,
  number    =  5,
  pages     = "225",
  month     =  nov,
  year      =  2024,
  copyright = "http://creativecommons.org/licenses/by/4.0/"
}

@ARTICLE{Dzib2018,
  title     = "Distances and kinematics of Gould belt star-forming regions with
               Gaia {DR2} results",
  author    = "Dzib, Sergio A and Loinard, Laurent and Ortiz-Le{\'o}n, Gisela N
               and Rodr{\'\i}guez, Luis F and Galli, Phillip A B",
  journal   = {\apj},
  publisher = "American Astronomical Society",
  volume    =  867,
  number    =  2,
  pages     = "151",
  month     =  nov,
  year      =  2018,
  copyright = "http://iopscience.iop.org/page/copyright"
}

@ARTICLE{Gaia_Collaboration2016,
  title     = "{The\textit{Gaia}mission}",
  author    = "{Gaia Collaboration} and Prusti, T and de Bruijne, J H J and
               Brown, A G A and Vallenari, A and Babusiaux, C and Bailer-Jones,
               C A L and Bastian, U and Biermann, M and Evans, D W and Eyer, L
               and Jansen, F and Jordi, C and Klioner, S A and Lammers, U and
               Lindegren, L and Luri, X and Mignard, F and Milligan, D J and
               Panem, C and Poinsignon, V and Pourbaix, D and Randich, S and
               Sarri, G and Sartoretti, P and Siddiqui, H I and Soubiran, C and
               Valette, V and van Leeuwen, F and Walton, N A and Aerts, C and
               Arenou, F and Cropper, M and Drimmel, R and H{\o}g, E and Katz,
               D and Lattanzi, M G and O'Mullane, W and Grebel, E K and
               Holland, A D and Huc, C and Passot, X and Bramante, L and
               Cacciari, C and Casta{\~n}eda, J and Chaoul, L and Cheek, N and
               De Angeli, F and Fabricius, C and Guerra, R and Hern{\'a}ndez, J
               and Jean-Antoine-Piccolo, A and Masana, E and Messineo, R and
               Mowlavi, N and Nienartowicz, K and Ord{\'o}{\~n}ez-Blanco, D and
               Panuzzo, P and Portell, J and Richards, P J and Riello, M and
               Seabroke, G M and Tanga, P and Th{\'e}venin, F and Torra, J and
               Els, S G and Gracia-Abril, G and Comoretto, G and
               Garcia-Reinaldos, M and Lock, T and Mercier, E and Altmann, M
               and Andrae, R and Astraatmadja, T L and Bellas-Velidis, I and
               Benson, K and Berthier, J and Blomme, R and Busso, G and Carry,
               B and Cellino, A and Clementini, G and Cowell, S and Creevey, O
               and Cuypers, J and Davidson, M and De Ridder, J and de Torres, A
               and Delchambre, L and Dell'Oro, A and Ducourant, C and
               Fr{\'e}mat, Y and Garc{\'\i}a-Torres, M and Gosset, E and
               Halbwachs, J-L and Hambly, N C and Harrison, D L and Hauser, M
               and Hestroffer, D and Hodgkin, S T and Huckle, H E and Hutton, A
               and Jasniewicz, G and Jordan, S and Kontizas, M and Korn, A J
               and Lanzafame, A C and Manteiga, M and Moitinho, A and Muinonen,
               K and Osinde, J and Pancino, E and Pauwels, T and Petit, J-M and
               Recio-Blanco, A and Robin, A C and Sarro, L M and Siopis, C and
               Smith, M and Smith, K W and Sozzetti, A and Thuillot, W and van
               Reeven, W and Viala, Y and Abbas, U and Abreu Aramburu, A and
               Accart, S and Aguado, J J and Allan, P M and Allasia, W and
               Altavilla, G and {\'A}lvarez, M A and Alves, J and Anderson, R I
               and Andrei, A H and Anglada Varela, E and Antiche, E and Antoja,
               T and Ant{\'o}n, S and Arcay, B and Atzei, A and Ayache, L and
               Bach, N and Baker, S G and Balaguer-N{\'u}{\~n}ez, L and
               Barache, C and Barata, C and Barbier, A and Barblan, F and
               Baroni, M and Barrado y Navascu{\'e}s, D and Barros, M and
               Barstow, M A and Becciani, U and Bellazzini, M and Bellei, G and
               Bello Garc{\'\i}a, A and Belokurov, V and Bendjoya, P and
               Berihuete, A and Bianchi, L and Bienaym{\'e}, O and Billebaud, F
               and Blagorodnova, N and Blanco-Cuaresma, S and Boch, T and
               Bombrun, A and Borrachero, R and Bouquillon, S and Bourda, G and
               Bouy, H and Bragaglia, A and Breddels, M A and Brouillet, N and
               Br{\"u}semeister, T and Bucciarelli, B and Budnik, F and
               Burgess, P and Burgon, R and Burlacu, A and Busonero, D and
               Buzzi, R and Caffau, E and Cambras, J and Campbell, H and
               Cancelliere, R and Cantat-Gaudin, T and Carlucci, T and
               Carrasco, J M and Castellani, M and Charlot, P and Charnas, J
               and Charvet, P and Chassat, F and Chiavassa, A and Clotet, M and
               Cocozza, G and Collins, R S and Collins, P and Costigan, G and
               Crifo, F and Cross, N J G and Crosta, M and Crowley, C and
               Dafonte, C and Damerdji, Y and Dapergolas, A and David, P and
               David, M and De Cat, P and de Felice, F and de Laverny, P and De
               Luise, F and De March, R and de Martino, D and de Souza, R and
               Debosscher, J and del Pozo, E and Delbo, M and Delgado, A and
               Delgado, H E and di Marco, F and Di Matteo, P and Diakite, S and
               Distefano, E and Dolding, C and Dos Anjos, S and Drazinos, P and
               Dur{\'a}n, J and Dzigan, Y and Ecale, E and Edvardsson, B and
               Enke, H and Erdmann, M and Escolar, D and Espina, M and Evans, N
               W and Eynard Bontemps, G and Fabre, C and Fabrizio, M and
               Faigler, S and Falc{\~a}o, A J and Farr{\`a}s Casas, M and Faye,
               F and Federici, L and Fedorets, G and
               Fern{\'a}ndez-Hern{\'a}ndez, J and Fernique, P and Fienga, A and
               Figueras, F and Filippi, F and Findeisen, K and Fonti, A and
               Fouesneau, M and Fraile, E and Fraser, M and Fuchs, J and
               Furnell, R and Gai, M and Galleti, S and Galluccio, L and
               Garabato, D and Garc{\'\i}a-Sedano, F and Gar{\'e}, P and
               Garofalo, A and Garralda, N and Gavras, P and Gerssen, J and
               Geyer, R and Gilmore, G and Girona, S and Giuffrida, G and
               Gomes, M and Gonz{\'a}lez-Marcos, A and
               Gonz{\'a}lez-N{\'u}{\~n}ez, J and Gonz{\'a}lez-Vidal, J J and
               Granvik, M and Guerrier, A and Guillout, P and Guiraud, J and
               G{\'u}rpide, A and Guti{\'e}rrez-S{\'a}nchez, R and Guy, L P and
               Haigron, R and Hatzidimitriou, D and Haywood, M and Heiter, U
               and Helmi, A and Hobbs, D and Hofmann, W and Holl, B and
               Holland, G and Hunt, J A S and Hypki, A and Icardi, V and Irwin,
               M and Jevardat de Fombelle, G and Jofr{\'e}, P and Jonker, P G
               and Jorissen, A and Julbe, F and Karampelas, A and Kochoska, A
               and Kohley, R and Kolenberg, K and Kontizas, E and Koposov, S E
               and Kordopatis, G and Koubsky, P and Kowalczyk, A and
               Krone-Martins, A and Kudryashova, M and Kull, I and Bachchan, R
               K and Lacoste-Seris, F and Lanza, A F and Lavigne, J-B and Le
               Poncin-Lafitte, C and Lebreton, Y and Lebzelter, T and Leccia, S
               and Leclerc, N and Lecoeur-Taibi, I and Lemaitre, V and
               Lenhardt, H and Leroux, F and Liao, S and Licata, E and
               Lindstr{\o}m, H E P and Lister, T A and Livanou, E and Lobel, A
               and L{\"o}ffler, W and L{\'o}pez, M and Lopez-Lozano, A and
               Lorenz, D and Loureiro, T and MacDonald, I and Magalh{\~a}es
               Fernandes, T and Managau, S and Mann, R G and Mantelet, G and
               Marchal, O and Marchant, J M and Marconi, M and Marie, J and
               Marinoni, S and Marrese, P M and Marschalk{\'o}, G and Marshall,
               D J and Mart{\'\i}n-Fleitas, J M and Martino, M and Mary, N and
               Matijevi{\v c}, G and Mazeh, T and McMillan, P J and Messina, S
               and Mestre, A and Michalik, D and Millar, N R and Miranda, B M H
               and Molina, D and Molinaro, R and Molinaro, M and Moln{\'a}r, L
               and Moniez, M and Montegriffo, P and Monteiro, D and Mor, R and
               Mora, A and Morbidelli, R and Morel, T and Morgenthaler, S and
               Morley, T and Morris, D and Mulone, A F and Muraveva, T and
               Musella, I and Narbonne, J and Nelemans, G and Nicastro, L and
               Noval, L and Ord{\'e}novic, C and Ordieres-Mer{\'e}, J and
               Osborne, P and Pagani, C and Pagano, I and Pailler, F and
               Palacin, H and Palaversa, L and Parsons, P and Paulsen, T and
               Pecoraro, M and Pedrosa, R and Pentik{\"a}inen, H and Pereira, J
               and Pichon, B and Piersimoni, A M and Pineau, F-X and Plachy, E
               and Plum, G and Poujoulet, E and Pr{\v s}a, A and Pulone, L and
               Ragaini, S and Rago, S and Rambaux, N and Ramos-Lerate, M and
               Ranalli, P and Rauw, G and Read, A and Regibo, S and Renk, F and
               Reyl{\'e}, C and Ribeiro, R A and Rimoldini, L and Ripepi, V and
               Riva, A and Rixon, G and Roelens, M and Romero-G{\'o}mez, M and
               Rowell, N and Royer, F and Rudolph, A and Ruiz-Dern, L and
               Sadowski, G and Sagrist{\`a} Sell{\'e}s, T and Sahlmann, J and
               Salgado, J and Salguero, E and Sarasso, M and Savietto, H and
               Schnorhk, A and Schultheis, M and Sciacca, E and Segol, M and
               Segovia, J C and Segransan, D and Serpell, E and Shih, I-C and
               Smareglia, R and Smart, R L and Smith, C and Solano, E and
               Solitro, F and Sordo, R and Soria Nieto, S and Souchay, J and
               Spagna, A and Spoto, F and Stampa, U and Steele, I A and
               Steidelm{\"u}ller, H and Stephenson, C A and Stoev, H and Suess,
               F F and S{\"u}veges, M and Surdej, J and Szabados, L and
               Szegedi-Elek, E and Tapiador, D and Taris, F and Tauran, G and
               Taylor, M B and Teixeira, R and Terrett, D and Tingley, B and
               Trager, S C and Turon, C and Ulla, A and Utrilla, E and
               Valentini, G and van Elteren, A and Van Hemelryck, E and van
               Leeuwen, M and Varadi, M and Vecchiato, A and Veljanoski, J and
               Via, T and Vicente, D and Vogt, S and Voss, H and Votruba, V and
               Voutsinas, S and Walmsley, G and Weiler, M and Weingrill, K and
               Werner, D and Wevers, T and Whitehead, G and Wyrzykowski, {\L}
               and Yoldas, A and {\v Z}erjal, M and Zucker, S and Zurbach, C
               and Zwitter, T and Alecu, A and Allen, M and Allende Prieto, C
               and Amorim, A and Anglada-Escud{\'e}, G and Arsenijevic, V and
               Azaz, S and Balm, P and Beck, M and Bernstein, H-H and Bigot, L
               and Bijaoui, A and Blasco, C and Bonfigli, M and Bono, G and
               Boudreault, S and Bressan, A and Brown, S and Brunet, P-M and
               Bunclark, P and Buonanno, R and Butkevich, A G and Carret, C and
               Carrion, C and Chemin, L and Ch{\'e}reau, F and Corcione, L and
               Darmigny, E and de Boer, K S and de Teodoro, P and de Zeeuw, P T
               and Delle Luche, C and Domingues, C D and Dubath, P and Fodor, F
               and Fr{\'e}zouls, B and Fries, A and Fustes, D and Fyfe, D and
               Gallardo, E and Gallegos, J and Gardiol, D and Gebran, M and
               Gomboc, A and G{\'o}mez, A and Grux, E and Gueguen, A and
               Heyrovsky, A and Hoar, J and Iannicola, G and Isasi Parache, Y
               and Janotto, A-M and Joliet, E and Jonckheere, A and Keil, R and
               Kim, D-W and Klagyivik, P and Klar, J and Knude, J and
               Kochukhov, O and Kolka, I and Kos, J and Kutka, A and Lainey, V
               and LeBouquin, D and Liu, C and Loreggia, D and Makarov, V V and
               Marseille, M G and Martayan, C and Martinez-Rubi, O and Massart,
               B and Meynadier, F and Mignot, S and Munari, U and Nguyen, A-T
               and Nordlander, T and Ocvirk, P and O'Flaherty, K S and Olias
               Sanz, A and Ortiz, P and Osorio, J and Oszkiewicz, D and
               Ouzounis, A and Palmer, M and Park, P and Pasquato, E and
               Peltzer, C and Peralta, J and P{\'e}turaud, F and Pieniluoma, T
               and Pigozzi, E and Poels, J and Prat, G and Prod'homme, T and
               Raison, F and Rebordao, J M and Risquez, D and Rocca-Volmerange,
               B and Rosen, S and Ruiz-Fuertes, M I and Russo, F and Sembay, S
               and Serraller Vizcaino, I and Short, A and Siebert, A and Silva,
               H and Sinachopoulos, D and Slezak, E and Soffel, M and
               Sosnowska, D and Strai{\v z}ys, V and ter Linden, M and Terrell,
               D and Theil, S and Tiede, C and Troisi, L and Tsalmantza, P and
               Tur, D and Vaccari, M and Vachier, F and Valles, P and Van
               Hamme, W and Veltz, L and Virtanen, J and Wallut, J-M and
               Wichmann, R and Wilkinson, M I and Ziaeepour, H and Zschocke, S",
  journal   = {\aap},
  publisher = "EDP Sciences",
  volume    =  595,
  pages     = "A1",
  month     =  nov,
  year      =  2016
}

@ARTICLE{Gaia_Collaboration2023,
  title     = "\textit{Gaia} Data Release 3",
  author    = "{Gaia Collaboration} and Vallenari, A and Brown, A G A and
               Prusti, T and de Bruijne, J H J and Arenou, F and Babusiaux, C
               and Biermann, M and Creevey, O L and Ducourant, C and Evans, D W
               and Eyer, L and Guerra, R and Hutton, A and Jordi, C and
               Klioner, S A and Lammers, U L and Lindegren, L and Luri, X and
               Mignard, F and Panem, C and Pourbaix, D and Randich, S and
               Sartoretti, P and Soubiran, C and Tanga, P and Walton, N A and
               Bailer-Jones, C A L and Bastian, U and Drimmel, R and Jansen, F
               and Katz, D and Lattanzi, M G and van Leeuwen, F and Bakker, J
               and Cacciari, C and Casta{\~n}eda, J and De Angeli, F and
               Fabricius, C and Fouesneau, M and Fr{\'e}mat, Y and Galluccio, L
               and Guerrier, A and Heiter, U and Masana, E and Messineo, R and
               Mowlavi, N and Nicolas, C and Nienartowicz, K and Pailler, F and
               Panuzzo, P and Riclet, F and Roux, W and Seabroke, G M and
               Sordo, R and Th{\'e}venin, F and Gracia-Abril, G and Portell, J
               and Teyssier, D and Altmann, M and Andrae, R and Audard, M and
               Bellas-Velidis, I and Benson, K and Berthier, J and Blomme, R
               and Burgess, P W and Busonero, D and Busso, G and C{\'a}novas, H
               and Carry, B and Cellino, A and Cheek, N and Clementini, G and
               Damerdji, Y and Davidson, M and de Teodoro, P and Nu{\~n}ez
               Campos, M and Delchambre, L and Dell'Oro, A and Esquej, P and
               Fern{\'a}ndez-Hern{\'a}ndez, J and Fraile, E and Garabato, D and
               Garc{\'\i}a-Lario, P and Gosset, E and Haigron, R and Halbwachs,
               J-L and Hambly, N C and Harrison, D L and Hern{\'a}ndez, J and
               Hestroffer, D and Hodgkin, S T and Holl, B and Jan{\ss}en, K and
               Jevardat de Fombelle, G and Jordan, S and Krone-Martins, A and
               Lanzafame, A C and L{\"o}ffler, W and Marchal, O and Marrese, P
               M and Moitinho, A and Muinonen, K and Osborne, P and Pancino, E
               and Pauwels, T and Recio-Blanco, A and Reyl{\'e}, C and Riello,
               M and Rimoldini, L and Roegiers, T and Rybizki, J and Sarro, L M
               and Siopis, C and Smith, M and Sozzetti, A and Utrilla, E and
               van Leeuwen, M and Abbas, U and {\'A}brah{\'a}m, P and Abreu
               Aramburu, A and Aerts, C and Aguado, J J and Ajaj, M and
               Aldea-Montero, F and Altavilla, G and {\'A}lvarez, M A and
               Alves, J and Anders, F and Anderson, R I and Anglada Varela, E
               and Antoja, T and Baines, D and Baker, S G and
               Balaguer-N{\'u}{\~n}ez, L and Balbinot, E and Balog, Z and
               Barache, C and Barbato, D and Barros, M and Barstow, M A and
               Bartolom{\'e}, S and Bassilana, J-L and Bauchet, N and Becciani,
               U and Bellazzini, M and Berihuete, A and Bernet, M and Bertone,
               S and Bianchi, L and Binnenfeld, A and Blanco-Cuaresma, S and
               Blazere, A and Boch, T and Bombrun, A and Bossini, D and
               Bouquillon, S and Bragaglia, A and Bramante, L and Breedt, E and
               Bressan, A and Brouillet, N and Brugaletta, E and Bucciarelli, B
               and Burlacu, A and Butkevich, A G and Buzzi, R and Caffau, E and
               Cancelliere, R and Cantat-Gaudin, T and Carballo, R and
               Carlucci, T and Carnerero, M I and Carrasco, J M and
               Casamiquela, L and Castellani, M and Castro-Ginard, A and
               Chaoul, L and Charlot, P and Chemin, L and Chiaramida, V and
               Chiavassa, A and Chornay, N and Comoretto, G and Contursi, G and
               Cooper, W J and Cornez, T and Cowell, S and Crifo, F and
               Cropper, M and Crosta, M and Crowley, C and Dafonte, C and
               Dapergolas, A and David, M and David, P and de Laverny, P and De
               Luise, F and De March, R and De Ridder, J and de Souza, R and de
               Torres, A and del Peloso, E F and del Pozo, E and Delbo, M and
               Delgado, A and Delisle, J-B and Demouchy, C and Dharmawardena, T
               E and Di Matteo, P and Diakite, S and Diener, C and Distefano, E
               and Dolding, C and Edvardsson, B and Enke, H and Fabre, C and
               Fabrizio, M and Faigler, S and Fedorets, G and Fernique, P and
               Fienga, A and Figueras, F and Fournier, Y and Fouron, C and
               Fragkoudi, F and Gai, M and Garcia-Gutierrez, A and
               Garcia-Reinaldos, M and Garc{\'\i}a-Torres, M and Garofalo, A
               and Gavel, A and Gavras, P and Gerlach, E and Geyer, R and
               Giacobbe, P and Gilmore, G and Girona, S and Giuffrida, G and
               Gomel, R and Gomez, A and Gonz{\'a}lez-N{\'u}{\~n}ez, J and
               Gonz{\'a}lez-Santamar{\'\i}a, I and Gonz{\'a}lez-Vidal, J J and
               Granvik, M and Guillout, P and Guiraud, J and
               Guti{\'e}rrez-S{\'a}nchez, R and Guy, L P and Hatzidimitriou, D
               and Hauser, M and Haywood, M and Helmer, A and Helmi, A and
               Sarmiento, M H and Hidalgo, S L and Hilger, T and H{\l}adczuk, N
               and Hobbs, D and Holland, G and Huckle, H E and Jardine, K and
               Jasniewicz, G and Jean-Antoine Piccolo, A and
               Jim{\'e}nez-Arranz, {\'O} and Jorissen, A and Juaristi Campillo,
               J and Julbe, F and Karbevska, L and Kervella, P and Khanna, S
               and Kontizas, M and Kordopatis, G and Korn, A J and
               K{\'o}sp{\'a}l, {\'A} and Kostrzewa-Rutkowska, Z and
               Kruszy{\'n}ska, K and Kun, M and Laizeau, P and Lambert, S and
               Lanza, A F and Lasne, Y and Le Campion, J-F and Lebreton, Y and
               Lebzelter, T and Leccia, S and Leclerc, N and Lecoeur-Taibi, I
               and Liao, S and Licata, E L and Lindstr{\o}m, H E P and Lister,
               T A and Livanou, E and Lobel, A and Lorca, A and Loup, C and
               Madrero Pardo, P and Magdaleno Romeo, A and Managau, S and Mann,
               R G and Manteiga, M and Marchant, J M and Marconi, M and Marcos,
               J and Marcos Santos, M M S and Mar{\'\i}n Pina, D and Marinoni,
               S and Marocco, F and Marshall, D J and Martin Polo, L and
               Mart{\'\i}n-Fleitas, J M and Marton, G and Mary, N and Masip, A
               and Massari, D and Mastrobuono-Battisti, A and Mazeh, T and
               McMillan, P J and Messina, S and Michalik, D and Millar, N R and
               Mints, A and Molina, D and Molinaro, R and Moln{\'a}r, L and
               Monari, G and Mongui{\'o}, M and Montegriffo, P and Montero, A
               and Mor, R and Mora, A and Morbidelli, R and Morel, T and
               Morris, D and Muraveva, T and Murphy, C P and Musella, I and
               Nagy, Z and Noval, L and Oca{\~n}a, F and Ogden, A and
               Ordenovic, C and Osinde, J O and Pagani, C and Pagano, I and
               Palaversa, L and Palicio, P A and Pallas-Quintela, L and Panahi,
               A and Payne-Wardenaar, S and Pe{\~n}alosa Esteller, X and
               Penttil{\"a}, A and Pichon, B and Piersimoni, A M and Pineau,
               F-X and Plachy, E and Plum, G and Poggio, E and Pr{\v s}a, A and
               Pulone, L and Racero, E and Ragaini, S and Rainer, M and
               Raiteri, C M and Rambaux, N and Ramos, P and Ramos-Lerate, M and
               Re Fiorentin, P and Regibo, S and Richards, P J and Rios Diaz, C
               and Ripepi, V and Riva, A and Rix, H-W and Rixon, G and
               Robichon, N and Robin, A C and Robin, C and Roelens, M and
               Rogues, H R O and Rohrbasser, L and Romero-G{\'o}mez, M and
               Rowell, N and Royer, F and Ruz Mieres, D and Rybicki, K A and
               Sadowski, G and S{\'a}ez N{\'u}{\~n}ez, A and Sagrist{\`a}
               Sell{\'e}s, A and Sahlmann, J and Salguero, E and Samaras, N and
               Sanchez Gimenez, V and Sanna, N and Santove{\~n}a, R and
               Sarasso, M and Schultheis, M and Sciacca, E and Segol, M and
               Segovia, J C and S{\'e}gransan, D and Semeux, D and Shahaf, S
               and Siddiqui, H I and Siebert, A and Siltala, L and Silvelo, A
               and Slezak, E and Slezak, I and Smart, R L and Snaith, O N and
               Solano, E and Solitro, F and Souami, D and Souchay, J and
               Spagna, A and Spina, L and Spoto, F and Steele, I A and
               Steidelm{\"u}ller, H and Stephenson, C A and S{\"u}veges, M and
               Surdej, J and Szabados, L and Szegedi-Elek, E and Taris, F and
               Taylor, M B and Teixeira, R and Tolomei, L and Tonello, N and
               Torra, F and Torra, J and Torralba Elipe, G and Trabucchi, M and
               Tsounis, A T and Turon, C and Ulla, A and Unger, N and Vaillant,
               M V and van Dillen, E and van Reeven, W and Vanel, O and
               Vecchiato, A and Viala, Y and Vicente, D and Voutsinas, S and
               Weiler, M and Wevers, T and Wyrzykowski, {\L} and Yoldas, A and
               Yvard, P and Zhao, H and Zorec, J and Zucker, S and Zwitter, T",
  journal   = {\aap},
  publisher = "EDP Sciences",
  volume    =  674,
  pages     = "A1",
  month     =  jun,
  year      =  2023,
  copyright = "https://creativecommons.org/licenses/by/4.0"
}

@ARTICLE{Galli2019,
  title     = "Structure and kinematics of the Taurus star-forming region from
               {Gaia-DR2} and {VLBI} astrometry",
  author    = "Galli, P A B and Loinard, L and Bouy, H and Sarro, L M and
               Ortiz-Le{\'o}n, G N and Dzib, S A and Olivares, J and Heyer, M
               and Hernandez, J and Rom{\'a}n-Z{\'u}{\~n}iga, C and Kounkel, M
               and Covey, K",
  journal   = {\aap},
  publisher = "EDP Sciences",
  volume    =  630,
  pages     = "A137",
  month     =  oct,
  year      =  2019,
  copyright = "http://creativecommons.org/licenses/by/4.0"
}

@ARTICLE{Galli2020b,
  title     = "Lupus {DANCe}",
  author    = "Galli, P A B and Bouy, H and Olivares, J and Miret-Roig, N and
               Vieira, R G and Sarro, L M and Barrado, D and Berihuete, A and
               Bertout, C and Bertin, E and Cuillandre, J-C",
  journal   = {\aap},
  publisher = "EDP Sciences",
  volume    =  643,
  pages     = "A148",
  month     =  nov,
  year      =  "2020b",
  copyright = "https://creativecommons.org/licenses/by/4.0"
}

@ARTICLE{Galli2021,
  title     = "Chamaeleon {DANCe}",
  author    = "Galli, P A B and Bouy, H and Olivares, J and Miret-Roig, N and
               Sarro, L M and Barrado, D and Berihuete, A and Bertin, E and
               Cuillandre, J-C",
  journal   = {\aap},
  publisher = "EDP Sciences",
  volume    =  646,
  pages     = "A46",
  month     =  feb,
  year      =  2021,
  copyright = "https://creativecommons.org/licenses/by/4.0"
}

@ARTICLE{Grosschedl2021,
  title     = "{3D} dynamics of the Orion cloud complex",
  author    = "Gro{\ss}schedl, Josefa E and Alves, Jo{\~a}o and Meingast,
               Stefan and Herbst-Kiss, Gabor",
  journal   = {\aap},
  publisher = "EDP Sciences",
  volume    =  647,
  pages     = "A91",
  month     =  mar,
  year      =  2021,
  copyright = "https://www.edpsciences.org/en/authors/copyright-and-licensing"
}

@ARTICLE{Hacar2016,
  title     = "{APOGEE} strings: A fossil record of the gas kinematic structure",
  author    = "Hacar, A and Alves, J and Forbrich, J and Meingast, S and
               Kubiak, K and Gro{\ss}schedl, J",
  journal   = {\aap},
  publisher = "EDP Sciences",
  volume    =  589,
  pages     = "A80",
  month     =  may,
  year      =  2016
}

@ARTICLE{Heiderman2010,
  title     = "The star formation rate and gas surface density relation in the
               milky way: Implications for extragalactic studies",
  author    = "Heiderman, Amanda and Evans, Neal J and Allen, Lori E and Huard,
               Tracy and Heyer, Mark",
  journal   = {\apj},
  publisher = "American Astronomical Society",
  volume    =  723,
  number    =  2,
  pages     = "1019--1037",
  month     =  nov,
  year      =  2010
}

@ARTICLE{Hourihane2023,
  title     = "The {\textit{Gaia}-ESO} Survey: Homogenisation of stellar
               parameters and elemental abundances",
  author    = "Hourihane, A and Fran{\c c}ois, P and Worley, C C and Magrini, L
               and Gonneau, A and Casey, A R and Gilmore, G and Randich, S and
               Sacco, G G and Recio-Blanco, A and Korn, A J and Allende Prieto,
               C and Smiljanic, R and Blomme, R and Bragaglia, A and Walton, N
               A and Van Eck, S and Bensby, T and Lanzafame, A and Frasca, A
               and Franciosini, E and Damiani, F and Lind, K and Bergemann, M
               and Bonifacio, P and Hill, V and Lobel, A and Montes, D and
               Feuillet, D K and Tautvai{\v s}ien{\.e}, G and Guiglion, G and
               Tabernero, H M and Gonz{\'a}lez Hern{\'a}ndez, J I and Gebran, M
               and Van der Swaelmen, M and Mikolaitis, {\v S} and Daflon, S and
               Merle, T and Morel, T and Lewis, J R and Gonz{\'a}lez Solares, E
               A and Murphy, D N A and Jeffries, R D and Jackson, R J and
               Feltzing, S and Prusti, T and Carraro, G and Biazzo, K and
               Prisinzano, L and Jofr{\'e}, P and Zaggia, S and Drazdauskas, A
               and Stonkut{\'e}, E and Marfil, E and Jim{\'e}nez-Esteban, F and
               Mahy, L and Guti{\'e}rrez Albarr{\'a}n, M L and Berlanas, S R
               and Santos, W and Morbidelli, L and Spina, L and Minkevi{\v
               c}i{\=u}t{\.e}, R",
  journal   = {\aap},
  publisher = "EDP Sciences",
  volume    =  676,
  pages     = "A129",
  month     =  aug,
  year      =  2023,
  copyright = "https://creativecommons.org/licenses/by/4.0"
}

@ARTICLE{Hunt2024,
  title     = "Improving the open cluster census",
  author    = "Hunt, Emily L and Reffert, Sabine",
  journal   = {\aap},
  publisher = "EDP Sciences",
  volume    =  686,
  pages     = "A42",
  month     =  jun,
  year      =  2024,
  copyright = "https://creativecommons.org/licenses/by/4.0"
}

@ARTICLE{Hunt2023,
  title     = "Improving the open cluster census",
  author    = "Hunt, Emily L and Reffert, Sabine",
  journal   = {\aap},
  publisher = "EDP Sciences",
  volume    =  673,
  pages     = "A114",
  month     =  may,
  year      =  2023,
  copyright = "https://creativecommons.org/licenses/by/4.0"
}

@ARTICLE{Katz2023,
  title     = "\textit{Gaia} Data Release 3",
  author    = "Katz, D and Sartoretti, P and Guerrier, A and Panuzzo, P and
               Seabroke, G M and Th{\'e}venin, F and Cropper, M and Benson, K
               and Blomme, R and Haigron, R and Marchal, O and Smith, M and
               Baker, S and Chemin, L and Damerdji, Y and David, M and Dolding,
               C and Fr{\'e}mat, Y and Gosset, E and Jan{\ss}en, K and
               Jasniewicz, G and Lobel, A and Plum, G and Samaras, N and
               Snaith, O and Soubiran, C and Vanel, O and Zwitter, T and
               Antoja, T and Arenou, F and Babusiaux, C and Brouillet, N and
               Caffau, E and Di Matteo, P and Fabre, C and Fabricius, C and
               Fragkoudi, F and Haywood, M and Huckle, H E and Hottier, C and
               Lasne, Y and Leclerc, N and Mastrobuono-Battisti, A and Royer, F
               and Teyssier, D and Zorec, J and Crifo, F and Jean-Antoine
               Piccolo, A and Turon, C and Viala, Y",
  journal   = {\aap},
  publisher = "EDP Sciences",
  volume    =  674,
  pages     = "A5",
  month     =  jun,
  year      =  2023,
  copyright = "https://creativecommons.org/licenses/by/4.0"
}

@ARTICLE{Jiang2024,
  title     = "The emerging stellar complex in Mon R2: Membership and optical
               variability classification",
  author    = "Jiang, Sally D and Hillenbrand, Lynne A",
  journal   = {\aj},
  publisher = "American Astronomical Society",
  volume    =  167,
  number    =  5,
  pages     = "221",
  month     =  may,
  year      =  2024,
  copyright = "http://creativecommons.org/licenses/by/4.0/"
}

@ARTICLE{Kounkel2018,
  title     = "The {APOGEE-2} survey of the Orion star-forming complex. {II}.
               Six-dimensional structure",
  author    = "Kounkel, Marina and Covey, Kevin and Su{\'a}rez, Genaro and
               Rom{\'a}n-Z{\'u}{\~n}iga, Carlos and Hernandez, Jesus and
               Stassun, Keivan and Jaehnig, Karl O and Feigelson, Eric D and
               Ram{\'\i}rez, Karla Pe{\~n}a and Roman-Lopes, Alexandre and Rio,
               Nicola Da and Stringfellow, Guy S and Kim, J Serena and
               Borissova, Jura and Fern{\'a}ndez-Trincado, Jos{\'e} G and
               Burgasser, Adam and Garc{\'\i}a-Hern{\'a}ndez, D A and Zamora,
               Olga and Pan, Kaike and Nitschelm, Christian",
  journal   = {\aj},
  publisher = "American Astronomical Society",
  volume    =  156,
  number    =  3,
  pages     = "84",
  month     =  aug,
  year      =  2018,
  copyright = "http://iopscience.iop.org/page/copyright"
}

@ARTICLE{Kounkel2022,
  title     = "Dynamical star-forming history of Per {OB2}",
  author    = "Kounkel, Marina and Deng, Tingyan and Stassun, Keivan G",
  journal   = {\aj},
  publisher = "American Astronomical Society",
  volume    =  164,
  number    =  2,
  pages     = "57",
  month     =  aug,
  year      =  2022,
  copyright = "http://creativecommons.org/licenses/by/4.0/"
}

@INPROCEEDINGS{Krumholz2011,
       author = {{Krumholz}, Mark R.},
        title = "{Star Formation in Molecular Clouds}",
    booktitle = {XV Special Courses at the National Observatory of Rio de Janeiro},
         year = 2011,
       editor = {{Telles}, Eduardo and {Dupke}, Renato and {Lazzaro}, Daniela},
       series = {American Institute of Physics Conference Series},
       volume = {1386},
        month = sep,
    publisher = {AIP},
        pages = {9-57},
          doi = {10.1063/1.3636038},
archivePrefix = {arXiv},
       eprint = {1101.5172},
 primaryClass = {astro-ph.GA},
       adsurl = {https://ui.adsabs.harvard.edu/abs/2011AIPC.1386....9K}
}

@ARTICLE{Lada2009,
  title     = "{THE} {CALIFORNIA} {MOLECULAR} {CLOUD}",
  author    = "Lada, Charles J and Lombardi, Marco and Alves, Jo{\~a}o F",
  journal   = {\apj},
  publisher = "American Astronomical Society",
  volume    =  703,
  number    =  1,
  pages     = "52--59",
  month     =  sep,
  year      =  2009
}

@ARTICLE{Larson1981,
  title     = "Turbulence and star formation in molecular clouds",
  author    = "Larson, R B",
  journal   = {\mnras},
  publisher = "Oxford University Press (OUP)",
  volume    =  194,
  number    =  4,
  pages     = "809--826",
  month     =  apr,
  year      =  1981
}

@ARTICLE{Lim2019,
  title     = "A Gaia view of the two {OB} associations Cygnus {OB2} and Carina
               {OB1}: the signature of their formation process",
  author    = "Lim, Beomdu and Naz{\'e}, Ya{\"e}l and Gosset, Eric and Rauw,
               Gregor",
  journal   = {\mnras},
  publisher = "Oxford University Press (OUP)",
  volume    =  490,
  number    =  1,
  pages     = "440--454",
  month     =  nov,
  year      =  2019,
  copyright = "https://academic.oup.com/journals/pages/open\_access/funder\_policies/chorus/standard\_publication\_model",
  language  = "en"
}

@ARTICLE{Lim2021,
  title     = "A kinematic perspective on the formation process of the stellar
               groups in the Rosette Nebula",
  author    = "Lim, Beomdu and Naz{\'e}, Ya{\"e}l and Hong, Jongsuk and Park,
               Byeong-Gon and Yun, Hyeong-Sik and Yi, Hee-Weon and Park,
               Sunkyung and Hwang, Narae and Lee, Jeong-Eun",
  journal   = {\aj},
  publisher = "American Astronomical Society",
  volume    =  162,
  number    =  2,
  pages     = "56",
  month     =  aug,
  year      =  2021,
  copyright = "https://iopscience.iop.org/page/copyright"
}

@ARTICLE{Lim2023,
  title     = "The kinematics of the young stellar population in the {W5}
               region of the Cassiopeia {OB6} association: Implication for the
               formation process of stellar associations",
  author    = "Lim, Beomdu and Hong, Jongsuk and Lee, Jinhee and Yun,
               Hyeong-Sik and Hwang, Narae and Park, Byeong-Gon",
  journal   = {\aj},
  publisher = "American Astronomical Society",
  volume    =  166,
  number    =  3,
  pages     = "97",
  month     =  sep,
  year      =  2023,
  copyright = "http://creativecommons.org/licenses/by/4.0/"
}

@ARTICLE{Lombardi2011,
  title     = "{2MASS} wide field extinction maps",
  author    = "Lombardi, M and Alves, J and Lada, C J",
  journal   = {\aap},
  publisher = "EDP Sciences",
  volume    =  535,
  pages     = "A16",
  month     =  nov,
  year      =  2011
}

@INCOLLECTION{Luhman2008,
       author = {{Luhman}, K.~L.},
        title = "{Chamaeleon}",
    booktitle = {Handbook of Star Forming Regions, Volume II},
         year = 2008,
       editor = {{Reipurth}, B.},
       volume = {5},
        pages = {169},
          doi = {10.48550/arXiv.0808.3207},
       adsurl = {https://ui.adsabs.harvard.edu/abs/2008hsf2.book..169L}
}

@ARTICLE{Maconi2025,
       author = {{Maconi}, E. and {Alves}, J. and {Swiggum}, C. and {Ratzenb{\"o}ck}, S. and {Gro{\ss}schedl}, J. and {K{\"o}hler}, P. and {Miret-Roig}, N. and {Meingast}, S. and {Konietzka}, R. and {Zucker}, C. and {Goodman}, A. and {Lombardi}, M. and {Knorr}, G. and {Lohmann}, G. and {Forbes}, J.~C. and {Burkert}, A. and {Opher}, M.},
        title = "{The Solar System's passage through the Radcliffe wave during the middle Miocene}",
      journal = {\aap},
         year = 2025,
        month = feb,
       volume = {694},
          eid = {A167},
        pages = {A167},
          doi = {10.1051/0004-6361/202452061},
archivePrefix = {arXiv},
       eprint = {2502.16360},
 primaryClass = {astro-ph.GA},
       adsurl = {https://ui.adsabs.harvard.edu/abs/2025A&A...694A.167M}
}

@ARTICLE{Marton2016,
  title     = "An all-sky support vector machine selection {ofWISEYSO}
               candidates",
  author    = "Marton, G and T{\'o}th, L V and Paladini, R and Kun, M and
               Zahorecz, S and McGehee, P and Kiss, Cs",
  journal   = {\mnras},
  publisher = "Oxford University Press (OUP)",
  volume    =  458,
  number    =  4,
  pages     = "3479--3488",
  month     =  jun,
  year      =  2016,
  language  = "en"
}

@ARTICLE{Marton2019,
  title     = "Identification of Young Stellar Object candidates in the Gaia
               {DR2} x {AllWISE} catalogue with machine learning methods",
  author    = "Marton, G and {\'A}brah{\'a}m, P and Szegedi-Elek, E and Varga,
               J and Kun, M and K{\'o}sp{\'a}l, {\'A} and Varga-Vereb{\'e}lyi,
               E and Hodgkin, S and Szabados, L and Beck, R and Kiss, Cs",
  journal   = {\mnras},
  publisher = "Oxford University Press (OUP)",
  volume    =  487,
  number    =  2,
  pages     = "2522--2537",
  month     =  aug,
  year      =  2019,
  copyright = "https://academic.oup.com/journals/pages/open\_access/funder\_policies/chorus/standard\_publication\_model",
  language  = "en"
}

@ARTICLE{Marton2023,
  title     = "{\textit{Gaia}Data} Release 3",
  author    = "Marton, G{\'a}bor and {\'A}brah{\'a}m, P{\'e}ter and Rimoldini,
               Lorenzo and Audard, Marc and Kun, M{\'a}ria and Nagy, Zs{\'o}fia
               and K{\'o}sp{\'a}l, {\'A}gnes and Szabados, L{\'a}szl{\'o} and
               Holl, Berry and Gavras, Panagiotis and Mowlavi, Nami and
               Nienartowicz, Krzysztof and de Fombelle, Gr{\'e}gory Jevardat
               and Lecoeur-Ta{\"\i}bi, Isabelle and Karbevska, Lea and Lario,
               Pedro Garcia and Eyer, Laurent",
  journal   = {\aap},
  publisher = "EDP Sciences",
  volume    =  674,
  pages     = "A21",
  month     =  jun,
  year      =  2023,
  copyright = "https://creativecommons.org/licenses/by/4.0"
}

@ARTICLE{McInnes2017,
  title     = "hdbscan: Hierarchical density based clustering",
  author    = "McInnes, Leland and Healy, John and Astels, Steve",
  journal   = "J. Open Source Softw.",
  publisher = "The Open Journal",
  volume    =  2,
  number    =  11,
  pages     = "205",
  month     =  mar,
  year      =  2017,
  copyright = "http://creativecommons.org/licenses/by/4.0/"
}

@ARTICLE{Muzic2022,
  title     = "Stellar population of the Rosette Nebula and {NGC} 2244",
  author    = "Mu{\v z}i{\'c}, K and Almendros-Abad, V and Bouy, H and Kubiak,
               K and Pe{\~n}a Ram{\'\i}rez, K and Krone-Martins, A and
               Moitinho, A and Concei{\c c}{\~a}o, M",
  journal   = {\aap},
  publisher = "EDP Sciences",
  volume    =  668,
  pages     = "A19",
  month     =  dec,
  year      =  2022,
  copyright = "https://creativecommons.org/licenses/by/4.0"
}

@ARTICLE{Orellana2021,
  title     = "New members of Cygnus {OB2} from Gaia {DR2}",
  author    = "Orellana, R B and De Biasi, M S and Pa{\'\i}z, L G",
  journal   = {\mnras},
  publisher = "Oxford University Press (OUP)",
  volume    =  502,
  number    =  4,
  pages     = "6080--6093",
  month     =  mar,
  year      =  2021,
  copyright = "https://academic.oup.com/journals/pages/open\_access/funder\_policies/chorus/standard\_publication\_model",
  language  = "en"
}

@ARTICLE{Ortiz-Leon2018,
  title     = "The Gould's belt distances survey ({GOBELINS)}. V. distances and
               kinematics of the Perseus molecular cloud",
  author    = "Ortiz-Le{\'o}n, Gisela N and Loinard, Laurent and Dzib, Sergio A
               and Galli, Phillip A B and Kounkel, Marina and Mioduszewski, Amy
               J and Rodr{\'\i}guez, Luis F and Torres, Rosa M and Hartmann,
               Lee and Boden, Andrew F and Evans, II, Neal J and Brice{\~n}o,
               Cesar and Tobin, John J",
  journal   = {\apj},
  publisher = "American Astronomical Society",
  volume    =  865,
  number    =  1,
  pages     = "73",
  month     =  sep,
  year      =  2018,
  copyright = "http://iopscience.iop.org/page/copyright"
}

@article{Pedregosa2012,
  title={Scikit-learn: Machine Learning in {P}ython},
  author={Pedregosa, F. and Varoquaux, G. and Gramfort, A. and Michel, V.
          and Thirion, B. and Grisel, O. and Blondel, M. and Prettenhofer, P.
          and Weiss, R. and Dubourg, V. and Vanderplas, J. and Passos, A. and
          Cournapeau, D. and Brucher, M. and Perrot, M. and Duchesnay, E.},
  journal={J. Mach. Learn. Res.},
  volume={12},
  pages={2825--2830},
  year={2011}
}

@ARTICLE{Piecka2021,
AUTHOR={Piecka, Martin  and Paunzen, Ernst },
           
TITLE={A Comparison of the Simulations and Observations for a Nearby Spiral Arm},
          
JOURNAL={Front. Astron. Space Sci.},
          
VOLUME={Volume 8},
  
YEAR={2021},
  
URL={https://www.frontiersin.org/journals/astronomy-and-space-sciences/articles/10.3389/fspas.2021.642776},
  
DOI={10.3389/fspas.2021.642776},
  
ISSN={2296-987X}}

@ARTICLE{Rahmah2016,
  title     = "Determination of optimal epsilon (eps) value on {DBSCAN}
               algorithm to clustering data on peatland hotspots in Sumatra",
  author    = "Rahmah, Nadia and Sitanggang, Imas Sukaesih",
  journal   = "IOP Conf. Ser. Earth Environ. Sci.",
  publisher = "IOP Publishing",
  volume    =  31,
  pages     = "012012",
  month     =  jan,
  year      =  2016,
  copyright = "http://creativecommons.org/licenses/by/3.0/"
}

@ARTICLE{Rice2016,
  title     = "A uniform catalog of molecular clouds in the milky way",
  author    = "Rice, Thomas S and Goodman, Alyssa A and Bergin, Edwin A and
               Beaumont, Christopher and Dame, T M",
  journal   = {\apj},
  publisher = "American Astronomical Society",
  volume    =  822,
  number    =  1,
  pages     = "52",
  month     =  may,
  year      =  2016,
  copyright = "http://iopscience.iop.org/page/copyright"
}

@ARTICLE{Sanchez-Sanjuan2024,
  title     = "Kinematic study of the Orion Complex: analysing the young
               stellar clusters from big and small structures",
  author    = "S{\'a}nchez-Sanju{\'a}n, Sergio and Hern{\'a}ndez, Jes{\'u}s and
               P{\'e}rez-Villegas, {\'A}ngeles and Rom{\'a}n-Z{\'u}{\~n}iga,
               Carlos and Aguilar, Luis and Ballesteros-Paredes, Javier and
               Bonilla-Barroso, Andrea",
  journal   = {\mnras},
  publisher = "Oxford University Press (OUP)",
  volume    =  534,
  number    =  3,
  pages     = "2566--2584",
  month     =  oct,
  year      =  2024,
  copyright = "https://creativecommons.org/licenses/by/4.0/",
  language  = "en"
}

@INPROCEEDINGS{Satopaa2011,
author = {Satopaa, Ville and Albrecht, Jeannie and Irwin, David and Raghavan, Barath},
title = {Finding a "Kneedle" in a Haystack: Detecting Knee Points in System Behavior},
year = {2011},
isbn = {9780769543864},
publisher = {IEEE Computer Society},
address = {USA},
url = {https://doi.org/10.1109/ICDCSW.2011.20},
doi = {10.1109/ICDCSW.2011.20},
booktitle = {Proceedings of the 2011 31st International Conference on Distributed Computing Systems Workshops},
pages = {166–171},
numpages = {6},
series = {ICDCSW '11}
}

@ARTICLE{Schonrich2010,
  title     = "Local kinematics and the local standard of rest",
  author    = "Sch{\"o}nrich, Ralph and Binney, James and Dehnen, Walter",
  journal   = {\mnras},
  publisher = "Oxford University Press (OUP)",
  volume    =  403,
  number    =  4,
  pages     = "1829--1833",
  month     =  apr,
  year      =  2010,
  language  = "en"
}

@ARTICLE{Steinmetz2020,
  title     = "The sixth Data Release of the Radial Velocity Experiment (rave).
               {II}. Stellar atmospheric parameters, chemical abundances, and
               distances",
  author    = "Steinmetz, Matthias and Guiglion, Guillaume and McMillan, Paul J
               and Matijevi{\v c}, Gal and Enke, Harry and Kordopatis, Georges
               and Zwitter, Toma{\v z} and Valentini, Marica and Chiappini,
               Cristina and Casagrande, Luca and Wojno, Jennifer and Anguiano,
               Borja and Bienaym{\'e}, Olivier and Bijaoui, Albert and Binney,
               James and Burton, Donna and Cass, Paul and de Laverny, Patrick
               and Fiegert, Kristin and Freeman, Kenneth and Fulbright, Jon P
               and Gibson, Brad K and Gilmore, Gerard and Grebel, Eva K and
               Helmi, Amina and Kunder, Andrea and Munari, Ulisse and Navarro,
               Julio F and Parker, Quentin and Ruchti, Gregory R and
               Recio-Blanco, Alejandra and Reid, Warren and Seabroke, George M
               and Siviero, Alessandro and Siebert, Arnaud and Stupar, Milorad
               and Watson, Fred and Williams, Mary E K and Wyse, Rosemary F G
               and Anders, Friedrich and Antoja, Teresa and Birko, Danijela and
               Bland-Hawthorn, Joss and Bossini, Diego and Garc{\'\i}a, Rafael
               A and Carrillo, Ismael and Chaplin, William J and Elsworth,
               Yvonne and Famaey, Benoit and Gerhard, Ortwin and Jofre, Paula
               and Just, Andreas and Mathur, Savita and Miglio, Andrea and
               Minchev, Ivan and Monari, Giacomo and Mosser, Benoit and Ritter,
               Andreas and Rodrigues, Thaise S and Scholz, Ralf-Dieter and
               Sharma, Sanjib and Sysoliatina, Kseniia and {(The Rave
               collaboration)}",
  journal   = {\aj},
  publisher = "American Astronomical Society",
  volume    =  160,
  number    =  2,
  pages     = "83",
  month     =  aug,
  year      =  2020,
  copyright = "https://iopscience.iop.org/page/copyright"
}

@ARTICLE{Szilagyi2021,
  title     = "The Gaia view of the Cepheus flare",
  author    = "Szil{\'a}gyi, M{\'a}t{\'e} and Kun, M{\'a}ria and
               {\'A}brah{\'a}m, P{\'e}ter",
  journal   = {\mnras},
  publisher = "Oxford University Press (OUP)",
  volume    =  505,
  number    =  4,
  pages     = "5164--5182",
  month     =  jun,
  year      =  2021,
  copyright = "https://academic.oup.com/journals/pages/open\_access/funder\_policies/chorus/standard\_publication\_model",
  language  = "en"
}

@ARTICLE{Tachihara2001,
  title     = "{12CO} molecular cloud survey and global star formation in lupus",
  author    = "Tachihara, Kengo and Toyoda, Shuichiro and Onishi, Toshikazu and
               Mizuno, Akira and Fukui, Yasuo and Neuh{\"a}user, Ralph",
  journal   = {\pasj},
  publisher = "Oxford University Press (OUP)",
  volume    =  53,
  number    =  6,
  pages     = "1081--1096",
  month     =  dec,
  year      =  2001,
  language  = "en"
}

@ARTICLE{Teixeira2020,
  title     = "A wide survey for circumstellar disks in the Lupus complex",
  author    = "Teixeira, P S and Scholz, A and Alves, J",
  journal   = {\aap},
  publisher = "EDP Sciences",
  volume    =  642,
  pages     = "A86",
  month     =  oct,
  year      =  2020,
  copyright = "https://www.edpsciences.org/en/authors/copyright-and-licensing"
}

@ARTICLE{Yan2019,
  title     = "Distances to molecular clouds at high galactic latitudes based
               on Gaia {DR2}",
  author    = "Yan, Qing-Zeng and Zhang, Bo and Xu, Ye and Guo, Sufen and
               Macquart, Jean-Pierre and Tang, Zheng-Hong and Walsh, Andrew
               John",
  journal   = {\aap},
  publisher = "EDP Sciences",
  volume    =  624,
  pages     = "A6",
  month     =  apr,
  year      =  2019,
  copyright = "https://www.edpsciences.org/en/authors/copyright-and-licensing"
}

@ARTICLE{Yang2025,
  title     = "Kinematics of young stellar objects under various stellar
               feedback",
  author    = "Yang, Longhui and Liu, Dejian and Hao, Chaojie and Lin, Zehao
               and Li, Yingjie and Dong, Yiwei and Lu, Zu-Jia and Liang, En-Wei
               and Xu, Y",
  journal   = {\apjs},
  publisher = "American Astronomical Society",
  volume    =  276,
  number    =  1,
  pages     = "22",
  month     =  jan,
  year      =  2025,
  copyright = "http://creativecommons.org/licenses/by/4.0/"
}

@ARTICLE{Zeidler2016,
  title     = "The {VISTA} Carina Nebula Survey",
  author    = "Zeidler, P and Preibisch, T and Ratzka, T and Roccatagliata, V
               and Petr-Gotzens, M G",
  journal   = {\aap},
  publisher = "EDP Sciences",
  volume    =  585,
  pages     = "A49",
  month     =  jan,
  year      =  2016
}

@ARTICLE{Zhang2023,
  title     = "Distances to nearby molecular clouds traced by young stars",
  author    = "Zhang, Miaomiao",
  journal   = {\apjs},
  publisher = "American Astronomical Society",
  volume    =  265,
  number    =  2,
  pages     = "59",
  month     =  apr,
  year      =  2023,
  copyright = "http://creativecommons.org/licenses/by/4.0/"
}

@ARTICLE{Zucker2019,
  title     = "A large catalog of accurate distances to local molecular clouds:
               The Gaia {DR2} edition",
  author    = "Zucker, Catherine and Speagle, Joshua S and Schlafly, Edward F
               and Green, Gregory M and Finkbeiner, Douglas P and Goodman,
               Alyssa A and Alves, Jo{\~a}o",
  journal   = {\apj},
  publisher = "American Astronomical Society",
  volume    =  879,
  number    =  2,
  pages     = "125",
  month     =  jul,
  year      =  2019,
  copyright = "https://iopscience.iop.org/page/copyright"
}

@ARTICLE{Zucker2020,
  title     = "A compendium of distances to molecular clouds in the Star
               Formation Handbook",
  author    = "Zucker, Catherine and Speagle, Joshua S and Schlafly, Edward F
               and Green, Gregory M and Finkbeiner, Douglas P and Goodman,
               Alyssa and Alves, Jo{\~a}o",
  journal   = {\aap},
  publisher = "EDP Sciences",
  volume    =  633,
  pages     = "A51",
  month     =  jan,
  year      =  2020,
  copyright = "https://www.edpsciences.org/en/authors/copyright-and-licensing"
}

@INPROCEEDINGS{1996kddm.conf..226E,
       author = {{Ester}, Martin and {Kriegel}, Hans-Peter and {Sander}, J{\"o}rg and {Xu}, Xiaowei},
        title = "{A Density-Based Algorithm for Discovering Clusters in Large Spatial Databases with Noise}",
    booktitle = {Second International Conference on Knowledge Discovery and Data Mining (KDD'96). Proceedings of a conference held August 2-4},
         year = 1996,
       editor = {{Pfitzner}, D.~W. and {Salmon}, J.~K.},
        month = jan,
        pages = {226–231},
       adsurl = {https://ui.adsabs.harvard.edu/abs/1996kddm.conf..226E}
}

@INCOLLECTION{2008hsf2.book..295C,
       author = {{Comer{\'o}n}, F.},
        title = "{The Lupus Clouds}",
    booktitle = {Handbook of Star Forming Regions, Volume II},
         year = 2008,
       editor = {{Reipurth}, B.},
       volume = {5},
        pages = {295},
       adsurl = {https://ui.adsabs.harvard.edu/abs/2008hsf2.book..295C}
}

@ARTICLE{Wolf1924,
       author = {{Wolf}, M.},
        title = "{Die Sternleeren beim Amerikanebel}",
      journal = {Astron. Nachr.},
         year = 1924,
        month = nov,
       volume = {223},
       number = {6},
        pages = {89},
          doi = {10.1002/asna.19242230602},
       adsurl = {https://ui.adsabs.harvard.edu/abs/1924AN....223...89W}
}

@ARTICLE{Wolf1923,
       author = {{Wolf}, M.},
        title = "{{\"U}ber den dunklen Nebel NGC 6960}",
      journal = {Astron. Nachr.},
         year = 1923,
        month = jul,
       volume = {219},
       number = {7},
        pages = {109},
          doi = {10.1002/asna.19232190702},
       adsurl = {https://ui.adsabs.harvard.edu/abs/1923AN....219..109W}
}

@ARTICLE{GregorioHetem1988,
       author = {{Gregorio Hetem}, J.~C. and {Sanzovo}, G.~C. and {Lepine}, J.~R.~D.},
        title = "{Star counts and IRAS sources in the southern dark clouds.}",
      journal = {\aaps},
         year = 1988,
        month = dec,
       volume = {76},
        pages = {347-363},
       adsurl = {https://ui.adsabs.harvard.edu/abs/1988A&AS...76..347G}
}

@ARTICLE{Bok1977,
       author = {{Bok}, Bart J.},
        title = "{Dark Nebulae, Globules, and Protostars}",
      journal = {\pasp},
         year = 1977,
        month = oct,
       volume = {89},
        pages = {597},
          doi = {10.1086/130172},
       adsurl = {https://ui.adsabs.harvard.edu/abs/1977PASP...89..597B}
}

@ARTICLE{Uranova1962,
       author = {{Uranova}, T.~A.},
        title = "{Methods of Investigation of Thin, Dark Nebulae by Means of Star Counts}",
      journal = {\sovast},
         year = 1962,
        month = dec,
       volume = {6},
        pages = {376},
       adsurl = {https://ui.adsabs.harvard.edu/abs/1962SvA.....6..376U}
}

@ARTICLE{2024Natur.628...62K,
       author = {{Konietzka}, Ralf and {Goodman}, Alyssa A. and {Zucker}, Catherine and {Burkert}, Andreas and {Alves}, Jo{\~a}o and {Foley}, Michael and {Swiggum}, Cameren and {Koller}, Maria and {Miret-Roig}, N{\'u}ria},
        title = "{The Radcliffe Wave is oscillating}",
      journal = {\nat},
         year = 2024,
        month = apr,
       volume = {628},
       number = {8006},
        pages = {62-65},
          doi = {10.1038/s41586-024-07127-3},
archivePrefix = {arXiv},
       eprint = {2402.12596},
 primaryClass = {astro-ph.GA},
       adsurl = {https://ui.adsabs.harvard.edu/abs/2024Natur.628...62K}
}

@ARTICLE{2014ApJ...791..131K,
       author = {{Koenig}, X.~P. and {Leisawitz}, D.~T.},
        title = "{A Classification Scheme for Young Stellar Objects Using the Wide-field Infrared Survey Explorer AllWISE Catalog: Revealing Low-density Star Formation in the Outer Galaxy}",
      journal = {\apj},
         year = 2014,
        month = aug,
       volume = {791},
       number = {2},
          eid = {131},
        pages = {131},
          doi = {10.1088/0004-637X/791/2/131},
archivePrefix = {arXiv},
       eprint = {1407.2262},
 primaryClass = {astro-ph.GA},
       adsurl = {https://ui.adsabs.harvard.edu/abs/2014ApJ...791..131K}
}

@ARTICLE{2018MNRAS.477.2068K,
       author = {{Kiminki}, Megan M. and {Smith}, Nathan},
        title = "{A radial velocity survey of the Carina Nebula's O-type stars}",
      journal = {\mnras},
         year = 2018,
        month = jun,
       volume = {477},
       number = {2},
        pages = {2068-2086},
          doi = {10.1093/mnras/sty748},
archivePrefix = {arXiv},
       eprint = {1803.07057},
 primaryClass = {astro-ph.SR},
       adsurl = {https://ui.adsabs.harvard.edu/abs/2018MNRAS.477.2068K}
}

@article{Luhman2022,
  author       = {Luhman, K. L.},
  title        = {A Census of Young Stars and Brown Dwarfs in Taurus with Gaia Data Release 3},
  journal      = {\apj},
  year         = {2022},
  volume       = {941},
  number       = {2},
  pages        = {170},
  doi          = {10.3847/1538-4357/aca29a}
}

@ARTICLE{Cantat-Gaudin2021,
       author = {{Cantat-Gaudin}, Tristan and {Brandt}, Timothy D.},
        title = "{Characterizing and correcting the proper motion bias of the bright Gaia EDR3 sources}",
      journal = {\aap},
         year = 2021,
        month = may,
       volume = {649},
          eid = {A124},
        pages = {A124},
          doi = {10.1051/0004-6361/202140807},
archivePrefix = {arXiv},
       eprint = {2103.07432},
 primaryClass = {astro-ph.GA},
       adsurl = {https://ui.adsabs.harvard.edu/abs/2021A&A...649A.124C}
}

@ARTICLE{Lindegren2021,
       author = {{Lindegren}, L. and {Bastian}, U. and {Biermann}, M. and {Bombrun}, A. and {de Torres}, A. and {Gerlach}, E. and {Geyer}, R. and {Hern{\'a}ndez}, J. and {Hilger}, T. and {Hobbs}, D. and {Klioner}, S.~A. and {Lammers}, U. and {McMillan}, P.~J. and {Ramos-Lerate}, M. and {Steidelm{\"u}ller}, H. and {Stephenson}, C.~A. and {van Leeuwen}, F.},
        title = "{Gaia Early Data Release 3. Parallax bias versus magnitude, colour, and position}",
      journal = {\aap},
         year = 2021,
        month = may,
       volume = {649},
          eid = {A4},
        pages = {A4},
          doi = {10.1051/0004-6361/202039653},
archivePrefix = {arXiv},
       eprint = {2012.01742},
 primaryClass = {astro-ph.IM},
       adsurl = {https://ui.adsabs.harvard.edu/abs/2021A&A...649A...4L}
}

@ARTICLE{MaizApellaniz2022,
       author = {{Ma{\'\i}z Apell{\'a}niz}, J.},
        title = "{An estimation of the Gaia EDR3 parallax bias from stellar clusters and Magellanic Clouds data}",
      journal = {\aap},
         year = 2022,
        month = jan,
       volume = {657},
          eid = {A130},
        pages = {A130},
          doi = {10.1051/0004-6361/202142365},
archivePrefix = {arXiv},
       eprint = {2110.01475},
 primaryClass = {astro-ph.IM},
       adsurl = {https://ui.adsabs.harvard.edu/abs/2022A&A...657A.130M}
}

@ARTICLE{Szilagyi2023,
       author = {{Szil{\'a}gyi}, M{\'a}t{\'e} and {Kun}, M{\'a}ria and {{\'A}brah{\'a}m}, P{\'e}ter and {Marton}, G{\'a}bor},
        title = "{The Gaia view of the Cepheus OB2 association}",
      journal = {\mnras},
         year = 2023,
        month = mar,
       volume = {520},
       number = {1},
        pages = {1390-1410},
          doi = {10.1093/mnras/stad027},
archivePrefix = {arXiv},
       eprint = {2301.02346},
 primaryClass = {astro-ph.SR},
       adsurl = {https://ui.adsabs.harvard.edu/abs/2023MNRAS.520.1390S}
}

@ARTICLE{Kuhn2019,
       author = {{Kuhn}, Michael A. and {Hillenbrand}, Lynne A. and {Sills}, Alison and {Feigelson}, Eric D. and {Getman}, Konstantin V.},
        title = "{Kinematics in Young Star Clusters and Associations with Gaia DR2}",
      journal = {\apj},
         year = 2019,
        month = jan,
       volume = {870},
       number = {1},
          eid = {32},
        pages = {32},
          doi = {10.3847/1538-4357/aaef8c},
archivePrefix = {arXiv},
       eprint = {1807.02115},
 primaryClass = {astro-ph.GA},
       adsurl = {https://ui.adsabs.harvard.edu/abs/2019ApJ...870...32K}
}

@ARTICLE{Costado2017,
       author = {{Costado}, M.~T. and {Alfaro}, E.~J. and {Gonz{\'a}lez}, M. and {Sampedro}, L.},
        title = "{Analysis of the kinematic structure of the Cygnus OB1 association}",
      journal = {\mnras},
         year = 2017,
        month = mar,
       volume = {465},
       number = {4},
        pages = {3879-3888},
          doi = {10.1093/mnras/stw2967},
archivePrefix = {arXiv},
       eprint = {1611.04398},
 primaryClass = {astro-ph.GA},
       adsurl = {https://ui.adsabs.harvard.edu/abs/2017MNRAS.465.3879C}
}

@ARTICLE{Rivera2015,
       author = {{Rivera}, Juana L. and {Loinard}, Laurent and {Dzib}, Sergio A. and {Ortiz-Le{\'o}n}, Gisela N. and {Rodr{\'\i}guez}, Luis F. and {Torres}, Rosa M.},
        title = "{Internal and Relative Motions of the Taurus and Ophiuchus Star-forming Regions}",
      journal = {\apj},
         year = 2015,
        month = jul,
       volume = {807},
       number = {2},
          eid = {119},
        pages = {119},
          doi = {10.1088/0004-637X/807/2/119},
archivePrefix = {arXiv},
       eprint = {1506.00921},
 primaryClass = {astro-ph.SR},
       adsurl = {https://ui.adsabs.harvard.edu/abs/2015ApJ...807..119R}
}

@ARTICLE{Feddersen2018,
       author = {{Feddersen}, Jesse R. and {Arce}, H{\'e}ctor G. and {Kong}, Shuo and {Shimajiri}, Yoshito and {Nakamura}, Fumitaka and {Hara}, Chihomi and {Ishii}, Shun and {Sasaki}, Kazushige and {Kawabe}, Ryohei},
        title = "{Expanding CO Shells in the Orion A Molecular Cloud}",
      journal = {\apj},
         year = 2018,
        month = aug,
       volume = {862},
       number = {2},
          eid = {121},
        pages = {121},
          doi = {10.3847/1538-4357/aacaf2},
archivePrefix = {arXiv},
       eprint = {1806.01893},
 primaryClass = {astro-ph.GA},
       adsurl = {https://ui.adsabs.harvard.edu/abs/2018ApJ...862..121F}
}

@ARTICLE{Pabst2020,
       author = {{Pabst}, C.~H.~M. and {Goicoechea}, J.~R. and {Teyssier}, D. and {Bern{\'e}}, O. and {Higgins}, R.~D. and {Chambers}, E.~T. and {Kabanovic}, S. and {G{\"u}sten}, R. and {Stutzki}, J. and {Tielens}, A.~G.~G.~M.},
        title = "{Expanding bubbles in Orion A: [C II] observations of M 42, M 43, and NGC 1977}",
      journal = {\aap},
         year = 2020,
        month = jul,
       volume = {639},
          eid = {A2},
        pages = {A2},
          doi = {10.1051/0004-6361/202037560},
archivePrefix = {arXiv},
       eprint = {2005.03917},
 primaryClass = {astro-ph.GA},
       adsurl = {https://ui.adsabs.harvard.edu/abs/2020A&A...639A...2P}
}

@article{gala,
       author = {{Price-Whelan}, Adrian M.},
        title = "{Gala: A Python package for galactic dynamics}",
      journal = {JOSS},
         year = 2017,
        month = oct,
       volume = {2},
          eid = {388},
        pages = {388},
          doi = {10.21105/joss.00388},
       adsurl = {https://ui.adsabs.harvard.edu/abs/2017JOSS....2..388P}
}

@misc{adrian_price_whelan_2020_4159870,
  author       = {Adrian Price-Whelan and
                  Brigitta Sipőcz and
                  Daniel Lenz and
                  Johnny Greco and
                  Nathaniel Starkman and
                  Dan Foreman-Mackey and
                  P. L. Lim and
                  Semyeong Oh and
                  Sergey Koposov and
                  Syrtis Major},
  title        = {adrn/gala: v1.3},
  month        = oct,
  year         = 2020,
  publisher    = {Zenodo},
  version      = {v1.3},
  doi          = {10.5281/zenodo.4159870},
  url          = {https://doi.org/10.5281/zenodo.4159870},
}

@ARTICLE{Akhmetov2024,
       author = {{Akhmetov}, V.~S. and {Bucciarelli}, B. and {Crosta}, M. and {Lattanzi}, M.~G. and {Spagna}, A. and {Re Fiorentin}, P. and {Bannikova}, E. Yu},
        title = "{A new kinematic model of the Galaxy: analysis of the stellar velocity field from Gaia Data Release 3}",
      journal = {\mnras},
         year = 2024,
        month = may,
       volume = {530},
       number = {1},
        pages = {710-729},
          doi = {10.1093/mnras/stae772},
archivePrefix = {arXiv},
       eprint = {2307.08527},
 primaryClass = {astro-ph.GA},
       adsurl = {https://ui.adsabs.harvard.edu/abs/2024MNRAS.530..710A}
}

@ARTICLE{Zhou2025,
       author = {{Zhou}, Ji-Xuan and {Li}, Guang-Xing and {Chen}, Bing-Qiu},
        title = "{Kinematics of the interstellar medium using Gaia: A catalogue of 102 young stellar object{\textendash}molecular cloud associations within 3.5 kpc of the Sun with 3D velocities}",
      journal = {\mnras},
         year = 2025,
        month = sep,
       volume = {542},
       number = {1},
        pages = {52-61},
          doi = {10.1093/mnras/staf1222},
       adsurl = {https://ui.adsabs.harvard.edu/abs/2025MNRAS.542...52Z}
}

@ARTICLE{Zhou2022,
       author = {{Zhou}, Ji-Xuan and {Li}, Guang-Xing and {Chen}, Bing-Qiu},
        title = "{Kinematics of the molecular interstellar medium probed by Gaia: steep velocity dispersion-size relation, isotropic turbulence, and location-dependent energy dissipation}",
      journal = {\mnras},
         year = 2022,
        month = jun,
       volume = {513},
       number = {1},
        pages = {638-647},
          doi = {10.1093/mnras/stac900},
archivePrefix = {arXiv},
       eprint = {2110.11595},
 primaryClass = {astro-ph.GA},
       adsurl = {https://ui.adsabs.harvard.edu/abs/2022MNRAS.513..638Z}
}

@ARTICLE{Zhou2024,
       author = {{Zhou}, Ji-Xuan and {Li}, Guang-Xing and {Chen}, Bing-Qiu},
        title = "{Gas content and evolution of a sample of YSO associations at d {\ensuremath{\lesssim}} 3.5 kpc from the Sun}",
      journal = {\mnras},
         year = 2024,
        month = apr,
       volume = {529},
       number = {2},
        pages = {1091-1103},
          doi = {10.1093/mnras/stae376},
archivePrefix = {arXiv},
       eprint = {2402.02393},
 primaryClass = {astro-ph.GA},
       adsurl = {https://ui.adsabs.harvard.edu/abs/2024MNRAS.529.1091Z}
}

@ARTICLE{Roquette2025,
       author = {{Roquette}, J. and {Audard}, M. and {Hernandez}, D. and {Gezer}, I. and {Marton}, G. and {Mas}, C. and {Madar{\'a}sz}, M. and {Dionatos}, O.},
        title = "{The NEMESIS catalogue of young stellar objects for the Orion star formation complex: I. General description of data curation}",
      journal = {\aap},
         year = 2025,
        month = oct,
       volume = {702},
          eid = {A63},
        pages = {A63},
          doi = {10.1051/0004-6361/202453588},
archivePrefix = {arXiv},
       eprint = {2501.08486},
 primaryClass = {astro-ph.SR},
       adsurl = {https://ui.adsabs.harvard.edu/abs/2025A&A...702A..63R}
}

@BOOK{Reipurth2008a,
       author = {{Reipurth}, B.},
        title = "{Handbook of Star Forming Regions, Volume I: The Northern Sky}",
         year = 2008,
       volume = {4},
       adsurl = {https://ui.adsabs.harvard.edu/abs/2008hsf1.book.....R}
}

@BOOK{Reipurth2008b,
       author = {{Reipurth}, B.},
        title = "{Handbook of Star Forming Regions, Volume II: The Southern Sky}",
         year = 2008,
       volume = {5},
       adsurl = {https://ui.adsabs.harvard.edu/abs/2008hsf2.book.....R}
}

@ARTICLE{Grossschedl2019,
       author = {{Gro{\ss}schedl}, Josefa Elisabeth and {Alves}, Jo{\~a}o and {Teixeira}, Paula S. and {Bouy}, Herv{\'e} and {Forbrich}, Jan and {Lada}, Charles J. and {Meingast}, Stefan and {Hacar}, {\'A}lvaro and {Ascenso}, Joana and {Ackerl}, Christine and {Hasenberger}, Birgit and {K{\"o}hler}, Rainer and {Kubiak}, Karolina and {Larreina}, Irati and {Linhardt}, Lorenz and {Lombardi}, Marco and {M{\"o}ller}, Torsten},
        title = "{VISION - Vienna survey in Orion. III. Young stellar objects in Orion A}",
      journal = {\aap},
         year = 2019,
        month = feb,
       volume = {622},
          eid = {A149},
        pages = {A149},
          doi = {10.1051/0004-6361/201832577},
archivePrefix = {arXiv},
       eprint = {1810.00878},
 primaryClass = {astro-ph.SR},
       adsurl = {https://ui.adsabs.harvard.edu/abs/2019A&A...622A.149G}
}

@ARTICLE{Ducourant2017,
       author = {{Ducourant}, C. and {Teixeira}, R. and {Krone-Martins}, A. and {Bontemps}, S. and {Despois}, D. and {Galli}, P.~A.~B. and {Bouy}, H. and {Le Campion}, J.~F. and {Rapaport}, M. and {Cuillandre}, J.~C.},
        title = "{Proper motion survey and kinematic analysis of the {\ensuremath{\rho}} Ophiuchi embedded cluster}",
      journal = {\aap},
         year = 2017,
        month = jan,
       volume = {597},
          eid = {A90},
        pages = {A90},
          doi = {10.1051/0004-6361/201527574},
archivePrefix = {arXiv},
       eprint = {1609.04963},
 primaryClass = {astro-ph.SR},
       adsurl = {https://ui.adsabs.harvard.edu/abs/2017A&A...597A..90D}
}

@ARTICLE{Galli2020a,
       author = {{Galli}, P.~A.~B. and {Bouy}, H. and {Olivares}, J. and {Miret-Roig}, N. and {Sarro}, L.~M. and {Barrado}, D. and {Berihuete}, A. and {Brandner}, W.},
        title = "{Corona-Australis DANCe. I. Revisiting the census of stars with Gaia-DR2 data}",
      journal = {\aap},
         year = "2020a",
        month = feb,
       volume = {634},
          eid = {A98},
        pages = {A98},
          doi = {10.1051/0004-6361/201936708},
archivePrefix = {arXiv},
       eprint = {2001.05190},
 primaryClass = {astro-ph.SR},
       adsurl = {https://ui.adsabs.harvard.edu/abs/2020A&A...634A..98G}
}

\begin{appendix} 

\section{Clustering of YSOs and OCs}\label{sec:appendix_clusters}

Figure \ref{fig:DBSCAN_clusters_complete} shows the comoving YSOs selected using the DBSCAN algorithm in the proper motion space for all the clouds, as explained in \S\ref{sec:method}. OC members are also shown for comparison.

\begin{figure*}
  \centering

  \begin{minipage}[b]{0.32\linewidth}
    \includegraphics[width=\linewidth]{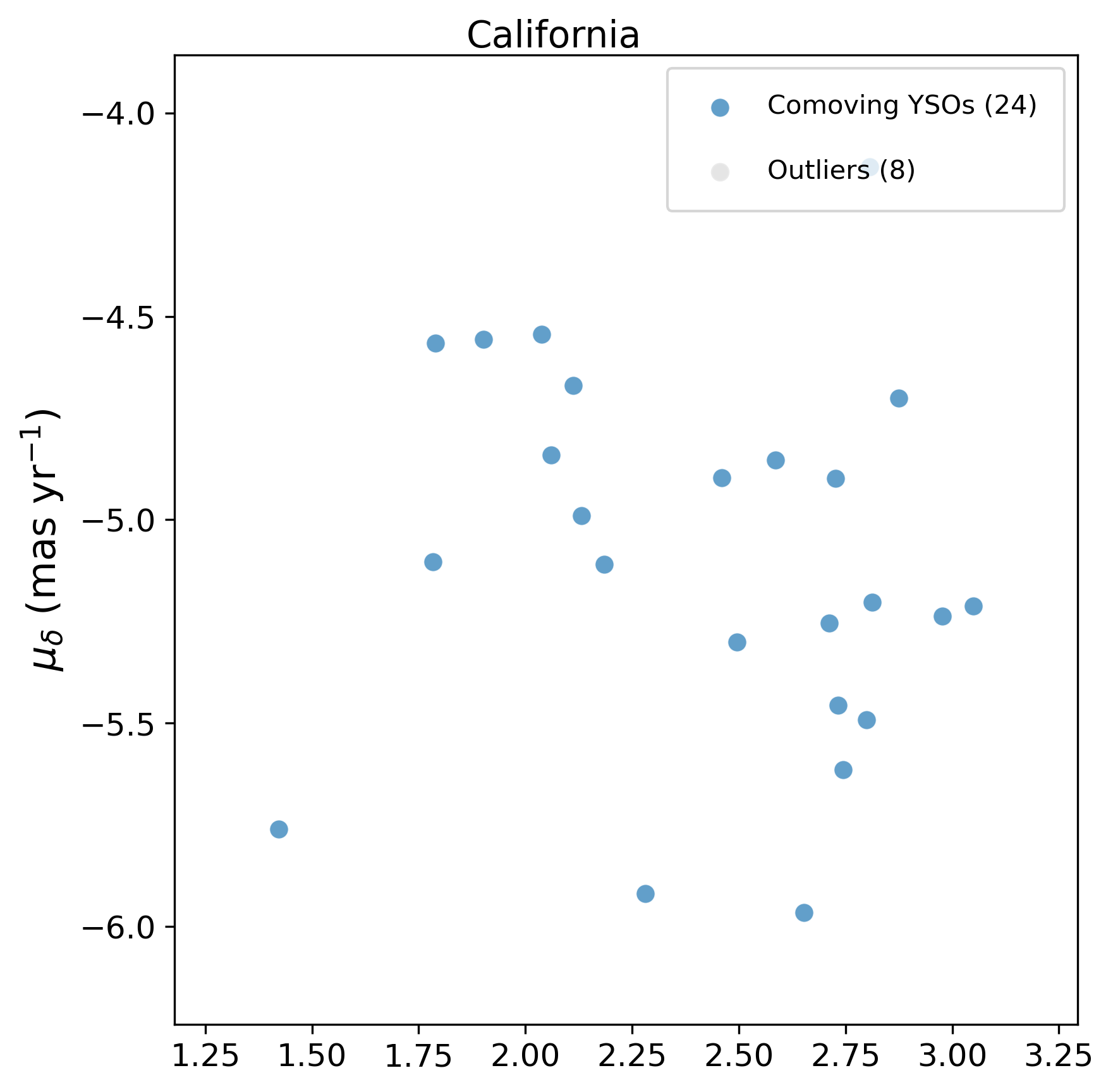}
  \end{minipage}\hfill
  \begin{minipage}[b]{0.32\linewidth}
    \includegraphics[width=\linewidth]{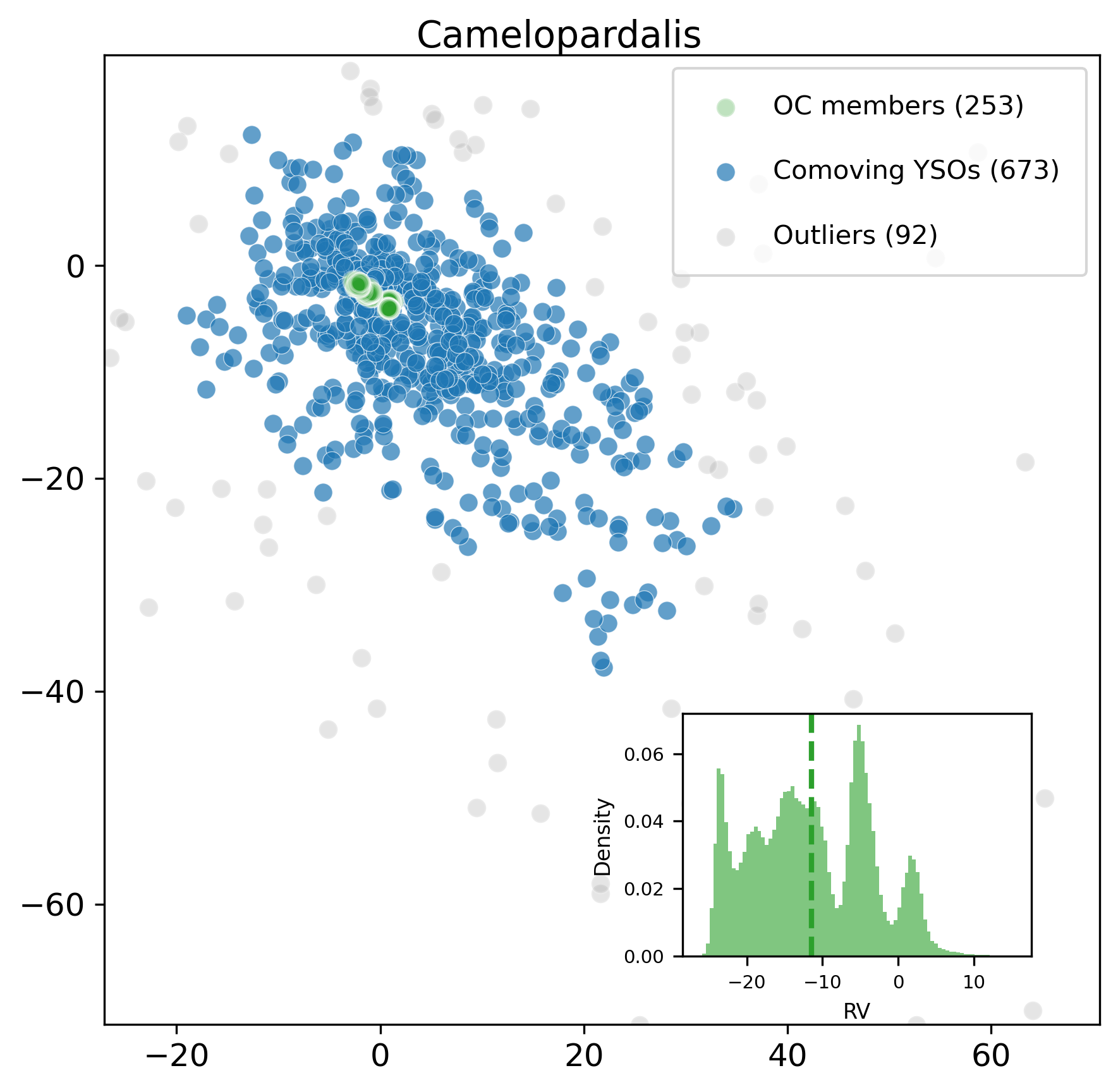}
  \end{minipage}\hfill
  \begin{minipage}[b]{0.32\linewidth}
    \includegraphics[width=\linewidth]{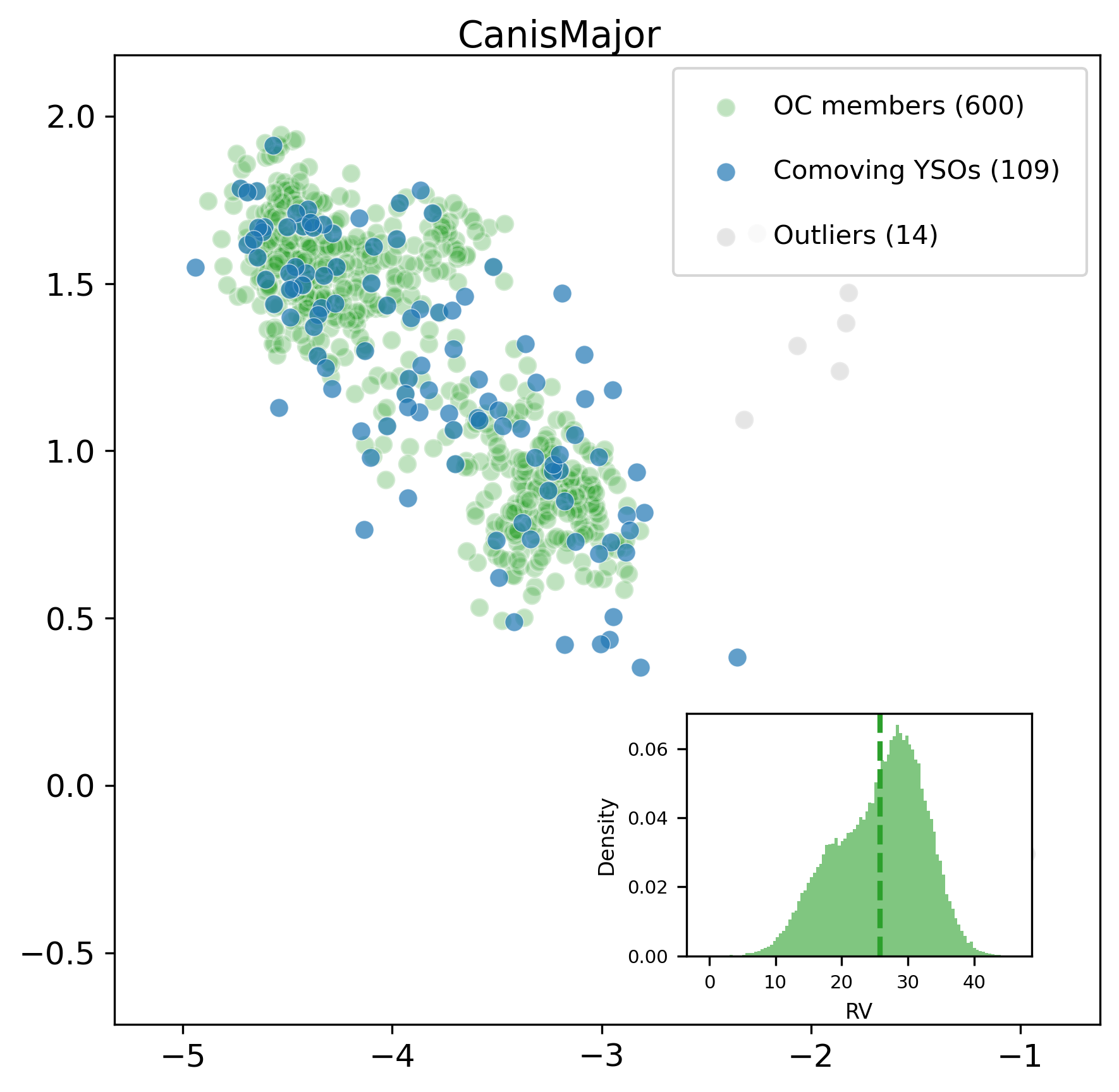}
  \end{minipage}

    \vspace{-0.5mm}

  \begin{minipage}[b]{0.32\linewidth}
    \includegraphics[width=\linewidth]{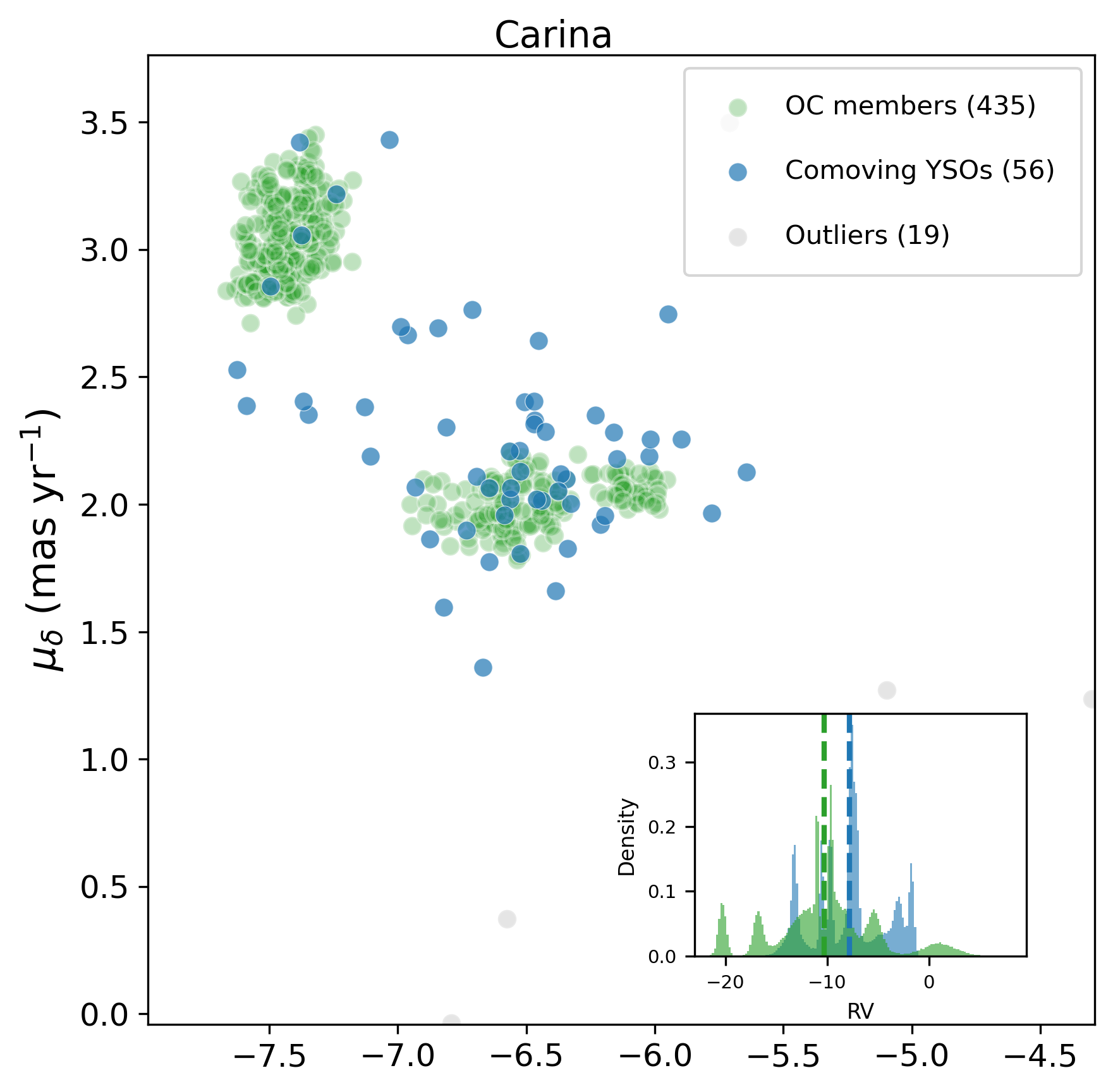}
  \end{minipage}\hfill
  \begin{minipage}[b]{0.32\linewidth}
    \includegraphics[width=\linewidth]{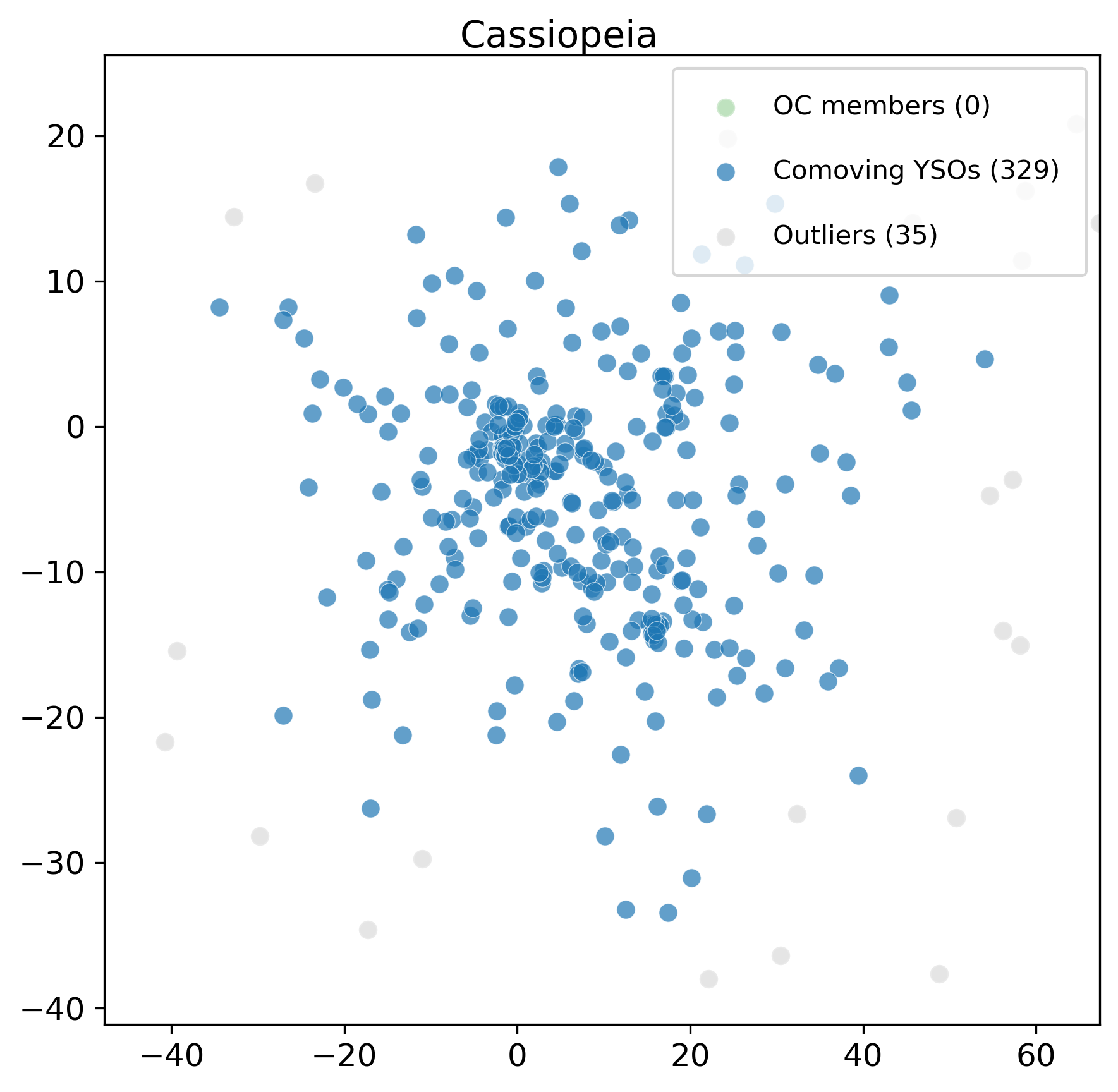}
  \end{minipage}\hfill
  \begin{minipage}[b]{0.32\linewidth}
    \includegraphics[width=\linewidth]{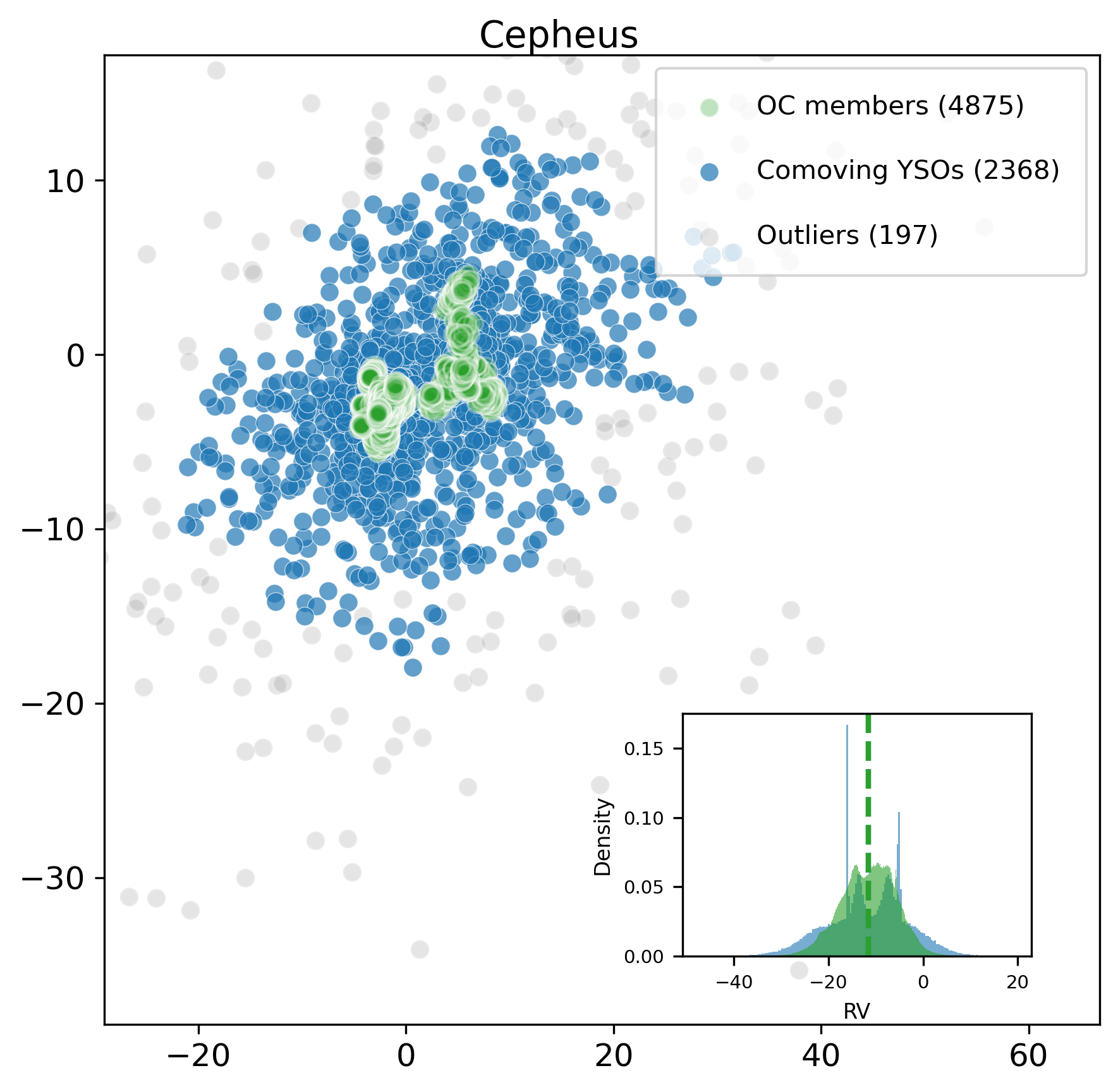}
  \end{minipage}
  
  \vspace{-0.5mm}

  \begin{minipage}[b]{0.32\linewidth}
    \includegraphics[width=\linewidth]{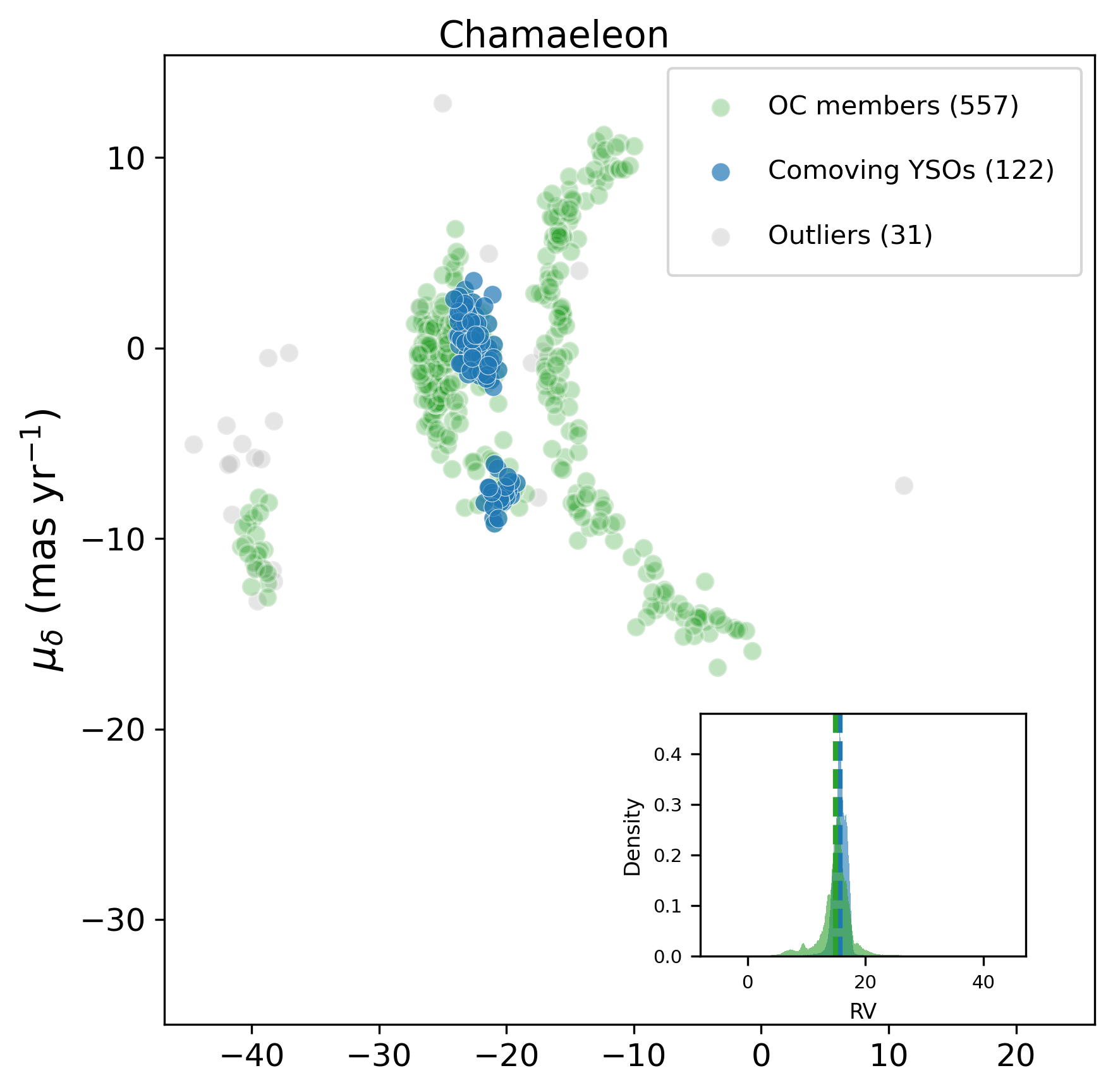}
  \end{minipage}\hfill
  \begin{minipage}[b]{0.32\linewidth}
    \includegraphics[width=\linewidth]{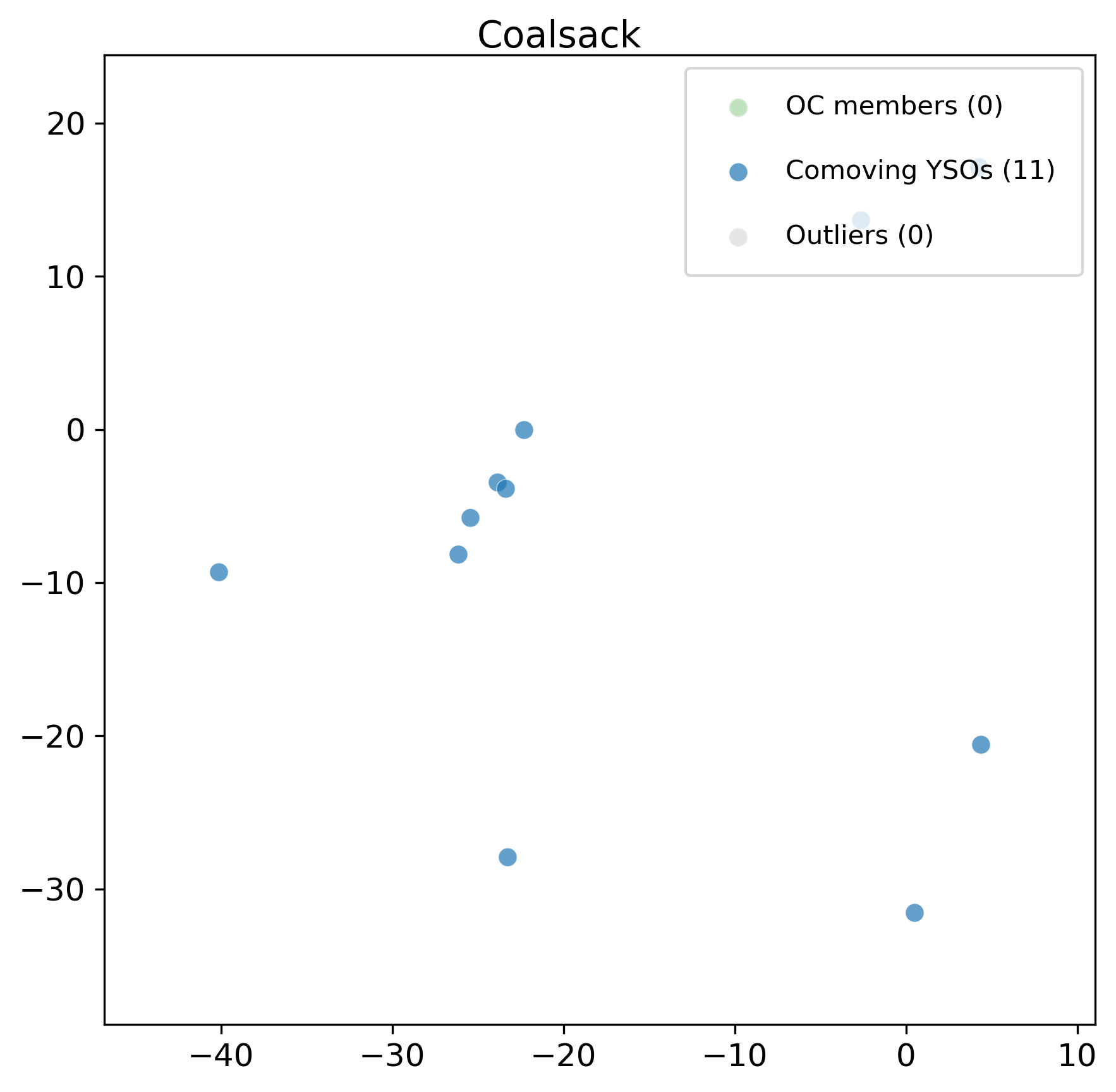}
  \end{minipage}\hfill
  \begin{minipage}[b]{0.32\linewidth}
    \includegraphics[width=\linewidth]{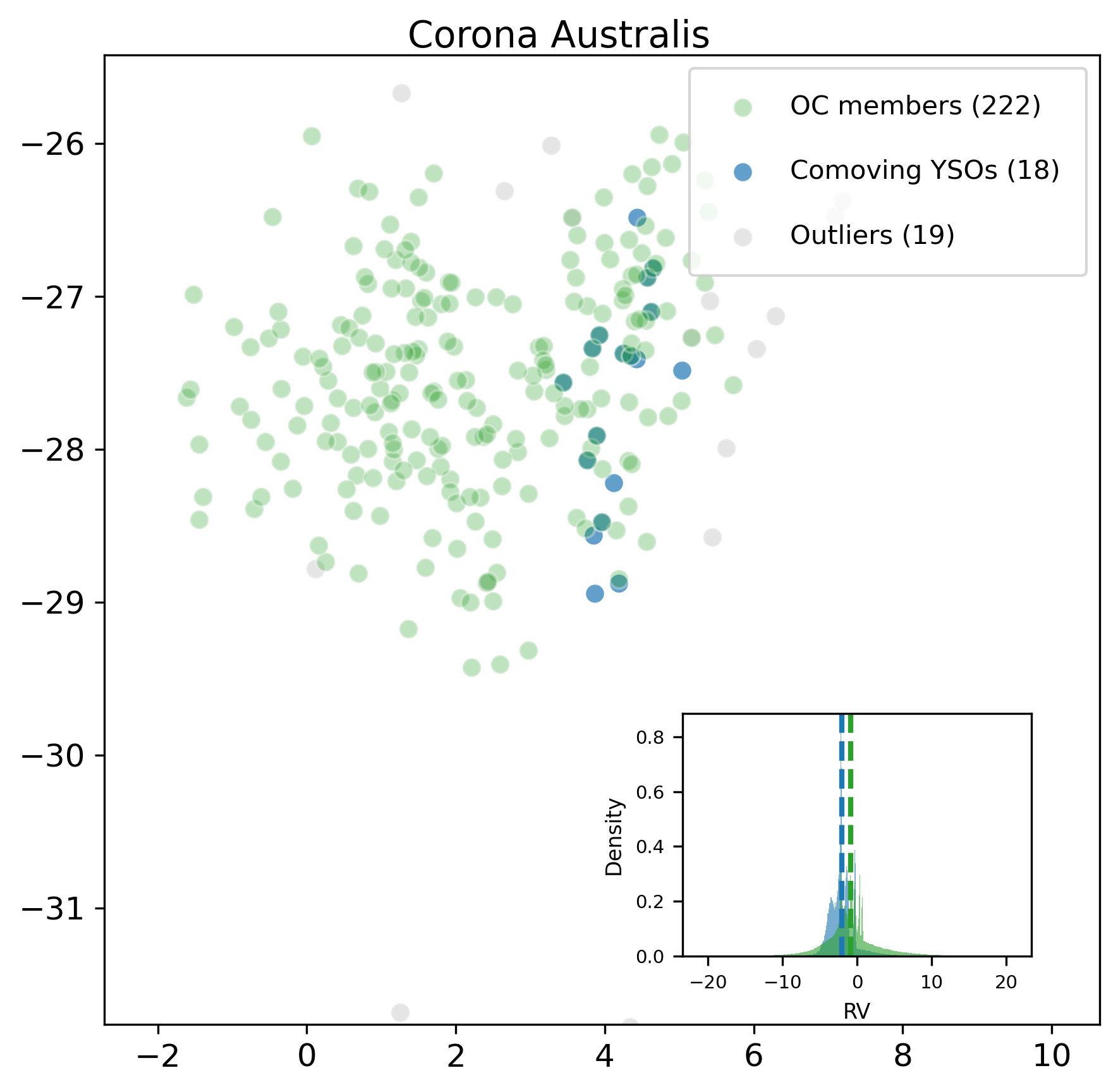}
  \end{minipage}

  \vspace{-0.5mm}

  \begin{minipage}[b]{0.32\linewidth}
    \includegraphics[width=\linewidth]{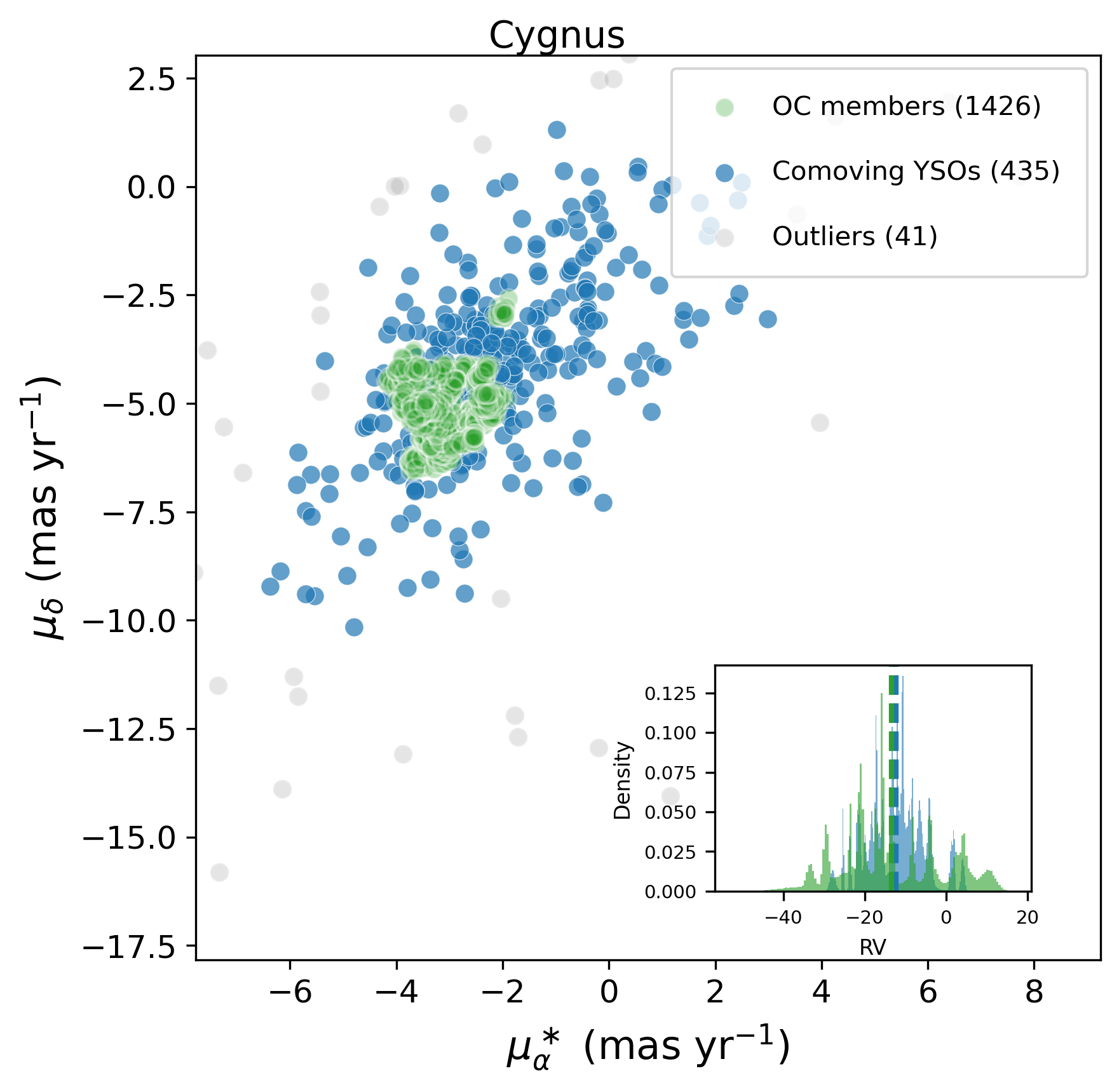}
  \end{minipage}\hfill
  \begin{minipage}[b]{0.32\linewidth}
    \includegraphics[width=\linewidth]{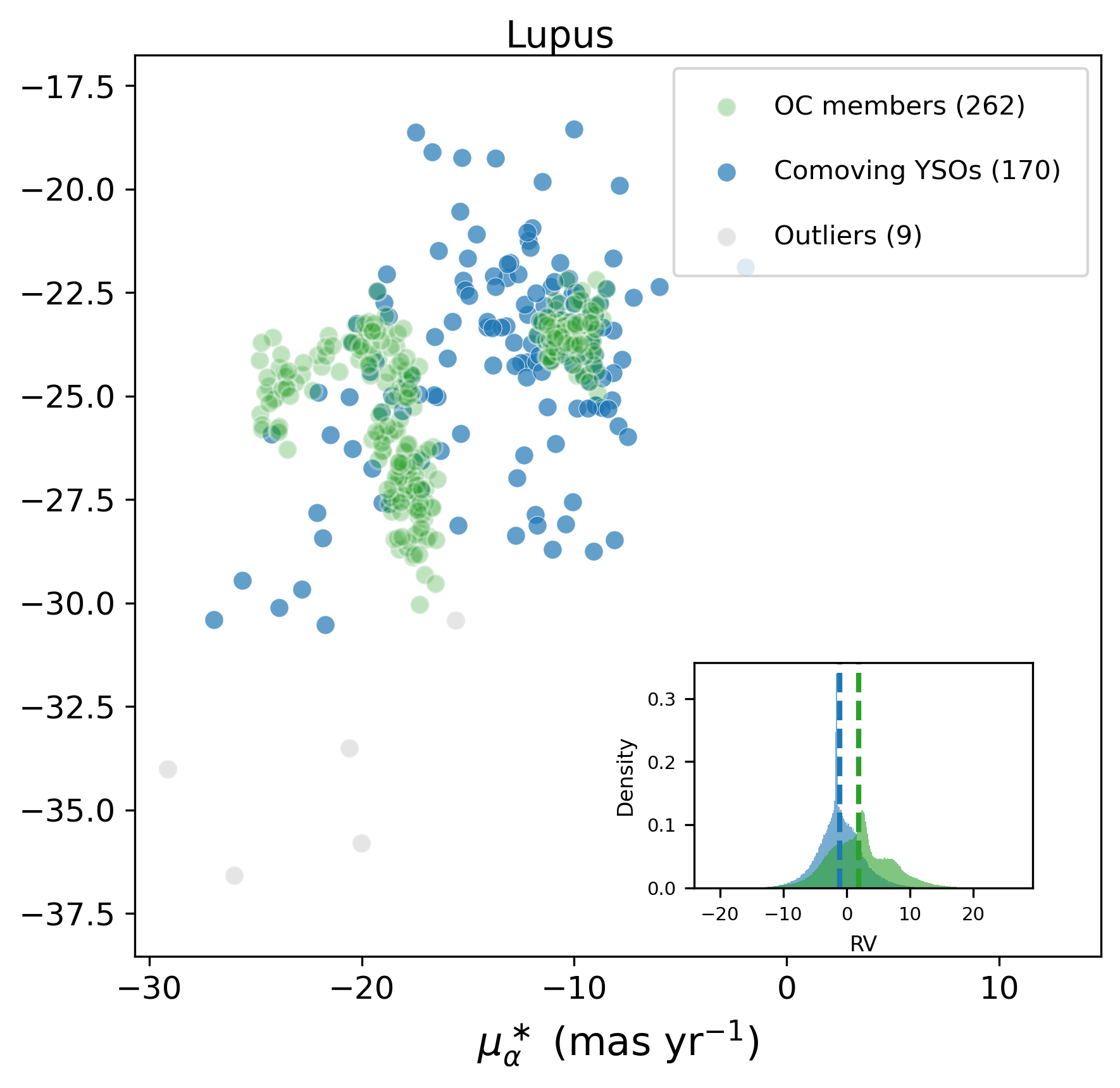}
  \end{minipage}\hfill
  \begin{minipage}[b]{0.32\linewidth}
    \includegraphics[width=\linewidth]{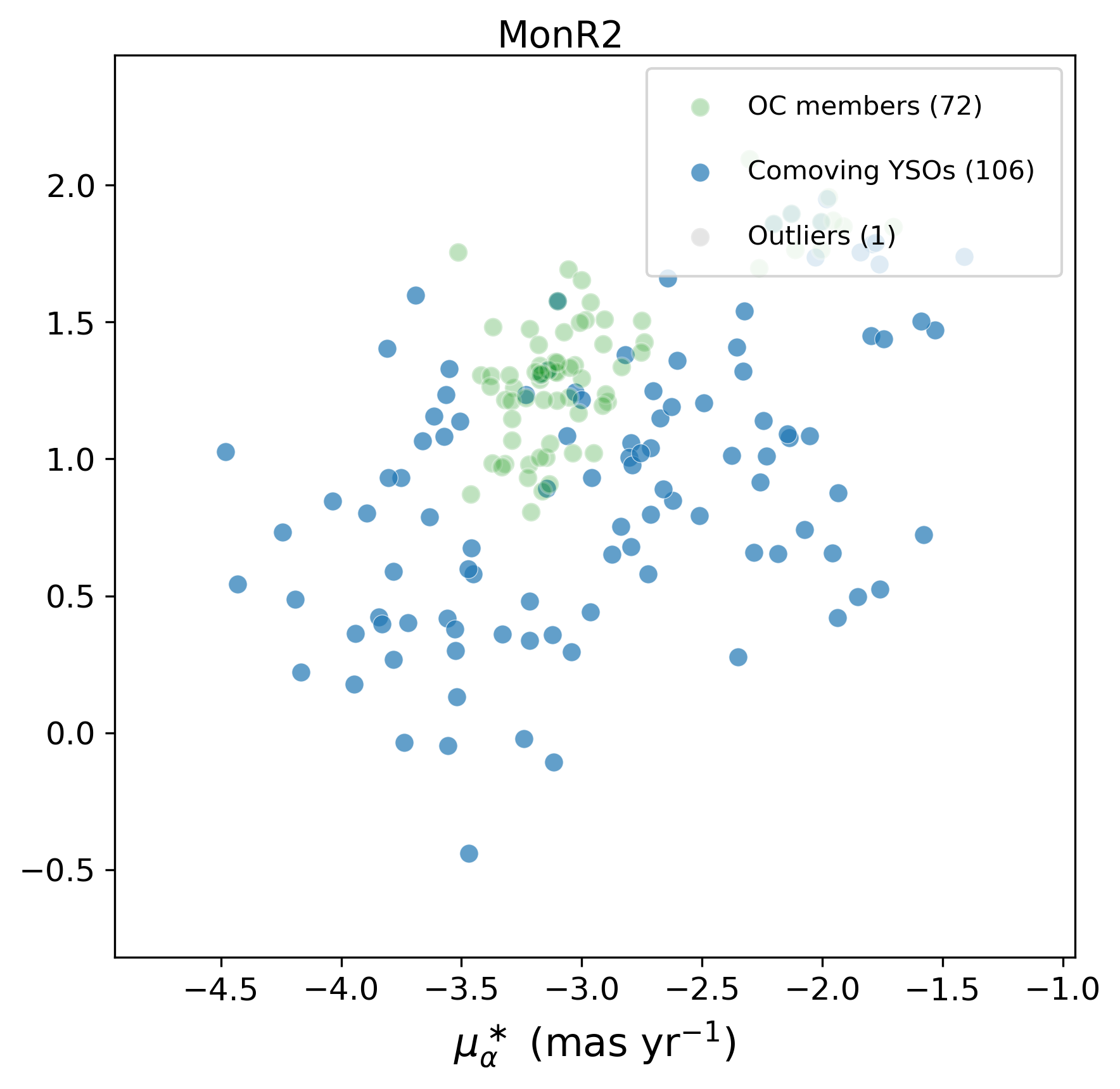}
  \end{minipage}

  \caption{Resulting clusters of comoving YSOs (blue) in the 2D ($\mu_{\alpha}^\ast$, $\mu_{\delta}$) parameter space for the remaining 15 MCs in our sample, with the OC members from the HR24 catalogue shown in green for comparison. When available, an inset in the lower-right corner of each plot shows the  RVs distributions of both populations based on $10^5$ Monte Carlo samplings; the dotted lines indicate the mean RV of the YSO and OC members.}
  \label{fig:DBSCAN_clusters_complete}
\end{figure*}
\begin{figure*}[t]
  \ContinuedFloat

  \begin{minipage}[b]{0.33\linewidth}
    \includegraphics[width=\linewidth]{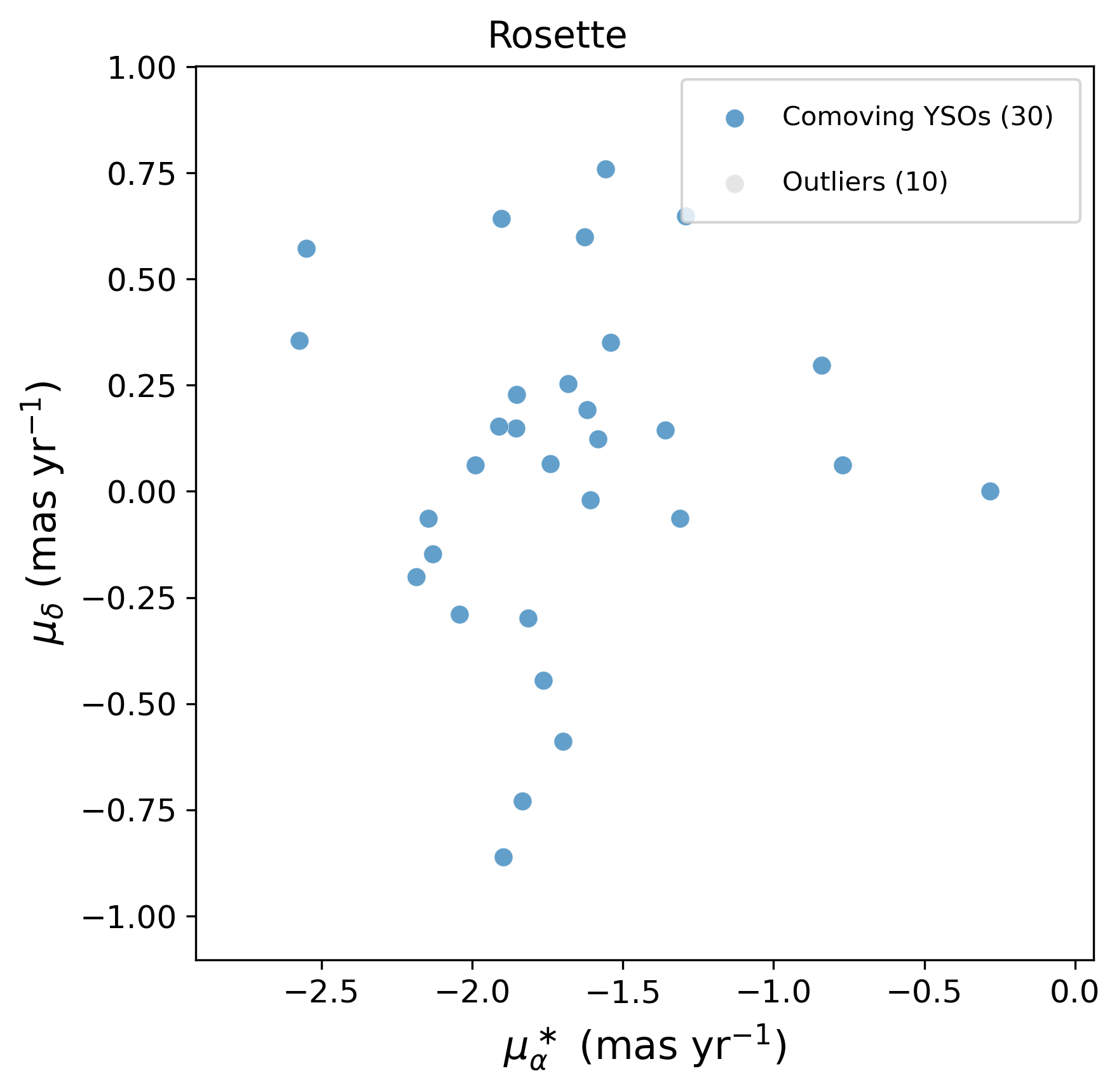}
  \end{minipage}\hfill
  \begin{minipage}[b]{0.33\linewidth}
    \includegraphics[width=\linewidth]{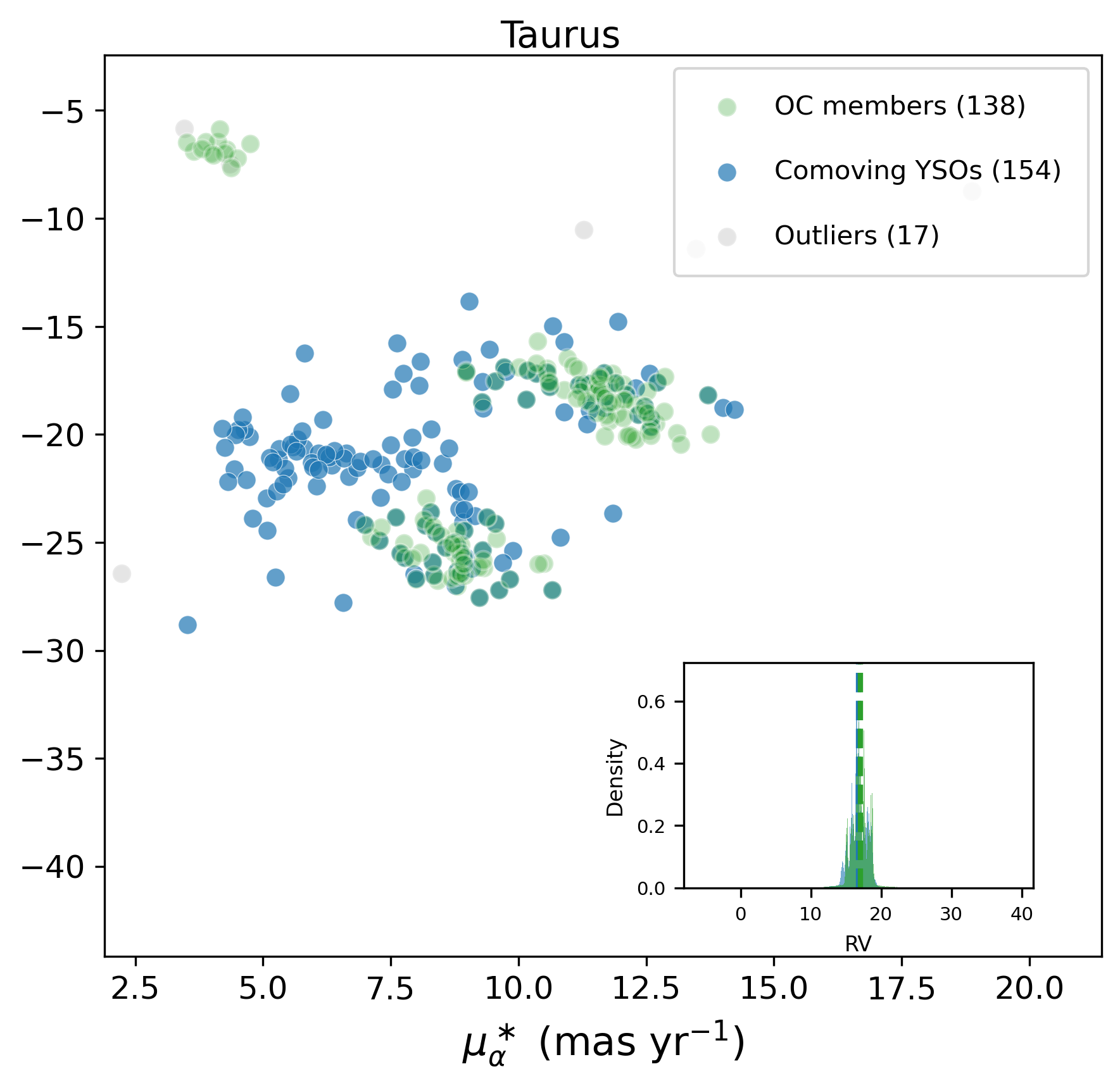}
  \end{minipage}\hfill
  \begin{minipage}[b]{0.33\linewidth}
    \includegraphics[width=\linewidth]{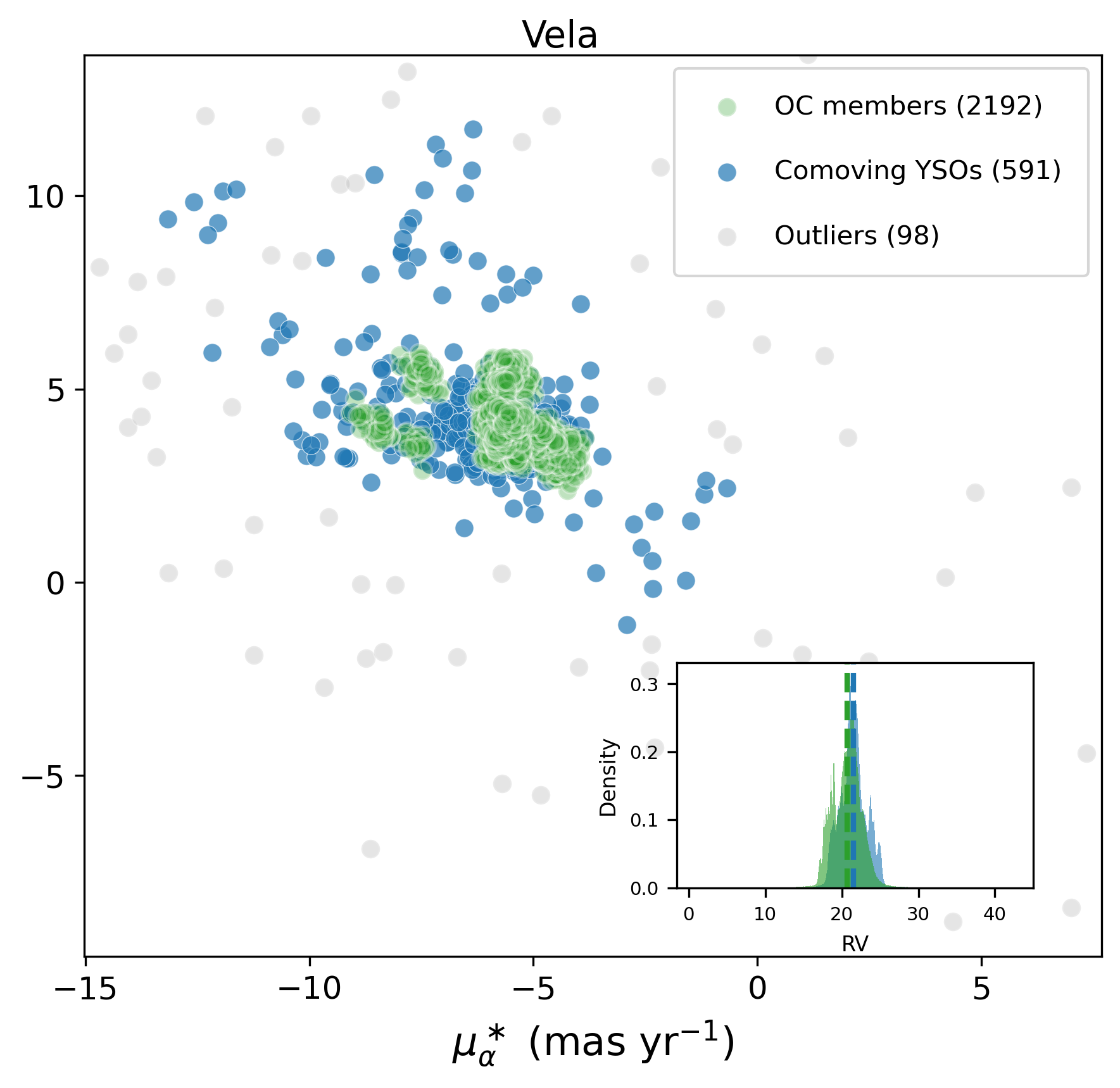}
  \end{minipage}

  \caption{Continued.}
\end{figure*}

 \section{Comparison with the literature}\label{appendix}
In this section we compare our 3D motion estimates for each cloud previously studied in the literature. A summary of the motions traced both from YSO and OC members is shown in Table \ref{tab:YSOvsOCmotions}.

\begin{table*}[ht!]
{\renewcommand{\arraystretch}{1.5}
{\fontsize{7.5}{9}\selectfont
\caption{\label{tab:YSOvsOCmotions} Mean proper motions and RV expressed as mean $\pm$ standard error (standard deviation) for 17 clouds of our sample computed by using YSOs and OCs as tracers.}
\centering
\setlength{\tabcolsep}{3pt}

\begin{tabular}{lcccccccc}
\toprule
Complex 
  & \multicolumn{2}{c}{$\overline \mu_{\alpha^\ast} \pm e_{\mu_{\alpha^\ast}}$ ($\sigma_{\mu_{\alpha^\ast}}$) mas yr$^{-1}$} 
  & \multicolumn{2}{c}{$\overline \mu_{\delta} \pm e_{\mu_{\delta}}$ ($\sigma_{\mu_\delta}$) mas yr$^{-1}$}
  & \multicolumn{2}{c}{$\overline{RV} \pm e_{RV}$ (km s$^{-1}$)}
  & \multicolumn{2}{c}{$N_\mu$ $\left[ N_{RV} \right]$}\\
\cmidrule(lr){2-3}\cmidrule(lr){4-5}\cmidrule(lr){6-7}\cmidrule(lr){8-9}
 & $\left<YSOs\right>$ & $\left<OCs\right>$
 & $\left<YSOs\right>$ & $\left<OCs\right>$
 & $\left<YSOs\right>$ & $\left<OCs\right>$
 & $\left<YSOs\right>$ & $\left<OC_{members}\right>$\\
\midrule
        \textbf{California} & 2.4 $\pm$ 0.1 (0.5) & - & -5.1 $\pm$ 0.1 (0.5) & - & 8.4 $\pm$ 0.9 (2) & - & 24 [7] & 0 [0] \\ 
        \textbf{Camelopardalis} & 3.6 $\pm$ 0.4 (9) & -1.00 $\pm$ 0.07 (1) & -6.3 $\pm$ 0.3 (8) & -2.57 $\pm$ 0.04 (0.7) & - & -11 $\pm$ 2 (8) & 673 [2] & 253 [14] \\ 
        Cam Local Layer & 5.2 $\pm$ 0.6 (10) & - & -8.7 $\pm$ 0.5 (10) & - & - & - & 339 [1] & 0 [0] \\ 
        Cam OB1 & 0.7 $\pm$ 0.4 (4) & -0.89 $\pm$ 0.08 (1) & -2.7 $\pm$ 0.4 (4) & -2.65 $\pm$ 0.05 (0.7) & - & -11 $\pm$ 2 (8) & 125 [0] & 205 [13] \\ 
        \textbf{Canis Major} & -3.83 $\pm$ 0.06 (0.6) & -3.94 $\pm$ 0.02 (0.6) & 1.23 $\pm$ 0.04 (0.4) & 1.32 $\pm$ 0.02 (0.4) & - & 26 $\pm$ 3 (7) & 109 [0] & 600 [7] \\ 
        \textbf{Carina} & -6.62 $\pm$ 0.06 (0.5) & -7.04 $\pm$ 0.02 (0.5) & 2.27 $\pm$ 0.06 (0.4) & 2.64 $\pm$ 0.03 (0.5) & -8 $\pm$ 1 (4) & -10 $\pm$ 1 (5) & 56 [12] & 435 [13] \\ 
        North & -6.56 $\pm$ 0.07 (0.5) & -6.98 $\pm$ 0.04 (0.5) & 2.28 $\pm$ 0.06 (0.4) & 2.58 $\pm$ 0.04 (0.5) & -8 $\pm$ 1 (4) & -14 $\pm$ 3 (5) & 42 [10] & 144 [3] \\ 
        Southern Pillars & -6.78 $\pm$ 0.07 (0.09) & -6.60 $\pm$ 0.02 (0.02) & 2.73 $\pm$ 0.05 (0.07) & 1.88 $\pm$ 0.02 (0.02) & - & - & 2 [0] & 1 [0] \\ 
        \textbf{Cassiopeia} & 6.0 $\pm$ 0.8 (10) & - & -4.6 $\pm$ 0.5 (8) & - & - & - & 329 [0] & 0 [0] \\ 
        \textbf{Cepheus} & 1.0 $\pm$ 0.1 (6) & 0.49 $\pm$ 0.05 (4) & -2.25 $\pm$ 0.08 (4) & -2.36 $\pm$ 0.03 (2) & -12 $\pm$ 2 (8) & -11.5 $\pm$ 0.4 (6) & 2368 [17] & 4875 [223] \\ 
        Cepheus OB2 & -1.65 $\pm$ 0.08 (2) & -2.04 $\pm$ 0.02 (0.8) & -3.59 $\pm$ 0.08 (2) & -3.28 $\pm$ 0.02 (1) & -9 $\pm$ 3 (5) & -15 $\pm$ 1 (7) & 566 [3] & 2158 [23] \\ 
        Cepheus Flare & 6.1 $\pm$ 0.2 (3) & 5.48 $\pm$ 0.03 (0.4) & 0.7 $\pm$ 0.2 (3) & 0.78 $\pm$ 0.07 (0.9) & -14 $\pm$ 3 (8) & -14 $\pm$ 2 (6) & 142 [8] & 164 [15] \\ 
        \textbf{Chamaeleon} & -22.1 $\pm$ 0.1 (1) & -21.0 $\pm$ 0.3 (7) & -1.3 $\pm$ 0.3 (3) & -1.7 $\pm$ 0.2 (5) & 15.7 $\pm$ 0.2 (1) & 15.0 $\pm$ 0.3 (3) & 122 [32] & 557 [79] \\ 
        Chamaeleon I & -22.48 $\pm$ 0.08 (0.8) & -22.58 $\pm$ 0.06 (0.8) & 0.3 $\pm$ 0.1 (1) & 0.45 $\pm$ 0.08 (1) & 15.7 $\pm$ 0.2 (1) & 15.3 $\pm$ 0.2 (2) & 97 [31] & 182 [62] \\ 
        Chamaeleon II & -20.4 $\pm$ 0.1 (0.6) & -20.3 $\pm$ 0.1 (0.9) & -7.8 $\pm$ 0.1 (0.6) & -7.7 $\pm$ 0.1 (0.6) & - & - & 20 [0] & 38 [0] \\ 
        \textbf{Corona Australis} & 4.2 $\pm$ 0.1 (0.4) & 2.2 $\pm$ 0.1 (2) & -27.7 $\pm$ 0.2 (0.7) & -27.55 $\pm$ 0.05 (0.7) & -2.1 $\pm$ 0.5 (2) & -0.9 $\pm$ 0.5 (3) & 18 [12] & 222 [38] \\ 
        \textbf{Cygnus} & -2.44 $\pm$ 0.07 (2) & -3.03 $\pm$ 0.01 (0.5) & -4.38 $\pm$ 0.09 (2) & -5.26 $\pm$ 0.02 (0.7) & -12 $\pm$ 1 (7) & -13 $\pm$ 2 (10) & 435 [46] & 1426 [33] \\ 
        Cygnus X North & -2.4 $\pm$ 0.2 (2) & -2.69 $\pm$ 0.02 (0.1) & -4.1 $\pm$ 0.2 (2) & -4.24 $\pm$ 0.02 (0.1) & -11 $\pm$ 4 (6) & - & 101 [3] & 38 [0] \\ 
        Cygnus X South & -2.46 $\pm$ 0.08 (1) & -3.03 $\pm$ 0.01 (0.5) & -4.47 $\pm$ 0.09 (2) & -5.29 $\pm$ 0.02 (0.6) & -12 $\pm$ 1 (7) & -13 $\pm$ 2 (10) & 334 [43] & 1388 [33] \\ 
        \textbf{Lupus} & -12.9 $\pm$ 0.3 (4) & -16.4 $\pm$ 0.3 (5) & -24.0 $\pm$ 0.2 (2) & -25.1 $\pm$ 0.1 (2) & -1.1 $\pm$ 0.8 (3) & 1.9 $\pm$ 0.6 (5) & 170 [18] & 262 [61] \\ 
        Lupus 1 & -15.7 $\pm$ 0.9 (4) & -17.7 $\pm$ 0.1 (0.2) & -23.3 $\pm$ 0.5 (2) & -24.7 $\pm$ 0.1 (0.3) & - & 4 $\pm$ 1 (3) & 18 [0] & 6 [4] \\ 
        Lupus 2 & -11.80 $\pm$ 0.03 (0.03) & - & -22.51 $\pm$ 0.02 (0.02) & - & - & - & 1 [0] & 0 [0] \\ 
        Lupus 3 & -10.1 $\pm$ 0.2 (1) & -11.1 $\pm$ 0.4 (3) & -23.7 $\pm$ 0.2 (1) & -24.2 $\pm$ 0.2 (2) & -1 $\pm$ 1 (4) & -1 $\pm$ 1 (4) & 45 [7] & 61 [10] \\ 
        Lupus 4 & -11.7 $\pm$ 0.5 (2) & -11.5 $\pm$ 0.4 (1) & -24.1 $\pm$ 0.4 (2) & -23.7 $\pm$ 0.2 (0.8) & - & - & 18 [2] & 17 [2] \\ 
        Lupus 5 & -10 $\pm$ 1 (2) & -17.29 $\pm$ 0.08 (0.08) & -23.8 $\pm$ 0.6 (1) & -28.83 $\pm$ 0.05 (0.05) & - & - & 4 [0] & 1 [0] \\ 
        \textbf{Mon R2} & -2.90 $\pm$ 0.07 (0.8) & -2.97 $\pm$ 0.05 (0.4) & 0.92 $\pm$ 0.05 (0.5) & 1.36 $\pm$ 0.04 (0.3) & - & - & 106 [5] & 72 [0] \\ 
        \textbf{Ophiuchus} & -9.2 $\pm$ 0.1 (3) & -10.8 $\pm$ 0.1 (3) & -24.4 $\pm$ 0.1 (2) & -24.3 $\pm$ 0.1 (3) & -5.9 $\pm$ 0.1 (2) & -5.6 $\pm$ 0.1 (2) & 462 [201] & 947 [388] \\ 
        \textbf{Orion} & 1.02 $\pm$ 0.02 (1) & 0.16 $\pm$ 0.03 (2) & -0.42 $\pm$ 0.02 (1) & -1.00 $\pm$ 0.03 (2) & 26.82 $\pm$ 0.08 (3) & 26.9 $\pm$ 0.1 (3) & 2895 [1108] & 4219 [970] \\ 
        Orion A & 1.08 $\pm$ 0.02 (0.8) & 0.65 $\pm$ 0.03 (0.9) & -0.15 $\pm$ 0.02 (1) & -0.12 $\pm$ 0.03 (0.9) & 26.69 $\pm$ 0.09 (3) & 27.4 $\pm$ 0.2 (3) & 2129 [877] & 1170 [276] \\ 
        Orion B & 0.40 $\pm$ 0.09 (2) & 0.2 $\pm$ 0.1 (2) & -0.73 $\pm$ 0.05 (0.9) & -0.52 $\pm$ 0.05 (0.7) & 28.0 $\pm$ 0.2 (2) & 28.4 $\pm$ 0.6 (4) & 305 [105] & 204 [48] \\ 
        $\lambda$ Orionis & 1.26 $\pm$ 0.05 (0.9) & 0.93 $\pm$ 0.03 (1) & -1.78 $\pm$ 0.06 (1) & -2.10 $\pm$ 0.03 (1) & 27.0 $\pm$ 0.2 (2) & 27.2 $\pm$ 0.1 (3) & 301 [117] & 1601 [401] \\ 
        \textbf{Perseus} & 4.89 $\pm$ 0.07 (1) & 4.98 $\pm$ 0.08 (2) & -6.79 $\pm$ 0.08 (2) & -7.1 $\pm$ 0.1 (2) & 15.15 $\pm$ 0.07 (1) & 15.9 $\pm$ 0.5 (4) & 370 [178] & 432 [79] \\ 
        NGC 1333 & 7.7 $\pm$ 0.3 (1) & 8.0 $\pm$ 0.1 (0.4) & -8.3 $\pm$ 0.3 (1) & -8.3 $\pm$ 0.1 (0.6) & - & 12 $\pm$ 1 (3) & 15 [1] & 17 [5] \\ 
        IC 348 & 4.63 $\pm$ 0.05 (0.8) & 6.68 $\pm$ 0.05 (0.4) & -6.56 $\pm$ 0.06 (0.9) & -9.42 $\pm$ 0.08 (0.6) & 15.22 $\pm$ 0.08 (0.9) & 16.3 $\pm$ 0.7 (3) & 263 [147] & 49 [12] \\ 
        \textbf{Rosette} & -1.70 $\pm$ 0.09 (0.5) & - & 0.06 $\pm$ 0.07 (0.4) & - & 29 $\pm$ 1 (4) & - & 30 [15] & 0 [0] \\ 
        \textbf{Taurus} & 8.4 $\pm$ 0.2 (2) & 9.8 $\pm$ 0.2 (2) & -21.2 $\pm$ 0.3 (3) & -19.9 $\pm$ 0.5 (6) & 16.8 $\pm$ 0.1 (1) & 17.0 $\pm$ 0.2 (1) & 154 [86] & 138 [75] \\ 
        \textbf{Vela} & -6.02 $\pm$ 0.06 (2) & -5.37 $\pm$ 0.02 (0.9) & 4.17 $\pm$ 0.06 (1) & 4.00 $\pm$ 0.02 (0.8) & 21.5 $\pm$ 0.2 (2) & 20.7 $\pm$ 0.1 (2) & 591 [72] & 2192 [257] \\ \bottomrule
    \end{tabular}
    }}
\end{table*}

\subsection*{B.1. Camelopardalis}

We divided this region in terms of the distance into the Cam Local Layer and the Cam OB1 layer following \citet{Dharmawardena2023}.
Indeed, \cite{Digel1996} concluded that the molecular clouds in Camelopardalis region comprises several layers at different distances and RVs, with velocities between –5 and +10~km~s$^{-1}$ for the local layer, and between –5 and –20 km/s for the Cam OB1 layer. This is broadly consistent with the mean value of $-11$~km~s$^{-1}$ with a dispersion of $8$~km s$^{-1}$ seen in this work. To our knowledge, there are no previous estimates of the proper motion of this cloud complex in the literature.

\subsection*{B.2. Canis Major}

The kinematics of seven subregions of the Canis Major MC complex were recently studied in \citet{Dong2024}, by using DBSCAN to cluster YSOs contained in each of them. As a result, \citet{Dong2024} obtained a mean $\mu_{\alpha^\ast} = -3.7 \pm 0.2$ mas yr$^{-1}$ and $\mu_\delta = 1.1 \pm 0.1$ mas yr$^{-1}$.

As can be seen in Table \ref{tab:YSOvsOCmotions}, they are in excellent agreement with the mean proper motions estimated with YSOs and with the OCs.

Regarding the RV, \citet{Dong2024} used $^{13}$CO to obtain an average RV of $15.1 \pm 0.5$ km s$^{-1}$ with respect to the Local Standard of Rest (LSR). To compare the heliocentric RV that we calculated from OCs with their value, we converted it to the LSR frame by adopting the solar motion as $\left(U_{\odot}, V_{\odot}, W_{\odot}\right)$ = $\left(11.1,12.24,7.25\right)$ km s$^{-1}$ from \citet{Schonrich2010} and using the galactic coordinates $\left(l, b\right)$ of the centroid of the Canis Major cloud according to \citet{Dharmawardena2023}. As a result, our $RV_{\rm LSR} \approx 9.3$,  shows an offset around $6$ km s$^{-1}$ with respect to \citet{Dong2024}.

\subsection*{B.3. Carina}

We divided the Carina cloud complex into North Carina and the Southern Pillars using the $l, b, d$ limits provided in \citet{Dharmawardena2023}. \citet{Berlanas2023} computed the mean proper motions of 7 OB associations in the Carina nebula. By using the mean motion of these associations, we find an overall proper motion from the authors of $\mu_{\alpha^\ast} = -6.71 \pm 0.04$ mas yr$^{-1}$ and $\mu_\delta = 2.37 \pm 0.04$ mas yr$^{-1}$.

These results are in excellent agreement with those obtained in our work, as can be seen in Table \ref{tab:YSOvsOCmotions}. Furthermore, the YSOs and the OCs allowed us to determine the mean RV between -7 and -10 km s$^{-1}$. They are within one standard deviation of the value reported by  \citet{2018MNRAS.477.2068K} (a mean RV of $0.6$ km s$^{-1}$ with a standard deviation of $9.1$ km s$^{-1}$, by using 40 O-type stars as tracers).

\subsection*{B.4. Cassiopeia}

As Fig. \ref{fig:DBSCAN_clusters_complete} shows, the Cassiopeia cloud shows a large dispersion along both proper motion axes. Furthermore, it does not contain any of the open clusters from \citet{Hunt2024} which, together with the absence of YSOs with available RVs in that region, makes it unsuitable for the aims of this work.

However, we highlight that \citet{Lim2023} analysed the kinematic properties of young stars in the massive SFR W5, the major part of the Cassiopeia OB6 association, using the \textit{Gaia} EDR3 data and high-resolution spectra. They identified eight stellar groups centred at the cavities of the giant H II regions, and three sparse groups at the border, and reported systemic proper motions of $-0.273$ and $-0.333$ mas yr$^{-1}$ for $\mu_{\alpha^\ast}$ and $\mu_\delta$ (within one standard deviation of our results), and a RV for the cloud in the range from $-45$ km s$^{-1}$ to $-35$ km s$^{-1}$.

\subsection*{B.5. Cepheus}
\citet{Dzib2018} analysed \textit{Gaia} DR2 data for YSOs in the Cepheus Flare and reported that proper motion dispersions were larger compared to other regions, possibly reflecting the formation of YSOs in different clouds, each with slightly different kinematics.
Therefore, we divided the cloud into two main groups: Cep OB2 and the Cepheus Flare. To that end, we used the $l$ and $b$ limits provided in \citet{Szilagyi2023} and \citet{Szilagyi2021} respectively, and a common $d= 650$ pc limit in the middle point between both substructures.
\citet{Szilagyi2021} used \textit{Gaia} EDR3 data to identify candidate pre-main-sequence stars in the Cepheus Flare region and found most of the subgroups with $\mu_{\alpha^\ast}\in[5.055, 8.041]$ mas yr$^{-1}$, $\mu_{\delta}\in [-1.699, 3.835]$ mas, and $RV \in [-7.6, 3.9]$ km s$^{-1}$; thus, consistent with our results.

\subsection*{B.6. Chamaeleon}

As shown in Fig. \ref{fig:DBSCAN_clusters_complete}, the Chamaeleon cloud shows three clearly distinct clusters in the 2D proper motion space. The presence of three different clouds in the Chamaeleon cloud complex has been studied in \citet{Luhman2008}, namely Chamaeleon I (Cha I), Chamaeleon II (Cha II), and Chamaeleon III. Only Cha I and II are thought to have ongoing star formation, and their structure and kinematics have been further described in \citet{Galli2021}. The proper motion values provided in the latter work are $\mu_\alpha^\ast$ = $-22.507 \pm 0.065$ mas yr$^{-1}$, and $\mu_\delta = 0.566 \pm 0.091$ mas yr$^{-1}$ for Cha I, and $\mu_\alpha^\ast$ = $-20.207 \pm 0.170$ mas yr$^{-1}$, and $\mu_\delta = -7.635 \pm 0.129$ mas yr$^{-1}$ for Cha II. Therefore, they align with our values.

\subsection*{B.7. Corona Australis}

\citet{Galli2020b} provide a census of 313 Class II and III YSOs in the environment of the Corona Australis cloud, 106 of which correspond to the inner part of the cloud. These members allowed the authors to provide mean proper motions of $\mu_{\alpha^\ast} = 4.28 \pm 0.08$ mas yr$^{-1}$ and $\mu_\delta = -27.18 \pm 0.09$ mas yr$^{-1}$, and thus in agreement with our values. 

Regarding the $RV$, the typical value adopted in the literature by \citet{James2006} of $-1.1 \pm 0.5$ km s$^{-1}$, although computed from only a few stars, agrees with our estimate.

\subsection*{B.8. Cygnus}

To compare the OC and YSO motions, we divided this region into North ($l \geq 81^\circ$) and South ($l < 81^\circ$) according to \citet{Dharmawardena2023}.
The proper motions of the Cygnus OB2 (that belongs to the southern region) have been derived in \citet{Orellana2021} by searching for overdensities of proper motions in a 2D $\mu_{\alpha^\ast}-\mu_\delta$ diagram extracted from \textit{Gaia} DR2. The authors obtained values of $\mu_{\alpha^\ast} = -2.71 \pm 0.02$ mas yr$^{-1}$ and $\mu_\delta = -4.24 \pm 0.02$ mas yr$^{-1}$. These values are mainly aligned with those obtained with the YSOs, while the OCs show a slightly larger difference.

The RV obtained in this work shows large dispersions for both OCs and YSOs, and within one standard deviation of the value of $-11 \pm 3$ km s$^{-1}$ reported by \citet{Lim2019} for Cygnus OB2. 

\subsection*{B.9. Lupus}
The Lupus complex has been described as being composed of nine clouds with individual distances varying  between 200 pc and 140 pc \citep{Zucker2019}. \citet{Teixeira2020} combined \textit{Gaia} DR2 proper motions and parallaxes with mid-infrared photometry to identify new disk-bearing YSOs in the Lupus complex.   They used \textit{Gaia} CMDs to derive stellar masses and ages  and found that YSOs with isochronal age less than 4 Myr are very likely part of Lupus, and  reported proper motions values of 
$\mu_{\alpha^\ast} = -10.2 \pm 1.0$ mas yr$^{-1}$ and $\mu_\delta = -23.4 \pm 0.8$ mas yr$^{-1}$.

\citet{Galli2020a} performed a comprehensive kinematic study identifying a sample of 113 high-confidence members of the Lupus association. This research used \textit{Gaia} DR2 proper motions and literature RV data. The mean proper motions reported for the Lupus full sample ($\mu_{\alpha^\ast} = -10.4 \pm 0.1$ mas yr$^{-1}$ and $\mu_\delta = -23.40 \pm 0.07$ mas yr$^{-1}$) are aligned with those in \citet{Teixeira2020}, but in this work the authors also provide mean motions for six of the nine clouds in Lupus (Lupus 1-6), finding a scatter of up to $2$ mas yr$^{-1}$ in $\overline \mu_{\alpha^\ast}$, that is also observed in Fig. \ref{fig:DBSCAN_clusters_complete}. As a result, our proper motions are shown to be in very good agreement with those calculated in these previous works.

Similarly, the mean RV reported by \citet{Galli2020a} for the global cloud is $1.4 \pm 0.5$ km s$^{-1}$, but ranging from $-0.7 \pm 2.2$ in Lupus 1 to $2.1 \pm 0.7$ km s$^{-1}$ in Lupus 4 (with RV data for Lupus 5 and 6 not available). These values are within one standard deviation of the RV calculated in our work using OCs and YSOs.

Since the Lupus subgroups are at the same distance \citep[$\sim 160$ pc;][]{Galli2020a}, we subdivided our YSOs and OCs into the different Lupus clouds using the $(l,b)$ coordinates provided for the nine Lupus clouds in \citet{Tachihara2001} and \citet{2008hsf2.book..295C}.

\subsection*{B.10. MonR2}

A recent study by \citet{Jiang2024} using \textit{Gaia} DR3 analysed the proper motions of YSOs in Monoceros R2 by identifying 959 highly probable MonR2 members with reliable proper motions and parallax measurements. The mean proper motions for these YSOs were found to be 
$\mu_{\alpha^\ast} = -3.16 \pm 0.78$ mas yr$^{-1}$ and $\mu_\delta = 0.62 \pm 0.53$ mas yr$^{-1}$. We found our values to be consistent with those results, and with the proper motions computed from \citet{Kuhn2019} using members of clusters and other associations. However, the lack of spectroscopic data for the members of these works and of our sample prevent us from providing a mean RV.

\subsection*{B.11. Ophiuchus}

\citet{Ducourant2017} performed a kinematic analysis of the L1688 dark cloud, the main cloud in the Ophiuchus cloud complex, by compiling a list of 68 YSOs and 14 YSO candidates. The authors used them to derive a mean proper motion for the cloud of $\mu_{\alpha^\ast} = -8.2 \pm 0.8$~mas~yr$^{-1}$ and $\mu_\delta =-24.3 \pm 0.8$~mas~yr$^{-1}$. Furthermore, \citet{Rivera2015} calculated the RV of four sources in Ophiuchus and determined a weighted mean of $\rm{RV} = -6.6 \pm 0.5$ km s$^{-1}$. As Table \ref{tab:YSOvsOCmotions} shows, both proper motions and $RV$ are in excellent agreement with the values provided in this work.

\subsection*{B.12. Orion}

The kinematics of the Orion cloud are one of the most studied in the literature \citep{Kounkel2018, Grosschedl2021, Sanchez-Sanjuan2024}, with a wide diversity of proper motions and RVs depending on which substructure is studied.

Therefore, to perform our analysis, we use the $(l,b)$ contours provided in \citet{Lombardi2011} for three main subregions: Orion A, Orion B, and $\lambda$ Orionis. As a result, the values obtained with the two tracers agree within one standard deviation. They also match the mean values computed by \citet{Grosschedl2021} for Orion A ($\mu_{\alpha^\ast} \approx 0.8$ mas yr$^{-1}$, $\mu_\delta \approx -0.5$ mas yr$^{-1}$, $RV\approx 24$ km s$^{-1}$), and Orion B ($\mu_{\alpha^\ast}\approx 1.6$ mas yr$^{-1}$, $\mu_\delta \approx -0.5$ mas yr$^{-1}$, $RV\approx 25$ km s$^{-1}$); and from \citet{Sanchez-Sanjuan2024} for $\lambda$ Orionis ($\mu_{\alpha^\ast} \approx 1.3$ mas yr$^{-1}$, $\mu_\delta \approx -2.1$ mas yr$^{-1}$, $RV\approx 27$ km s$^{-1}$).

\subsection*{B.13. Perseus}
The Perseus molecular cloud is a prominent star-forming region containing several young stellar clusters. A study using data from \textit{Gaia} and other surveys \citep{Ortiz-Leon2018} has provided insights into the proper motions and RVs of YSOs within the two main subregions in this cloud: IC 348 and NGC 1333. They found the following values for IC 348: $\mu_{\alpha^\ast} = 4.35 \pm 0.03$ mas yr$^{-1}$ and $\mu_\delta = -6.76 \pm 0.01$ mas yr$^{-1}$; and for NGC 1333: $\mu_{\alpha^\ast} = 7.34 \pm 0.05$ mas yr$^{-1}$ and $\mu_\delta = -9.90 \pm 0.03$ mas yr$^{-1}$. 

Indeed, the fact that the Perseus cloud is more complex than a single comoving structure can be seen in Fig. \ref{fig:DBSCAN_clusters}. Therefore, we used the limits $(l,b)$ of each subregion provided by \citet{Dharmawardena2023} to verify that the results of the YSOs and the OC restricted to these subregions align very well with those of \citet{Ortiz-Leon2018}.

The same study published mean Galactic ($U$, $V$, $W$) velocities for these subregions. Using the centroid coordinates $(l, b)$ of each of them, we estimated that the $RV$s found in their work for IC 348 and NGC 1333 are $\sim 16$ km s$^{-1}$ and $\sim 15$ km s$^{-1}$, respectively. Therefore, they are in excellent agreement with our mean $RV$, as shown in Table \ref{tab:YSOvsOCmotions}.

\subsection*{B.14. Rosette}

The kinematics of stellar groups in the Rosette nebula have been recently studied by \citet{Lim2021} and \citet{Muzic2022}. Both studies paint a coherent picture of a star-forming region around the young ($\leq 2$ Myr), expanding, and likely unbound cluster NGC 2244.
\citet{Lim2021} used the \textit{Gaia} astrometry of 403 OB stars, disk-bearing YSOs, and X-ray sources to trace the proper motion of NGC 2244. The values provided in their work are $\overline \mu_{\alpha^\ast} = -1.731$ mas yr$^{-1}$ and $\overline \mu_{\delta} = 0.312$ mas yr$^{-1}$. On the other hand, \citet{Muzic2022} used 688 X-ray sources and infrared excess candidates to provide a mean $\overline \mu_{\alpha^\ast} = -1.75 \pm 0.41$ mas yr$^{-1}$, and $\overline \mu_\delta = 0.25 \pm 0.44$ mas yr$^{-1}$. As can be seen, in both cases, the proper motions of those authors match those provided here. 

{We note, however, that the NGC 2244 OC is not included in our sample, since it is located at a distance of 1415~pc \citep{Hunt2024}, while the distance boundaries of Rosette as provided in \citet{Dharmawardena2023} go from 1150 to 1329~pc.}

On the other hand, \citet{Lim2021} found a systemic $RV_{\rm LSR} = 13.4 \pm 2.7$ km s$^{-1}$. Consequently, our $RV$ value of $27 \pm 2$ km s$^{-1}$, after being transformed to the LSR ($RV_{\rm LSR} \approx 11.3$ km s$^{-1}$), agrees with it.

\subsection*{B.15. Taurus}

Recent work has clarified the kinematics of the Taurus Molecular Cloud using \textit{Gaia} DR3 and VLBA astrometry. For instance, \citet{Galli2019} found, using \textit{Gaia} DR2 and VLBI astrometry for 519 Taurus members, mean proper motion components of $\mu_{\alpha^\ast} = 8.1$ mas yr$^{-1}$, and $\mu_\delta = -21.0$ mas yr$^{-1}$, with $RV = 16.15$ km s$^{-1}$, and thus in excellent consonance with the values reported in our work. More recent work based on \textit{Gaia} DR3 \citep{Luhman2022} further identifies about 13 kinematic subgroups; while median values per group are provided, typical values cluster around $\mu_{\alpha^\ast} = 5 \text{ to }14$ mas yr$^{-1}$ and  $\mu_\delta = -12 \text{ to }-26$ mas yr$^{-1}$, thus agreeing with the previous one.

\end{appendix}

\end{document}